**Breakdown of the zeroth law of thermodynamics as a consequence of diffraction of diffuse radiation at multidimensional regular structures**

V.V. Savukov [*]

[*] D. I. Ustinov Voenmekh Baltic State Technical University, St. Petersburg

This article is devoted to the complete results of an experimental test of the theoretical assumption that the basic axiomatic postulate of statistical physics according to which it is equally probable for a closed system to reside in any of the microstates accessible to it may be invalid for nonergodic cases. In the course of photometric experiments for the purpose of recording the predicted loss of isotropy by a diffuse light field when it came into contact with a two-dimensional phase-type diffraction grating, a significant deviation from Lambert's law was detected when the diffuse photon gas was scattered by the grating surface. This caused angular anisotropy of the radiation fluxes to appear in the initially homogeneous light field. These results provide a basis for revising the determination of the most probable macrostate of a closed system.

## РЕФЕРАТ

Статья посвящена полному описанию результатов экспериментальной проверки теоретического предположения, согласно которому базовый аксиоматический постулат статистической физики о равновероятности пребывания замкнутой системы в любом из доступных ей микросостояний может быть несправедлив для неэргодических случаев. В ходе фотометрических экспериментов, призванных зафиксировать прогнозируемую потерю изотропности диффузным световым полем вследствие его контакта с двумерной дифракционной решёткой фазового типа, выявилось значимое отклонение от закона Ламберта при рассеянии диффузного фотонного газа поверхностью решётки. Это приводило к появлению угловой анизотропии потоков излучения в изначально однородном световом поле. Полученные результаты дают основание для ревизии определения наиболее вероятного макросостояния замкнутой системы [10].

**Ключевые слова:** дифракция, Ламберт, индикатриса, рассеяние, диффузный.

## Breakdown of the zeroth law of thermodynamics as a consequence of diffraction of diffuse radiation at multidimensional regular structures

Vladimir V. Savukov

## ABSTRACT

This article is devoted to the complete results of an experimental test of the theoretical assumption that the basic axiomatic postulate of statistical physics according to which it is equally probable for a closed system to reside in any of the microstates accessible to it may be invalid for nonergodic cases. In the course of photometric experiments for the purpose of recording the predicted loss of isotropy by a diffuse light field when it came into contact with a two-dimensional phase-type diffraction grating, a significant deviation from Lambert's law was detected when the diffuse photon gas was scattered by the grating surface. This caused angular anisotropy of the radiation fluxes to appear in the initially homogeneous light field. These results provide a basis for revising the determination of the most probable macrostate of a closed system [10].

**Key words:** diffraction, Lambert, indicatrix, scattering, diffuse.

**OCIS:** 000.2658, 000.6590, 050.1940, 290.0290, 290.5820, 290.5825

**PACS:** 05.90.+m ; 42.25.Fx ; 42.25.Hz

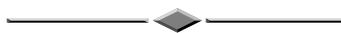

---

[1] URL: http://www.savukov.ru/viniti_0507_b2009_full_rus.pdf

# Введение

Статистическая физика равновесных систем базируется на постулате о равновероятности всех микросостояний, доступных замкнутой системе. Данный постулат имеет характер априорно декларируемой правдоподобной гипотезы, принятие которой позволяет утверждать, что равновесное состояние у исследуемой системы может быть только одно, причём жёстко регламентируются и параметры этого единственного состояния. В термодинамике функциональным аналогом данного постулата служит **«Нулевое начало»**.

Ещё в 1913 г. Геде [1], а затем и другими [2, 3] исследователями впервые было строго доказано, что диффузный газ[1] может находиться в состоянии динамического равновесия с некоторой отражающей поверхностью, если только среднеинтегральная вероятностная характеристика угловых направлений движения частиц уже рассеянного этой поверхностью газа[2] описывается законом Ламберта[3] или законом Кнудсена[4] [4, 5]. Данные законы имеют тождественную математическую форму и характеризуют идентичные параметры рассеяния частиц различной природы на некоторой поверхности:

2D-системы: $\quad f(\beta) = \cos(\beta) > 0 \quad$ для $\quad \beta \in [0, \pi/2],\ \varphi = Const$;  (1)

3D-системы: $\quad f(\beta) = \sin(2\beta) \geq 0 \quad$ для $\quad \beta \in [0, \pi/2],\ d\varphi \in (0, 2\pi]$.  (2)

где $f(\beta)$ – плотность вероятности того, что частица покидает рассеивающую поверхность под углом отражения $\beta \in [0, \pi/2]$; $\varphi$ – азимутальный угол рассеяния частицы: для двумерного (2D) случая этот угол фиксирован, для трёхмерных (3D) моделей рассеяния он интегрально обобщён для конечного сектора азимутальных углов $d\varphi > 0$.

Таким образом, согласно постулатам статистической физики и термодинамики, для диффузного фотонного газа[5] любого спектрального состава[6] в ходе равновесного рассеяния его частиц на каком-либо объекте – обязан выполняться закон Ламберта[7].

---

[1] Для каждой частицы диффузного газа — равновероятна любая угловая ориентация вектора её импульса в геометрическом пространстве (изотропность импульсов).

[2] В данной работе рассматриваются *изоэнергетические* процессы рассеяния. Поэтому релаксация импульса частицы газа сводится лишь к изменению угловой направленности вектора этого импульса, скалярная величина которого остаётся постоянной.

[3] Такое название используют в оптике, когда рассматривают взаимодействие потока фотонов электромагнитного излучения с какой-либо поверхностью.

[4] Это название обычно применяют в молекулярной динамике или в иных случаях, когда частицы изучаемого газа имеют ненулевую массу покоя.

[5] Под фотонным диффузным газом в настоящей работе понимается неполяризованное некогерентное электромагнитное излучение, для отдельных фотонов которого с равной вероятностью реализуется любая возможная угловая ориентация в трёхмерном геометрическом пространстве их волновых **k** - векторов.

[6] Это очевидно, так как принцип детального равновесия предполагает сохранение свойства изотропности для каждой отдельно взятой спектральной компоненты излучения.

[7] Напротив, если рассеиванию подвергается фотонный газ с изначально неизотропными свойствами, то в такой ситуации нарушение закона Ламберта встречается даже чаще, чем его соблюдение (скажем, при зеркальном отражении анизотропного излучения).

## Цель работы

Целью работы является прямая экспериментальная проверка ранее высказанного автором теоретического предположения [6], согласно которому базовый аксиоматический постулат статистической физики может быть несправедлив для неэргодического случая дифракционного рассеяния[1] диффузного (изотропного) газа квантовых частиц[2] на поверхности специального дифракционного оптического элемента (ДОЭ). В частности, прогнозировалась потеря изотропности диффузным световым полем вследствие его контакта с двумерной дифракционной решёткой фазового типа. Это означало бы наличие изоэнергетических процессов переноса, целенаправленно искажающих заранее "приготовленное" равномерное (т. е. соответствующего равновесному) распределение состояний отдельных фотонов внутри доступных им областей фазового пространства.

## Методология проведения работы

Вышеуказанная проверка выполнялась в ходе фотометрических экспериментов, призванных зафиксировать прогнозируемую потерю изотропности диффузным световым полем вследствие его контакта с двумерной фазовой дифракционной решёткой.

В общих чертах процесс осуществления каждого из экспериментов выглядел так: дифракционный оптический элемент (ДОЭ) с цилиндрической формой поверхности помещался в центральную часть объёма фотометрической камеры. Внутри этой камеры создавалось диффузное световое поле со спектральным составом видимого диапазона. Через малое отверстие в корпусе фотометрической камеры выполнялось фотографирование ДОЭ на фоне её внутренней стенки. Зафиксированная картина предполагаемого искажения изначально диффузного излучения поверхностью ДОЭ обрабатывалась необходимыми методами дисперсионного и регрессионного анализа (фотографическая фотометрия), выявляющими наличие или отсутствие ожидаемых явлений анизотропии.

## Описание экспериментальных установок и их элементов

С целью контроля воспроизводимости получаемых результатов, работы независимо проводились на трёх экспериментальных установках, которые отличались друг от друга конструкциями фотометрических камер, источниками первичного излучения и методами его стохастизации, а также параметрами регистрирующей фотоаппаратуры:

– **Экспериментальная установка №1:** выполнена на основе цилиндрической фотометрической камеры и внешнего локального источника первичного светового излучения.

– **Экспериментальная установка №2:** выполнена на основе цилиндрической камеры и особого внутреннего источника уже стохастизированного светового излучения.

– **Экспериментальная установка №3:** имеет сферическую камеру и внутренний источник.

---

[1] Это становится принципиально возможным при наличии не равной нулю дивергенции потока фазовых траекторий в некоторых частях фазового пространства системы.

[2] Нарушения указанной аксиоматики могут быть характерны для некоторых видов квантовых систем в таких предметных областях, как физическая оптика, твёрдотельная электроника и молекулярная динамика, когда имеет место упругое волновое рассеяние (дифракция) газа квантовых частиц (фотонов электромагнитного излучения, электронов проводимости или молекул газа) на границах заполняемого ими объёма.

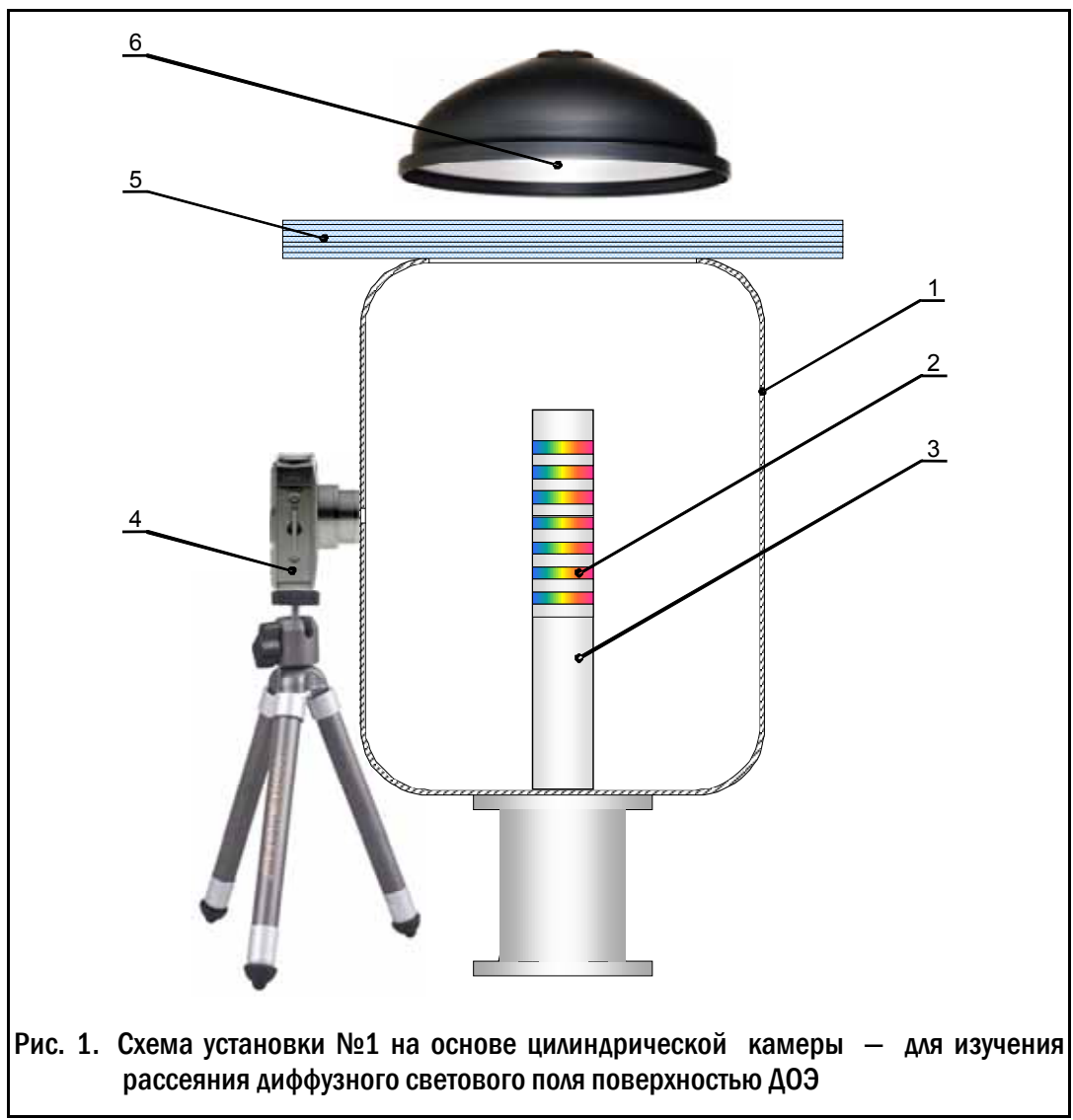

**Рис. 1. Схема установки №1 на основе цилиндрической камеры — для изучения рассеяния диффузного светового поля поверхностью ДОЭ**

**Описание элементов экспериментальной установки №1 на основе цилиндрической камеры**

1. Цилиндрическая фотометрическая камера, внутренняя поверхность которой имеет покрытие, обеспечивающее диффузное рассеивание вводимого в камеру излучения.

2. Гибкая реплика дифракционной решётки **(ДОЭ)**, закреплённая на опорном цилиндре 3.

3. Опорный цилиндр, служащий для придания нужной макроскопической формы гибкой реплике дифракционной решётки 2 (ДОЭ), фиксируемой на его поверхности.

4. Фотографический аппарат. Предназначен для получения снимков образца ДОЭ на фоне излучения, диффузно рассеиваемого внутренней поверхностью камеры 1.

5. Рассеивающий фильтр из стопки молочных акриловых листов[1] со специально матированной поверхностью. Суммарная толщина всего фильтра – от 15 до 80 мм.

6. Внешний источник первичного (не стохастизированного) светового излучения.

---

[1] Акрил ТУ2216-031-55856863-2004 марки **«Acryma® 72 O»** (Opal, "молочное"); размер одного листа: 400×400×5 мм; для света видимого спектра коэфф. пропускания: $\cong 23\%$.

**Описание фотометрических камер и их отличий   –   для всех экспериментальных установок**

1. **Фотометрическая камера №1 цилиндрической формы,** излучение в которую вводится от внешнего источника – через массивный фильтр из молочных акриловых плит с матированной поверхностью, размещённый на открытом торце камеры. Такая конструкция практически не накладывает ограничений на размер, тип и мощность источника излучения, что удобно при варьировании различных видов этих источников:
   – Форма: цилиндрическая (рабочее положение оси симметрии – вертикальное).
   – Материал и толщина стенок: алюминий, 2 мм.
   – Диаметр камеры (внутренний): 260 мм.
   – Длина вдоль оси симметрии (внешняя): 185 мм.
   – Диаметр отверстия для объектива фотоаппарата: 5.0 мм ($\cong 0.00768\%$ площади всей внутренней поверхности камеры); центр отверстия для объектива отстоит от нижней поверхности камеры на 85 мм.
   – Материал покрытия внутренней поверхности камеры: ярко-белая матовая краска на акриловой основе. Имеет усреднённое для видимой области значение коэффициента спектрального отражения $\approx 0.97$; блеск по ISO 2813, 85° – не более 10%.

2. **Фотометрическая камера №2 цилиндрической формы.** Источник светового излучения в ней – равномерно распределён по всей внутренней поверхности:
   – Форма: цилиндрическая (рабочее положение оси симметрии – вертикальное).
   – Материал и толщина стенок: алюминий, 2 мм.
   – Диаметр камеры (внутренний): 220 мм.
   – Длина вдоль оси симметрии (внешняя): 155 мм.
   – Диаметр отверстия для объектива фотоаппарата: 6.5 мм ($\cong 0.01826\%$ площади всей внутренней поверхности камеры); центр отверстия для объектива отстоит от нижней поверхности камеры на 85 мм.
   – Покрытие внутренней поверхности камеры: светорассеивающая подложка из матовой ярко-белой краски с коэффициентом диффузного отражения для видимого спектра $\approx 0.97$, покрытая слоем твёрдого коллоидного $\simeq 10\%$ раствора порошка самосветящегося люминофора марки **«Пента Л-1»** в прозрачном матовом лаке.

Перед началом каждого эксперимента особое светоаккумулирующее покрытие боковых и торцевых стенок камеры (см. описание внутреннего покрытия) длительно облучалось люминесцентной лампой "накачки", после чего внутренняя поверхность камеры приблизительно восемь часов испускала синий видимый свет с доминирующей длиной волны около $\lambda \approx 420$ нм. В течение первых 30 минут яркости этого света было достаточно для выполнения любых необходимых фотометрических работ.

Несмотря на громоздкий характер подготовки каждого эксперимента и фиксированный спектр получаемого света, изотропность излучения, формируемого в этой камере №2, была практически идеальной[1]. Пылинки мелкодисперсного люминофора, внедрённые в объём прозрачного матового лака, представляли собой огромное множество элементарных источников света, равномерно распределённых по всей внутренней поверхности камеры. А низкая концентрация частиц люминофора в составе коллоидной "взвеси" не мешала дополнительной стохастизации испускаемого этими частицами излучения матово-белой подложкой описанного двухслойного покрытия.

---

[1] Традиционные методы формирования диффузного излучения, основанные на использовании локализованных в пространстве макроскопических источников света, уступали описанному по эффективности даже при использовании такого "классического" покрытия, как сульфат бария $BaSO_4$ ( $\rho_\lambda \approx 0.987 \div 0.992$ для $\lambda \approx 0.400 \div 0.750$ мкм ).

3. **Фотометрическая камера №3 сферической формы.** Это "классическая" по своей форме камера из числа тех, которые обычно и применяют для создания диффузных световых полей. Источник светового излучения расположен непосредственно в центре камеры, – так, чтобы образец ДОЭ "загораживал" этот источник от фотоаппарата:

   – Форма камеры: сферическая.
   – Диаметр камеры (внутренний): 400 мм.
   – Диаметры отверстий для объектива фотоаппарата: 32, 16 и 8 мм (соответственно 0.16000%, 0.04000% и 0.01000% площади всей внутренней поверхности камеры).
   – Материал покрытия внутренней поверхности камеры: ярко-белая матовая краска на акриловой основе. Имеет усреднённое для видимой области значение коэффициента спектрального отражения $\approx 0.97$; блеск по ISO 2813, 85º – не более 10%.

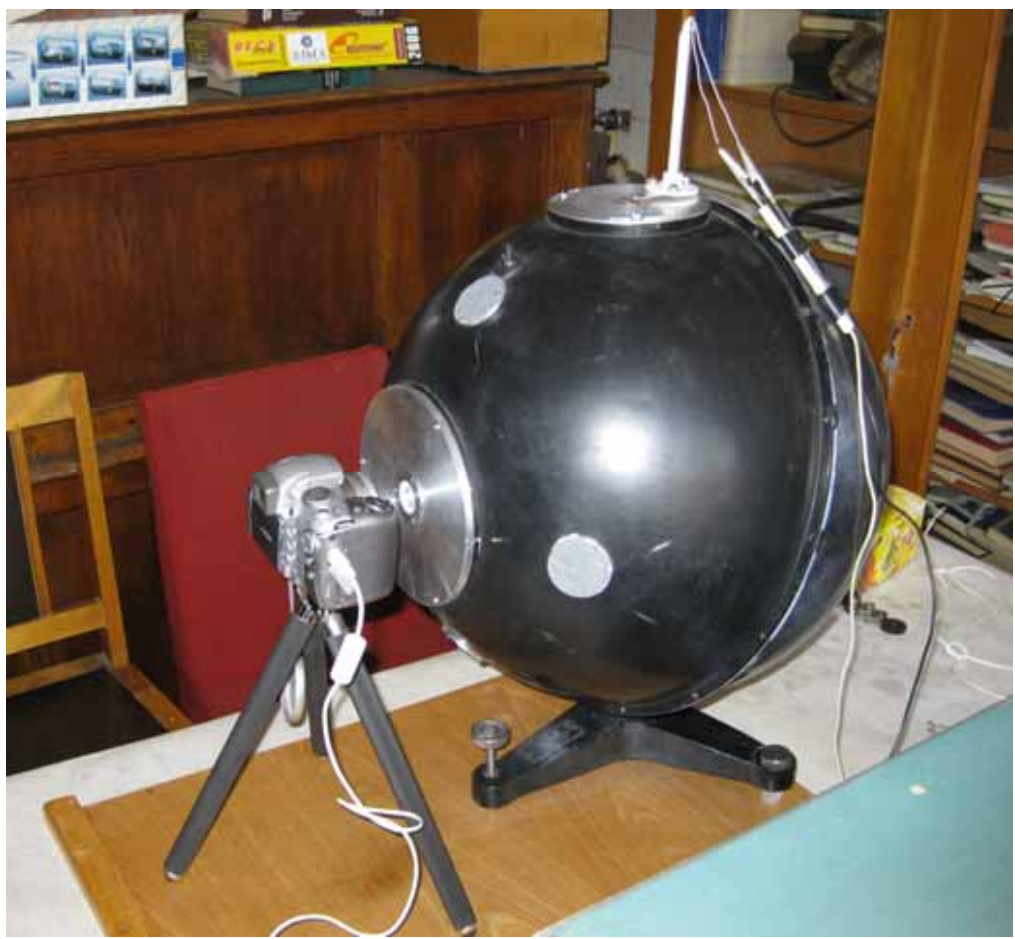

**Рис. 2.** Внешний вид установки для изучения рассеяния диффузного светового поля поверхностью ДОЭ — на основе фотометрической камеры №3

## Первичные источники электромагнитного излучения (свет видимого диапазона)

1. В составе установки №1 использовались следующие первичные источники света:

   – Галогенная лампа накаливания мощностью 35 Вт с непрерывным ("планковским") спектром излучения (цветовая температура $T \approx 3000°K$). Установлена (цоколь **G4**) в светильнике марки **Camelion KD-321** с рефлектором особой овальной формы.

   – Люминесцентная газоразрядная лампа мощностью 20 Вт с линейчатым ("трёхполосным") спектром "тёплого" излучения, – цветовая температура $T \approx 2700°K$. Установлена (цоколь **E27**) в светильнике **Camelion KD-313** с круглым рефлектором.

   – Люминесцентная газоразрядная лампа мощностью 2×9 Вт с линейчатым ("трёхполосным") спектром "холодного" излучения, – цветовая температура $T \approx 4200°K$. Установлена (цоколь **G23**) в светильнике **Camelion KD-022** с плоским рефлектором.

   – Люминесцентная газоразрядная лампа мощностью 24 Вт с линейчатым ("трёхполосным") спектром "холодного" излучения, – цветовая температура $T \approx 6700°K$. Установлена (цоколь **E27**) в светильнике **Camelion KD-313** с круглым рефлектором.

   – Светодиодная лампа мощностью 1 Вт с узким спектром квазимонохромного излучения, соответствующего длине волны $\lambda \approx 700 \pm 5$ нм (красный цвет). Установлена (цоколь **E27**) в светильнике марки **Camelion KD-313** с круглым рефлектором.

2. В составе установки №2 использовался особый источник света на базе люминофора **«Пента Л-1»**, который был подробно описан в подразделе **Фотометрическая камера №2**.

3. В составе установки №3 использовалась бесцокольная миниатюрная галогенная лампа накаливания **«КГМ-12В»** капсульного типа мощностью 40 Вт, с непрерывным ("планковским") спектром излучения, – цветовая температура около $T \approx 2700°K$.

## Образцы дифракционных оптических элементов (ДОЭ)

Исследования проводились на шести образцах цилиндрической формы, рабочие рассеивающие поверхности которых представляли собой ДОЭ, изготовленные в виде гибких реплик дифракционной решётки нижеописанного вида (см. рис. 3):

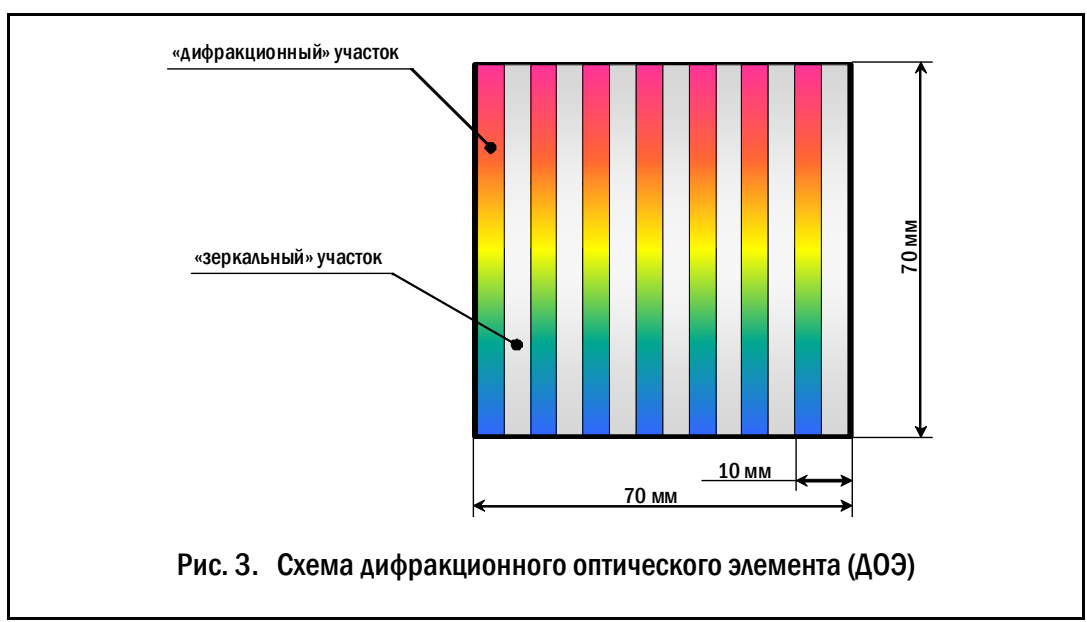

Рис. 3. Схема дифракционного оптического элемента (ДОЭ)

Рабочая часть[1] ДОЭ имеет квадратную форму с размерами сторон 70×70 мм. Лист ДОЭ представляет собой выполненную с матрицы гибкую реплику, допускающую равномерный (круговой) изгиб вокруг горизонтальной оси на угол до 360º. Поверхность отдельного ДОЭ состоит из 14 или 7 равновеликих прямоугольных полос шириной, соответственно, 5 мм или 10 мм. Все полосы имеют отражающее покрытие из алюминия[2]. На половине этих полос сформирован микрорельеф с заданными параметрами («дифракционные» участки, представляющие собой фазовую решётку), а разделяющие их полосы обладают зеркально гладкой поверхностью (см. рис. 3 на стр. 8).

Благодаря наличию в составе ДОЭ чередующихся (смежных) дифракционных и зеркальных участков – появляется объективная возможность контроля однородности исходного светового поля. Действительно, такие смежные участки ДОЭ одномоментно находятся в идентичных условиях освещения. Поэтому отсутствие изначальной изотропности светового поля станет очевидным по его отражению от зеркальных участков.

Геометрия микрорельефа дифракционных участков ДОЭ имеет квазисинусоидальный профиль, характеризуемый стандартными для синусоидальной функции пропорциями между шагом $S_o$ (периодом) и амплитудой вариации высоты $A$:

шаг решётки: $S_o \approx 417, 833, 1667$ и $10000$ нм, размах амплитуды: $A = \pm \dfrac{S_o}{2\pi}$ ;

Размерность геометрии микрорельефа дифракционных участков отдельных ДОЭ была либо одномерной (линейной), либо двумерной (ортогональная перекрёстная решётка с пересечением штрихов под взаимным углом 90º), – см. рис. 4. По отношению к сторонам квадратной рабочей области ДОЭ линии штрихов разных дифракционных решёток были ориентированы ("наклонены") либо под углом 45º, либо под углом 90º.

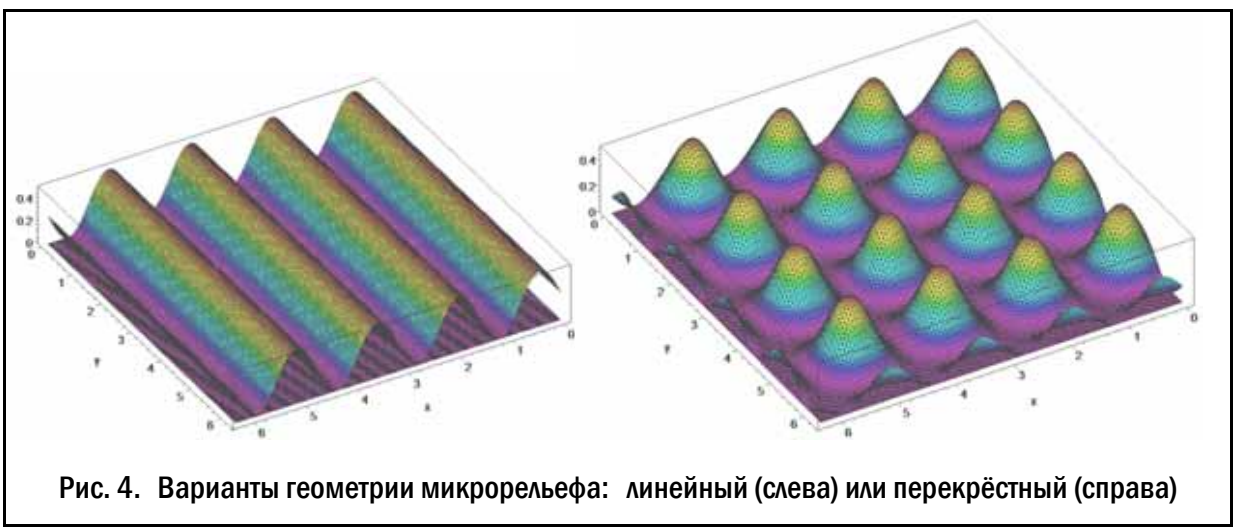

**Рис. 4.** Варианты геометрии микрорельефа: линейный (слева) или перекрёстный (справа)

В таблице 1, приведённой далее, изложена сводная информация о параметрах тех образцов дифракционных оптических элементов[3], которые использовались в работе.

---

[1] Размеры этого листа с репликой превышали размеры рабочей части ДОЭ на 15-30 мм.

[2] Какие-либо защитные прозрачные покрытия поверх плёнки Al – не применялись.

[3] Изготовление гибких реплик ДОЭ было осуществлено фирмой "ХолоГрэйт" (Holograte™). Первый образец ("D-833/90") был любезно предоставлен на безвозмездной основе.

| Таблица 1. Варьируемые параметры дифракционных оптических элементов (ДОЭ) | | | | | | |
|---|---|---|---|---|---|---|
| Номер п/п | Наименование реплики ДОЭ | Размерность микрорельефа | Наклон штрихов | Шаг $S_o$ (нм) | Количество линий/мм | Ширина и число полос в ДОЭ |
| 01 | D-833/90 | двумерная | 90° | ≈ 833 | 1200 | 7 полос по 10 мм |
| 02 | D-417/45 | двумерная | 45° | ≈ 417 | 2400 | 14 полос по 5 мм |
| 03 | D-833/45 | двумерная | 45° | ≈ 833 | 1200 | 14 полос по 5 мм |
| 04 | M-833/45 | одномерная | 45° | ≈ 833 | 1200 | 14 полос по 5 мм |
| 05 | D-1667/45 | двумерная | 45° | ≈ 1667 | 600 | 14 полос по 5 мм |
| 06 | D-10000/45 | двумерная | 45° | 10 000 | 100 | 14 полос по 5 мм |
| Примечание: все гибкие реплики ДОЭ выполнены на основе полипропиленовой плёнки толщиной 50 мкм | | | | | | |

### Опорные цилиндры для фиксации дифракционных оптических элементов (ДОЭ)

Опорный цилиндр обеспечивает фиксацию гибкой реплики ДОЭ в определённом месте внутренней рабочей полости установки. Наличие цилиндра должно в минимальной степени влиять на общую картину распределения диффузного излучения внутри фотометрической камеры. Поэтому оптимальным решением были признан массивный металлический цилиндр с зеркальной боковой и матово-белой торцевой поверхностью.

Каждый дифракционный оптический элемент закреплялся на опорном цилиндре таким образом, чтобы при вертикальном положении цилиндра дифракционные и зеркальные полоски ДОЭ были горизонтальны (см. на рис. 5 примеры этой конструкции):

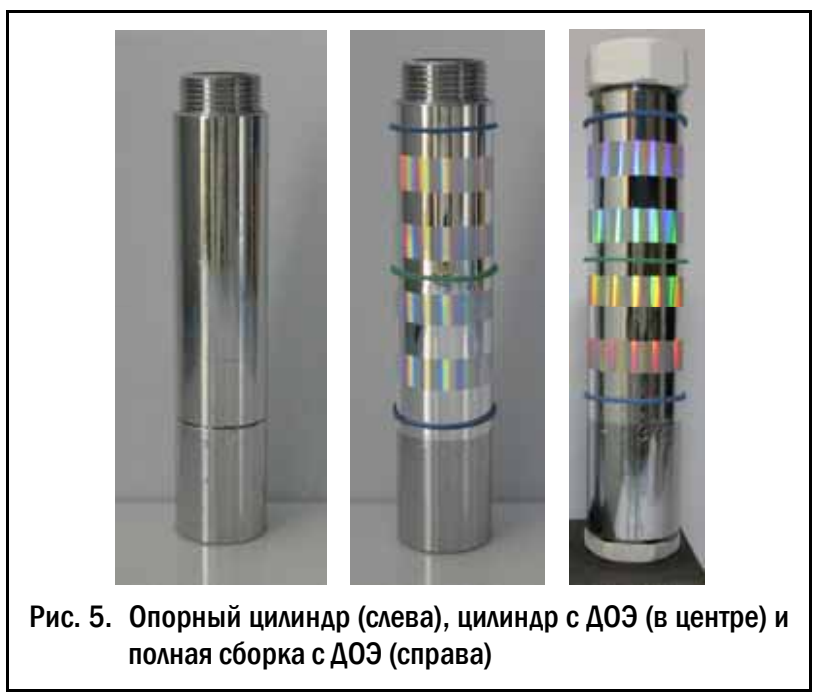

**Рис. 5.** Опорный цилиндр (слева), цилиндр с ДОЭ (в центре) и полная сборка с ДОЭ (справа)

Следует обратить внимание на то, что закрепление гибкой реплики ДОЭ на опорном цилиндре с помощью тонких резиновых колец позволяет в дальнейшем размонтировать собранную конструкцию без повреждения ДОЭ. Кроме того, наличие данных колец в поле зрения – облегчает работу системы автофокусировки при фотосъёмке.

Габариты ДОЭ (см. выше) определили размеры опорного цилиндра следующим образом: диаметр 30 мм; длина участка, укрываемого гибкой репликой ДОЭ, составила 100 мм. Длина свободной от ДОЭ части цилиндра – могла по мере необходимости варьироваться в зависимости от его положения внутри фотометрической камеры.

### Регистрирующая аппаратура

В состав экспериментальных установок №1 и №2 (с цилиндрическими камерами) входил ультракомпактный цифровой фотоаппарат фирмы Canon™ **«Digital Ixus 90 IS»** со следующими параметрами, существенными при выполнении настоящей работы:

– 1/2.3-дюймовый CCD-датчик изображения: $\approx 10 \cdot 10^6$ эффективных пикселей.

– Максимальное разрешение кадра: 3648×2736 пикселей.

– Эквивалентная чувствительность ISO: 80, 100, 200, 400, 800, 1600, 3200.

– Тип цветового фильтра: шаблон Байера по "основным цветам" (sRGB).

– Выдержка: от 1/1500 до 15 сек.

– Максимальное оптическое увеличение объектива (ZOOM): 3.0×

– Минимальное расстояние фокусировки при макросъёмке: 3 см от края объектива.

– Формат записи и хранения результирующего кадра: JPEG.

В состав экспериментальной установки №3 (со сферической камерой) входил компактный цифровой фотоаппарат фирмы Canon™ **«PowerShot S2 IS»**, который имел следующие параметры, существенные при выполнении настоящей работы:

– 1/2.5-дюймовый CCD-датчик изображения: $\approx 5 \cdot 10^6$ эффективных пикселей.

– Максимальное разрешение кадра: 2592×1944 пикселей.

– Эквивалентная чувствительность ISO: 50, 100, 200, 400.

– Тип цветового фильтра: шаблон Байера по "основным цветам" (sRGB).

– Выдержка: от 1/3200 до 15 сек.

– Максимальное оптическое увеличение объектива (ZOOM): 12.0×

– Минимальное расстояние фокусировки при макросъёмке: 0 см от края объектива.

– Формат записи и хранения результирующего кадра: JPEG.

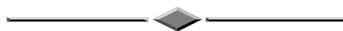

**ПРИМЕЧАНИЕ:** Все работы, связанные с экспериментальными установками №1 и №2, были осуществлены автором настоящей статьи. Работы на установке №3 выполнялись в заказном порядке в **ФГУП НПК «ГОИ им. С. И. Вавилова»**.

# Результаты работы

## Проверка изотропности исходного светового поля и отражающих свойств ДОЭ

Проверка изотропности исходного светового поля производилась на основании дисперсионного анализа фотоснимков дифракционных оптических элементов, размещаемых внутри тестируемых фотометрических камер в разных угловых положениях.

Для примера на рис. 6 приведена фотография «инфракрасного» дифракционного оптического элемента[1] "D-10000/45" в диффузном световом поле. В правой части даны различные спектральные изображения одного из сегментов боковой поверхности ДОЭ[2], а именно - той его "зеркальной полоски" (см. рис. 3), которая ближе всего находится к оптической оси объектива фотоаппарата. Пикселы изображения сегмента равномерно распределены по узким участкам, ориентированным вдоль образующих цилиндра. Угол наблюдения каждого участка – можно считать неизменным для всей его поверхности.

| Исходная фотография ДОЭ | Выбранный для анализа сегмент изображения |
|---|---|
| 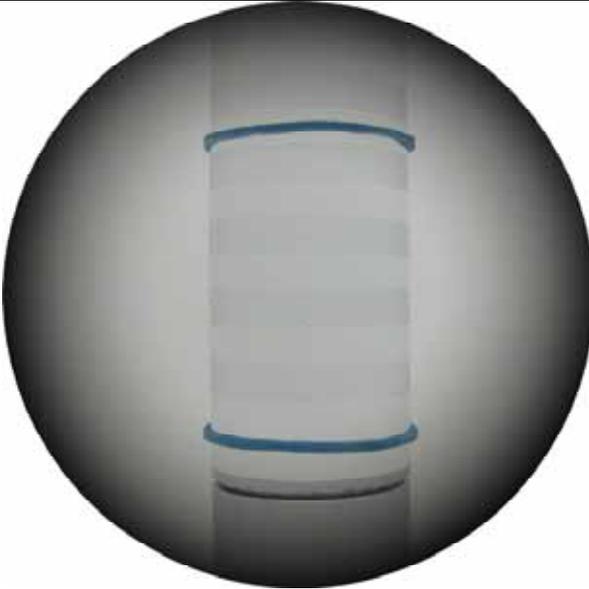 | **Исходная цветовая гамма** 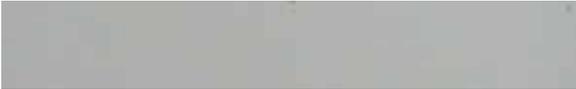 |
| | «Красный» участок спектра ($\lambda \approx 700 \pm 20$ нм) 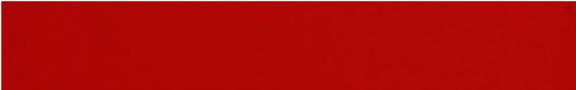 |
| | «Зелёный» участок спектра ($\lambda \approx 530 \pm 20$ нм) 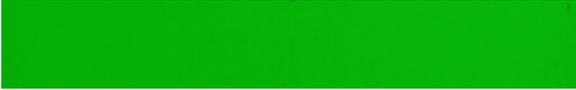 |
| | «Синий» участок спектра ($\lambda \approx 420 \pm 20$ нм) 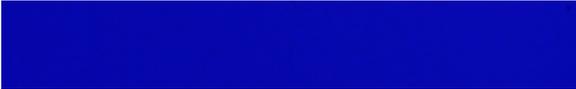 |

Рис. 6. Фотографические изображения дифракционного оптического элемента "D-10000/45" и анализируемого сегмента его поверхности, – в диффузном световом поле камеры №1. Источник излучения: люминесцентная лампа с цветовой температурой $T \approx 4200°K$

---

[1] Шаг решётки этого ДОЭ (≈10 мкм) слишком велик для того, чтобы прогнозируемые эффекты анизотропии проявили себя в видимой части спектра ($\lambda \approx 380 \div 760$ нм).

[2] Разрешающая способность выявления относительного перепада яркости у точек изображения, видимых одновременно (в одном кадре), приблизительно на 2-3 порядка выше, чем возможность фиксирования изменений в абсолютных значениях яркости различных (находящихся в отдельных кадрах) объектов. Поэтому эффективность анализа изображения цилиндрического ДОЭ, элементы поверхности которого одновременно видны под разными углами, более высока, чем, например, в том случае, когда сравниваются отдельные снимки по-разному ориентированного плоского образца.

Формула угла наблюдения $\beta$ каждого из участков разбиения анализируемого сегмента:

$$\beta = \arcsin\left(\frac{L/R_C}{\sqrt{1 + \frac{R^2}{r^2}\cdot\left(\frac{L^2}{R_C^2} - 1\right)}}\right) \qquad (3)$$

где: $R_C$ - радиус опорного цилиндра ДОЭ, ось симметрии которого перпендикулярна оптической оси объектива фотоаппарата и находится на расстоянии $L$ от края этого объектива; $R$ - визуальный размер радиуса опорного цилиндра, измеренный на плоскости фотографии, $r$ - аналогичное визуальное расстояние на фотографическом изображении от оси симметрии цилиндра до наблюдаемой точки боковой поверхности ДОЭ.

Для случая $L \gg R_C$ приемлема упрощённая формула:

$$\beta \approx \arcsin\left(\frac{r}{R}\right) \quad \text{при} \quad L \gg R_C \qquad (4)$$

На рис. 7 для зеркального сегмента боковой поверхности ДОЭ "D-10000/45" представлен график математических ожиданий относительной яркости и их дисперсионных оценок (ось ординат), как функции угла наблюдения (ось абсцисс), – в "синем" ($\lambda \approx 420 \pm 20$ нм) участке спектра. Яркость оптимально масштабируется относительно диапазона своей вариации в текущем фотокадре. Угол наблюдения даётся в градусах: нулевое значение соответствует нормали к поверхности. Горизонтальным пунктиром отмечен среднеинтегральный уровень для всех математических ожиданий яркости.

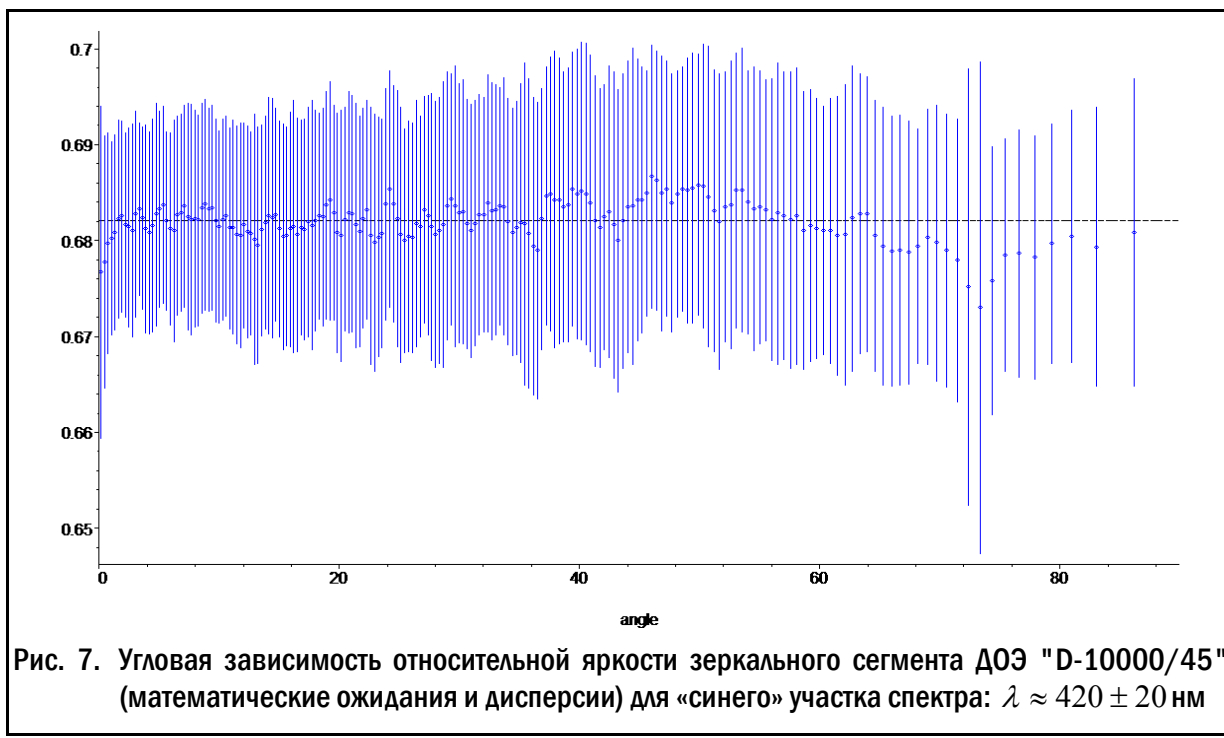

**Рис. 7.** Угловая зависимость относительной яркости зеркального сегмента ДОЭ "D-10000/45" (математические ожидания и дисперсии) для «синего» участка спектра: $\lambda \approx 420 \pm 20$ нм

В данном конкретном случае анализируемый сегмент изображения образца ДОЭ "D-10000/45" содержит 824×139=114536 пикселов (по 3 sRGB-субпиксела). Этот сегмент делится на 200 групп (участков наблюдения) по $\approx 573$ пиксела в каждом участке.

На рис. 8 для зеркального сегмента "D-10000/45" представлен аналогичный график математических ожиданий относительной яркости и их дисперсионных оценок, как функции угла наблюдения, – в "зелёном" ($\lambda \approx 530 \pm 20$ нм) участке спектра.

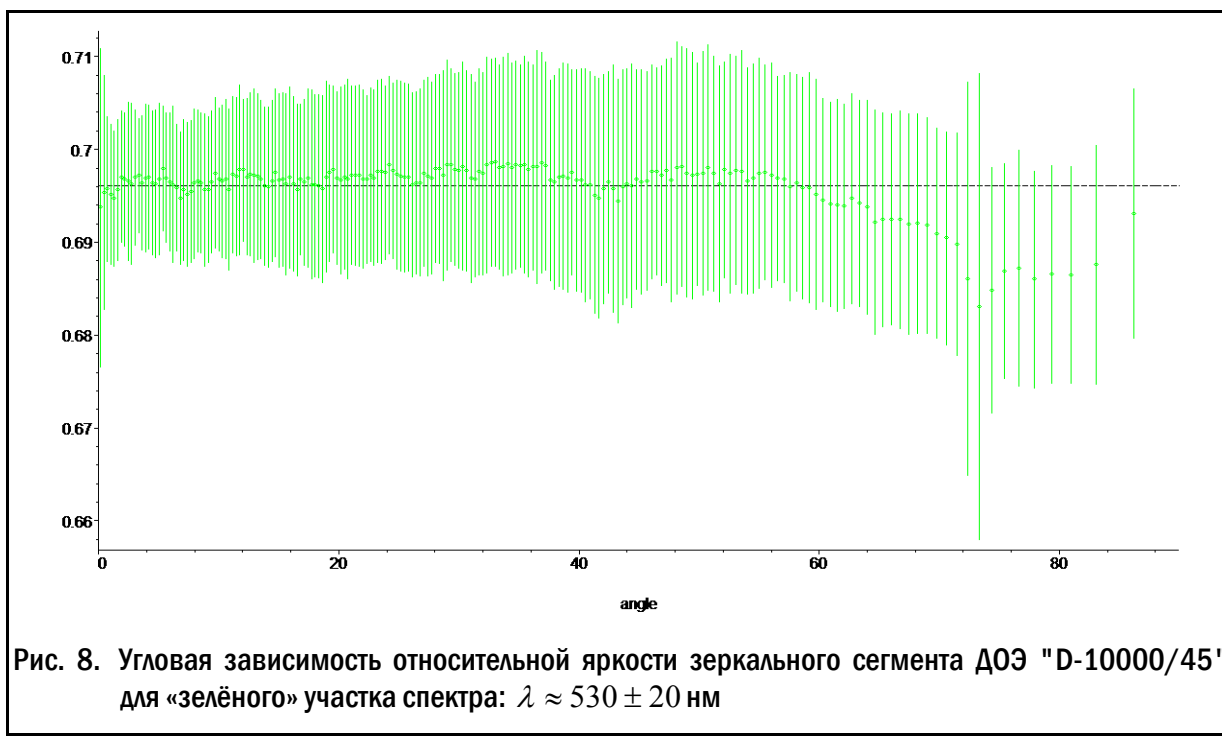

**Рис. 8.** Угловая зависимость относительной яркости зеркального сегмента ДОЭ "D-10000/45" для «зелёного» участка спектра: $\lambda \approx 530 \pm 20$ нм

На рис. 9 для зеркального сегмента "D-10000/45" также представлен аналогичный график математических ожиданий относительной яркости и их дисперсионных оценок, как функции угла наблюдения, – в "красном" ($\lambda \approx 700 \pm 20$ нм) участке спектра.

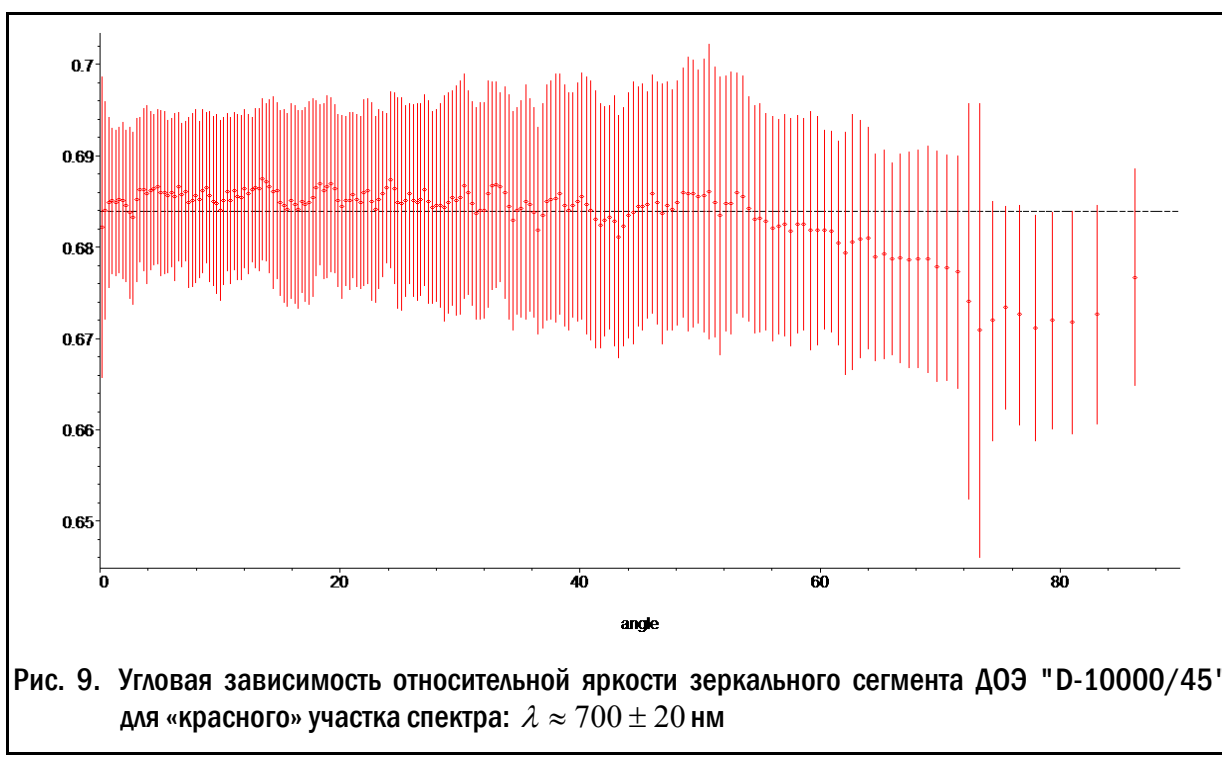

**Рис. 9.** Угловая зависимость относительной яркости зеркального сегмента ДОЭ "D-10000/45" для «красного» участка спектра: $\lambda \approx 700 \pm 20$ нм

Подобное тестирование было проведено для всех конструктивно выполнимых сочетаний дифракционных оптических элементов, фотометрических камер и источников излучения. Полученные результаты всегда качественно соответствовали вышеприведённым, т. е. имели воспроизводимый характер. Это позволяет сделать такие выводы:

— Применённый фотометрический метод обладает весьма высокой чувствительностью. Об этом можно судить, например, по такой детали: незначительный дефект исследуемого сегмента ДОЭ (см. рис. 6 на стр. 12, правый верхний угол изображения сегмента) вызывает заметный провал математического ожидания относительной яркости (угол наблюдения $\approx 73°$) при одновременном возрастании дисперсии, сигнализирующем о снижении достоверности данного значения математического ожидания. Таким образом, использованный метод позволяет по анализу яркостных характеристик зеркальных участков ДОЭ надёжно зафиксировать любое значимое отклонение от изотропности у исходного светового поля, поскольку наличие этого отклонения станет очевидным по результатам соответствующего дисперсионного исследования.

— Как показали тестовые результаты, для диапазона световых волн $\lambda \approx 400 \div 440$ нм относительные вариации яркости различных точек сегмента боковой поверхности ДОЭ не превышают статистической погрешности измерений, которая составляет менее 1% от величины среднеинтегрального уровня данного параметра. Очевидно, что для "синего" диапазона излучения спектральный коэффициент отражения алюминиевого покрытия практически не зависит от угла падения света, причём это характерно как для зеркально гладких поверхностей, так и для поверхностей дифракционных решёток. Интересно заметить, что именно для диапазона волн $\lambda \approx 400 \div 500$ нм спектральный коэффициент отражения алюминия имеет максимально высокое (для видимой части спектра) и, одновременно, весьма стабильное значение:

$$\rho_\lambda = 0.93 \approx Const \quad \text{для} \quad \forall \lambda \in (400 \text{ нм}, 500 \text{ нм})$$

— В диапазоне длин световых волн $\lambda > 500$ нм наблюдается некоторое снижение значения коэффициента спектрального отражения от зеркальных участков ДОЭ по мере увеличения углов отражения: $\beta \geq 55° \div 60°$. Максимальное снижение яркости для "зелёного" участка спектра достигает величины ~1%, а для "красного" участка ~2% (при углах отражения $\beta > 70°$ — см. рис. 8 и 9). Такого рода зависимость приводит, например, к визуально наблюдаемому снижению яркости контура поверхности ДОЭ цилиндрической формы, т. е. при больших (скользящих) углах отражения (см. рис 6). Это явление хорошо выражено при использовании источников излучения с низкой цветовой температурой. В случае коротковолнового излучения (например, для самосветящегося люминофора с длиной волны $\lambda \approx 420$ нм) данный эффект отсутствует.

— Всё вышесказанное позволяет утверждать, что методы стохастизации световых полей, применённые в фотометрических камерах №1-3, определённо достигли своей цели: формируемый в этих камерах фотонный газ действительно обладает пространственно изотропными свойствами, т. е. является диффузным. Для коротковолновых участков ($\lambda \approx 400 \div 500$ нм) спектральных диапазонов используемого излучения данный вывод может быть принят без оговорок. В случае же использования света с бо́льшей длиной волны ($\lambda > 500$ нм) для точной интерпретации результатов экспериментов может понадобиться учёт вышеописанной зависимости спектрального коэффициента отражения от соответствующего угла. Впрочем, такого рода поправки едва ли будут иметь существенное значение ввиду своей явной незначительности.

## Ламбертовское рассеяние в ортогональных плоскостях дифракционных решёток

Дадим определение ортогональной плоскости: ортогональной будем называть такую плоскость, которая ориентирована перпендикулярно и к формообразующей макроскопической поверхности дифракционной решётки, и к линиям её штрихов.

Рассмотрим рассеяние диффузного фотонного газа на одномерной или двумерной дифракционной решётке, фиксируя яркость её поверхности в какой-либо из ортогональных плоскостей. Известный на данный момент опыт говорит о том, что заметных нарушений изотропности светового поля при этом не происходит, а значит и закон Ламберта соблюдается вполне. Однако, такой результат, всё же, заранее не очевиден.

Индикатриса плотности вероятности[1] рассеяния фотона синусоидальной решёткой для случая его падения (и отражения) в ортогональной плоскости определяется формулой:

$$I_m(\alpha, \beta, \lambda, S_o, N) = \left( \frac{\sec(Z/2) \cdot \sin(N \cdot Z)}{Z} \right)^2 \quad \text{при:} \quad Z = \frac{\pi \cdot S_o \cdot (\sin(\alpha) + \sin(\beta))}{\lambda} \quad (5)$$

где $\alpha$ – угол падения фотона на решётку: $\forall \alpha \in (-\pi/2, +\pi/2)$;

$\beta$ – угол отражения фотона от решётки: $\forall \beta \in (-\pi/2, +\pi/2)$;

$\lambda$ – длина волны рассеиваемого излучения: $\lambda > 0$;

$S_o$ – шаг (период) решётки вдоль направления, ортогонального штрихам: $S_o \geq \lambda$;

$N$ – число регулярных элементов (штрихов) решётки: $N = 1, 2, ..., \infty$.

Для примера на рис. 10 приведён график вероятностного рассеяния (в ортогональной плоскости) для фотона с длиной волны $\lambda = 420$ нм, отвесно ($\alpha = 0°$) падающего на поверхность фазовой решётки синусоидального профиля с шагом $S_o = 833$ нм:

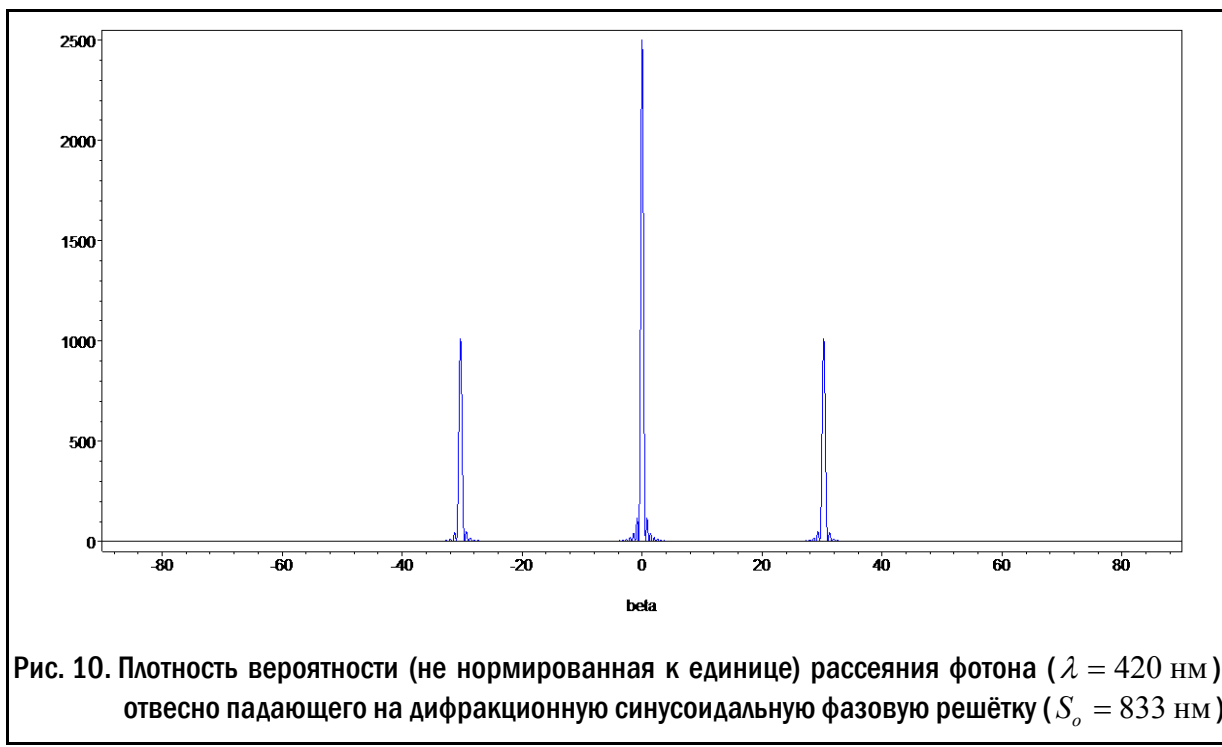

**Рис. 10.** Плотность вероятности (не нормированная к единице) рассеяния фотона ($\lambda = 420$ нм), отвесно падающего на дифракционную синусоидальную фазовую решётку ($S_o = 833$ нм)

---

[1] С точностью до коэффициента нормировки.

В данном случае индикатриса рассеяния имеет три главных максимума: нулевого порядка (так называемый "зеркальный" максимум, для которого всегда $\beta \equiv -\alpha$) и двух первых порядков. На угловое направление, соответствующее "зеркальному" максимуму, здесь приходится около 47% всей вероятности дифракционного рассеяния. Вообще же доля "зеркального" максимума обычно варьируется в диапазоне от 15% до 95%. Например, "зеркальная" компонента резко уменьшается при таких значениях угла падения $\alpha$, когда направление $\beta$ на какой-либо из ненулевых (т. е. нечётных[1]) максимумов начинает приближаться к "скользящему": $\beta \to \pm 90°$. При этом удельная доля, приходящаяся на данный максимум, сильно возрастает за счёт вероятностей рассеяния в других угловых направлениях. В частности, в приведённом на рис. 11 случае на долю "зеркального" максимума достаётся уже менее 20% от общей вероятности рассеяния.

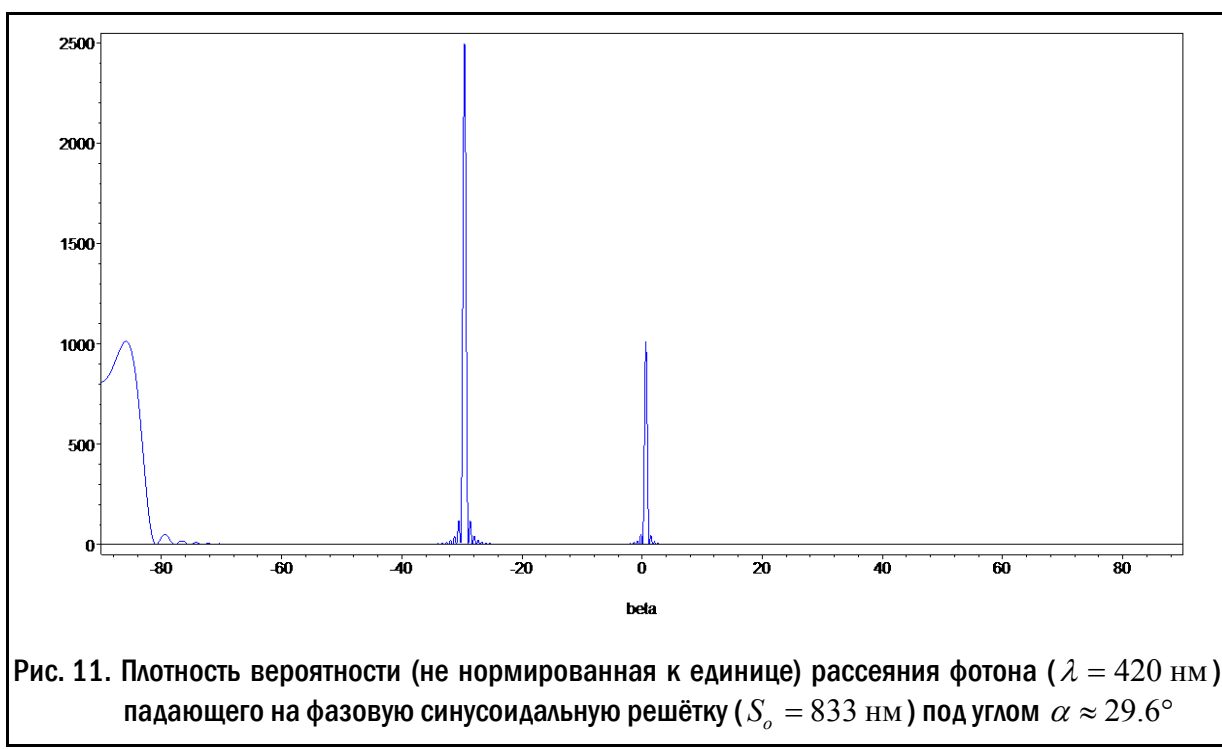

**Рис. 11.** Плотность вероятности (не нормированная к единице) рассеяния фотона ($\lambda = 420$ нм), падающего на фазовую синусоидальную решётку ($S_o = 833$ нм) под углом $\alpha \approx 29.6°$

Таким образом, индикатриса дифракционного рассеяния, характерная для фиксированного угла падения, совсем не похожа ни на диффузную ламбертовскую индикатрису вида (1) или (2), ни, например, на импульсную функцию Дирака, которая используется для описания одноканального когерентного (т. е. зеркального) отражения.

Удивительно, что столь сложно выглядящая осцилляционная зависимость, как (5), реализуемая в сочетании с ламбертовской плотностью вероятности взаимодействия фотонов диффузного газа с поверхностью дифракционной решётки, тем не менее, даёт интегральный отклик опять-таки в виде ламбертовской индикатрисы, но уже рассеяния.

Результаты эксперимента, иллюстрирующего всё вышесказанное, приведены на стр. 18. На рис. 12 даются снимки ДОЭ "D-833/90", помещённого в фотометрическую камеру №2. Поскольку в спектре источника излучения присутствуют лишь коротковолновые компоненты, изучалось лишь изображение анализируемого сегмента, получаемое "синими" субпикселами матрицы фотоаппарата, – см. график на рис. 13.

---

[1] Для синусоидальных решёток – чётные максимумы отсутствуют (кроме нулевого).

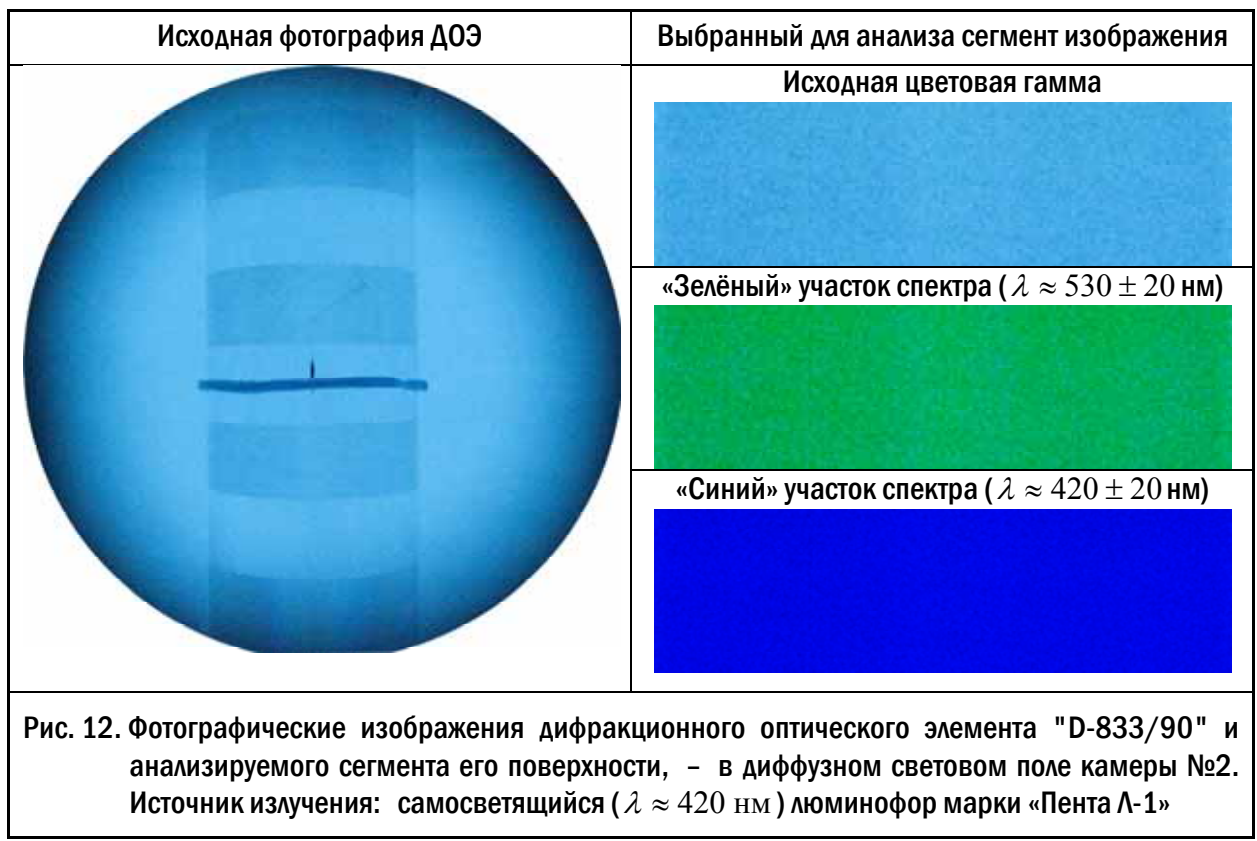

**Рис. 12.** Фотографические изображения дифракционного оптического элемента "D-833/90" и анализируемого сегмента его поверхности, – в диффузном световом поле камеры №2. Источник излучения: самосветящийся ($\lambda \approx 420$ нм) люминофор марки «Пента Л-1»

В данном конкретном случае анализируемый сегмент изображения образца ДОЭ "D-833/90" содержит 945×270=255150 пикселов (по 3 sRGB-субпиксела). Этот сегмент делится на 200 групп (участков наблюдения) по $\approx 1276$ пикселов в каждом участке.

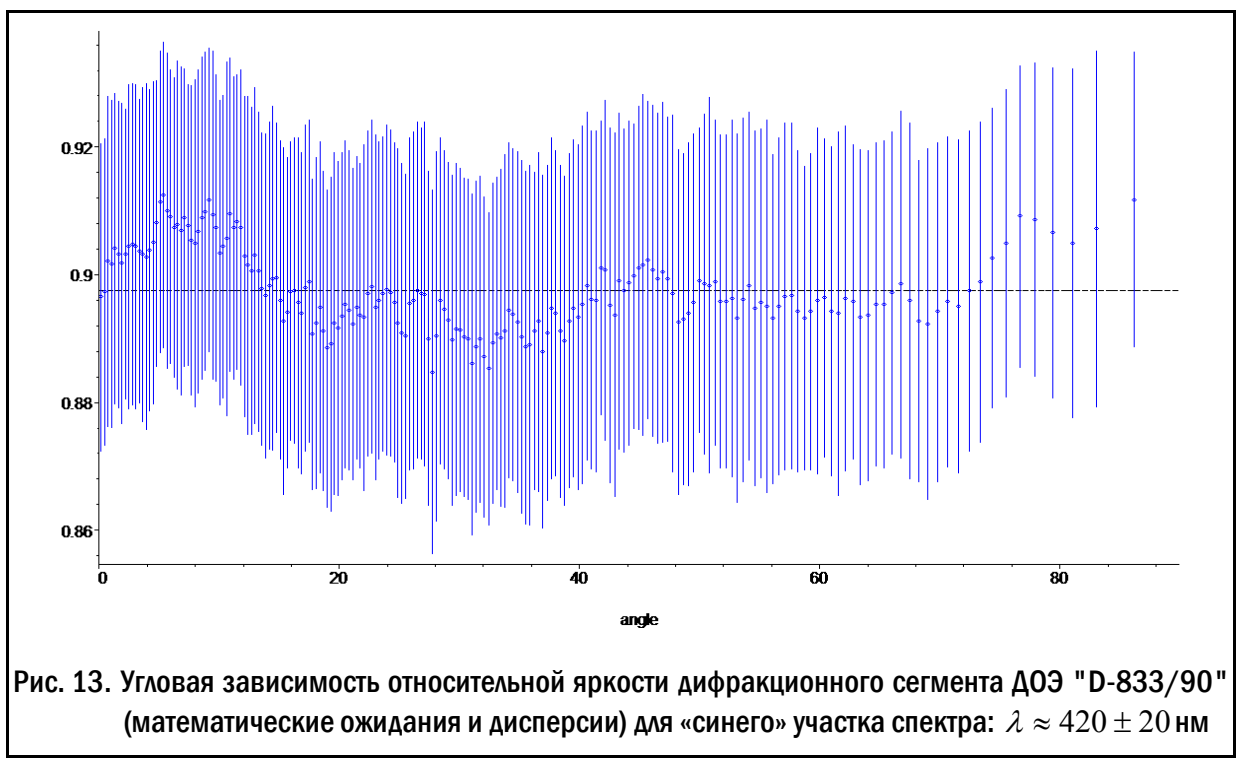

**Рис. 13.** Угловая зависимость относительной яркости дифракционного сегмента ДОЭ "D-833/90" (математические ожидания и дисперсии) для «синего» участка спектра: $\lambda \approx 420 \pm 20$ нм

**Характер рассеяния в диагональных плоскостях двумерной дифракционной решётки**

Ещё раз посмотрим на фотографию ДОЭ "D-833/90" (рис. 12). Мы видим изображения четырёх дифракционных участков образца, имеющих соответствующий микрорельеф. Эти участки чередуются с зеркальными, выглядящими более светлыми.

Обратим внимание на два дифракционных участка в центре снимка, которые находятся выше и ниже зеркального участка с отражением объектива фотокамеры (по его поверхности проходит крепёжное резиновое кольцо). Яркость поверхности двух указанных дифракционных участков выглядит вполне однородной. Собственно, один из них (нижний) и являлся анализируемым сегментом изображения ДОЭ (см. рис. 13).

Посмотрим теперь на два оставшихся дифракционных участка ДОЭ: самый верхний и самый нижний. Яркость поверхности этих участков – отнюдь не однородна и имеет одинаково выраженные градиенты, симметричные относительно вертикальной оси. Однако данный снимок не даёт возможности для анализа природы этих аномалий:

– изображения интересующих нас объектов находятся на периферии фотокадра;
– сами объекты (дифракционные участки ДОЭ) находятся не в центре фотометрической камеры, а возле её верхней и нижней границ, имеющих форму плоских кругов.

Первый из перечисленных факторов даёт основания предположить, что наблюдаемые аномалии яркости могут быть следствием несовершенства оптической системы использованного фотоаппарата. Влияние второго фактора гипотетически может заключаться в наличии неких ранее не замеченных нарушений изотропности светового поля в непосредственной близости от стенок фотометрической камеры ("краевой эффект").

Между тем, характер изображения говорит, что аномалии яркости наблюдаются не в ортогональных, а в диагональных плоскостях, которые ориентированы перпендикулярно к поверхности решётки[1] ДОЭ и пересекают линии её штрихов под углом в 45°.

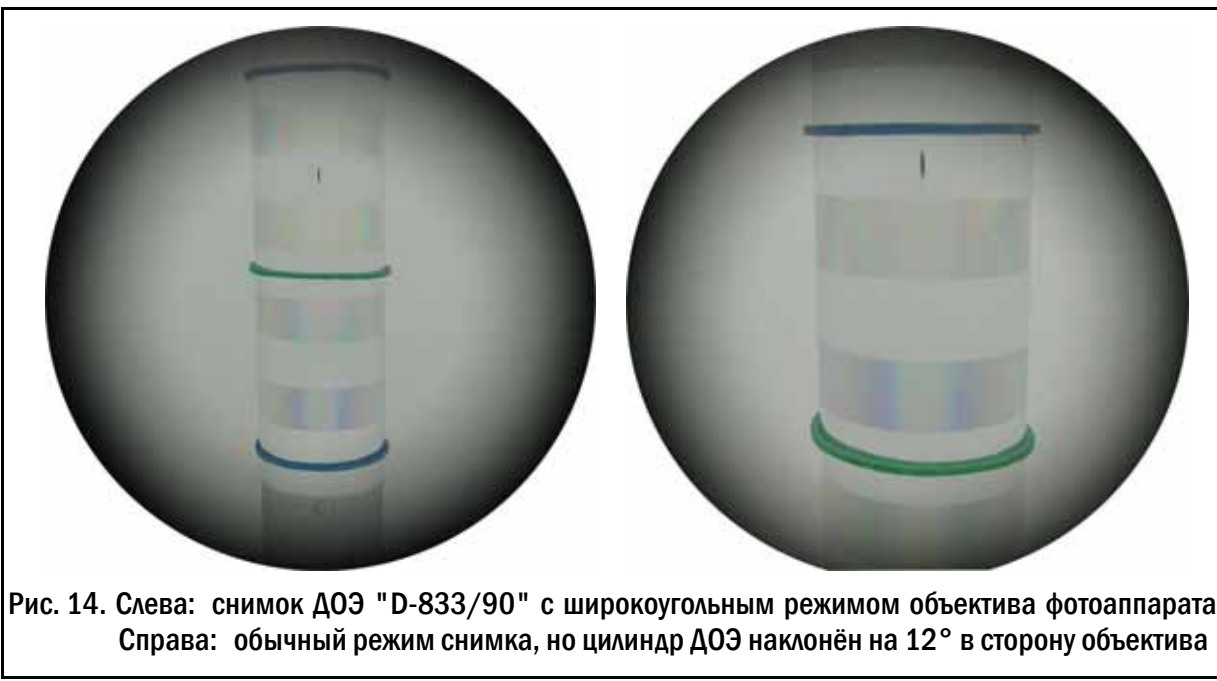

Рис. 14. Слева: снимок ДОЭ "D-833/90" с широкоугольным режимом объектива фотоаппарата;
Справа: обычный режим снимка, но цилиндр ДОЭ наклонён на 12° в сторону объектива

---

[1] Для двумерных решёток пересечение диагональной плоскостью всех их штрихов под углом в 45° возможно лишь тогда, когда сами эти штрихи – взаимоортогональны.

На рис. 14 приведены снимки ДОЭ "D-833/90" в фотометрической камере №1, свидетельствующие в пользу такого предположения. Если оно действительно верно, то поворот гибкой реплики дифракционной решётки на поверхности опорного цилиндра на угол 45° (см. рис. 15) должен "переместить" визуальное проявление предполагаемого эффекта с периферии кадра в его центр, – в район оптической оси фотообъектива.

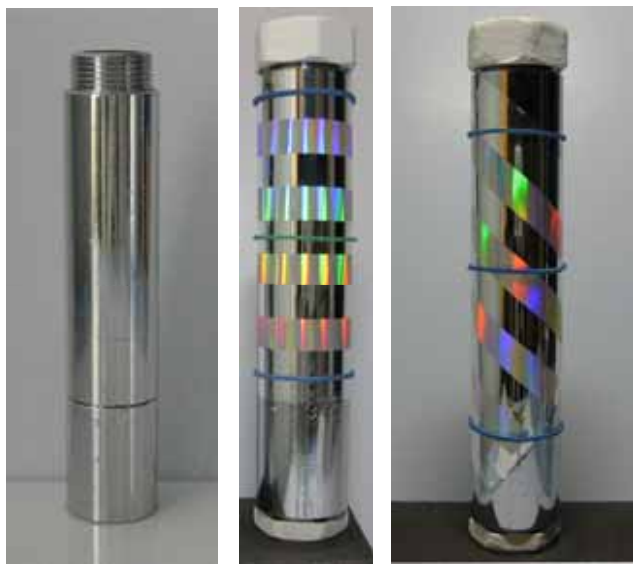

**Рис. 15.** Опорный цилиндр (слева), ДОЭ "D-833/90" с ортогонально ориентированной репликой дифракционной решётки (в центре) и с диагонально ориентированной репликой (справа)

Проверка показала (рис. 16), что в диагональных плоскостях двумерных фазовых решёток действительно имеет место анизотропное рассеяние диффузного фотонного газа.

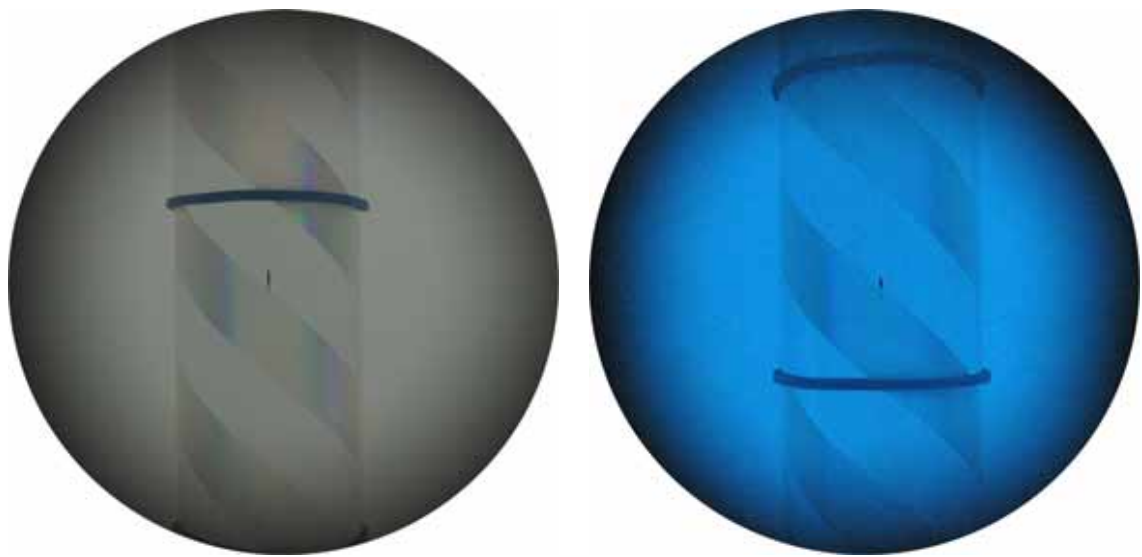

**Рис. 16.** Образец "D-833/90" с репликой, диагонально ориентированной на опорном цилиндре: в фотометрической камере №1 (снимок слева) и в камере №2 (снимок справа)

Апробация воспроизводимости выявленного эффекта была реализована для всех конструктивно выполнимых сочетаний дифракционных оптических элементов, фотометрических камер и источников излучения. Часть этих результатов представлена далее.

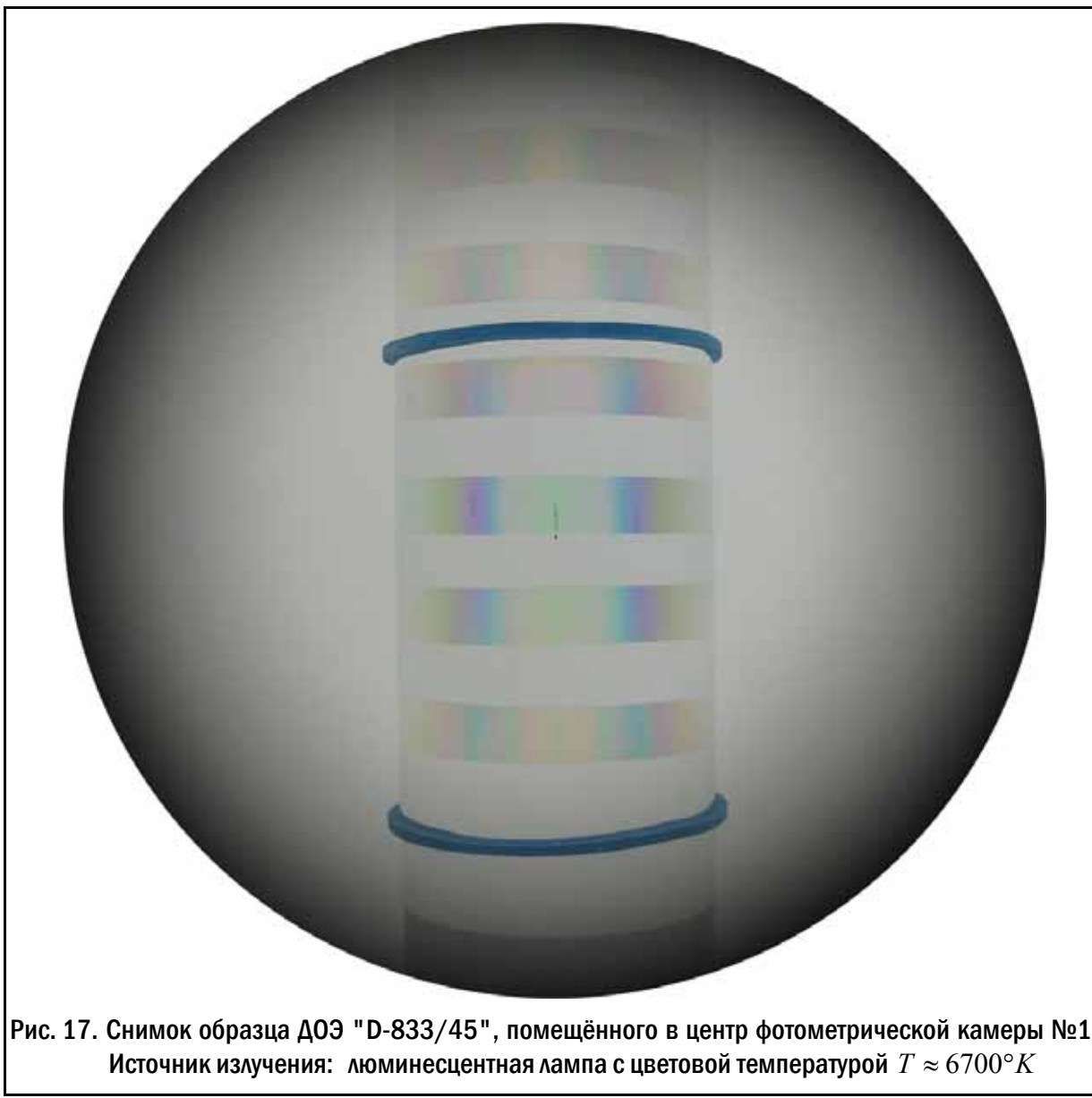

**Рис. 17.** Снимок образца ДОЭ "D-833/45", помещённого в центр фотометрической камеры №1; Источник излучения: люминесцентная лампа с цветовой температурой $T \approx 6700°K$

## Параметры проведения эксперимента

1. **Фотометрическая камера №1:** цилиндрической формы, с внешним локальным источником первичного светового излучения — см. стр. 5 и далее.

2. **Источник первичного светового излучения:** люминесцентная газоразрядная лампа 24 Вт с "трёхполосным" спектром излучения (цветовая температура $T \approx 6700°K$).

3. **Регистрирующая аппаратура:** ультракомпактный фотоаппарат Canon™ «Digital Ixus 90 IS».

4. **Размер, форма и положение дифракционного оптического образца (ДОЭ):** гибкая реплика дифракционной решётки закреплена на опорном цилиндре с диаметром 30 мм. Цилиндр помещён в центре камеры, расстояние от его оси до фотообъектива L=132 мм.

5. **Параметры дифракционной решётки ДОЭ:** фазовая, отражающая, двумерная, с квазисинусоидальным профилем штрихов, имеющих шаг $S_o \approx 833$ нм. Штрихи взаимно ортогональны и составляют 45° со сторонами рабочей области гибкой реплики.

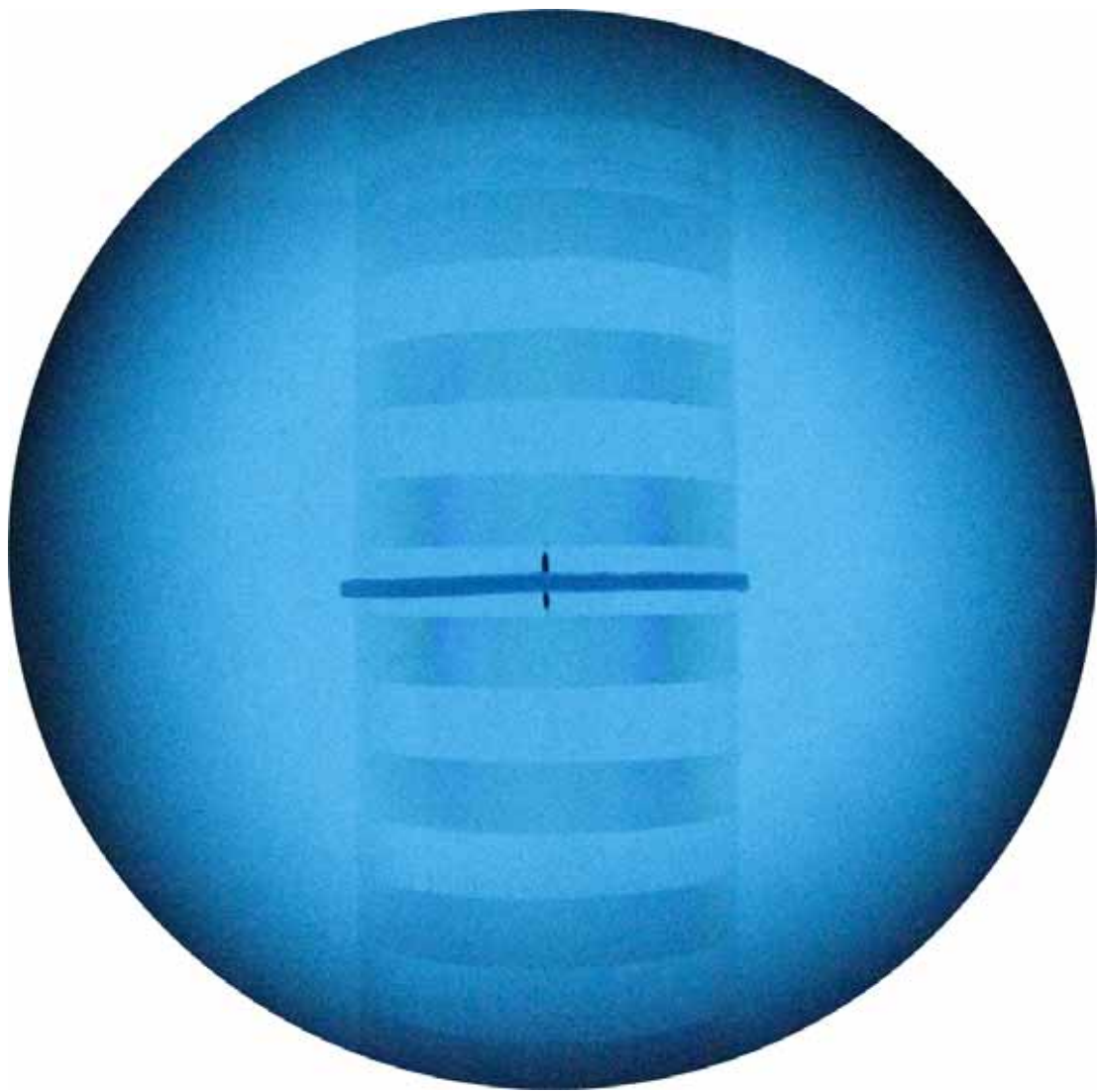

**Рис. 18.** Снимок образца ДОЭ "D-833/45", помещённого в центр фотометрической камеры №2; Источник излучения: самосветящийся ($\lambda \approx 420$ нм) люминофор марки «Пента Л-1»

## Параметры проведения эксперимента

1. **Фотометрическая камера №2:** цилиндрической формы, с распределённым по внутренней поверхности источником первичного светового излучения – см. стр. 5 и далее.

2. **Источник первичного светового излучения:** светорассеивающая подложка, покрытая слоем твёрдого прозрачного раствора самосветящегося люминофора ($\lambda \approx 420$ нм).

3. **Регистрирующая аппаратура:** ультракомпактный фотоаппарат Canon™ «Digital Ixus 90 IS».

4. **Размер, форма и положение дифракционного оптического образца (ДОЭ):** гибкая реплика дифракционной решётки закреплена на опорном цилиндре с диаметром 30 мм. Цилиндр помещён в центре камеры, расстояние от его оси до фотообъектива L=112 мм.

5. **Параметры дифракционной решётки ДОЭ:** фазовая, отражающая, двумерная, с квазисинусоидальным профилем штрихов, имеющих шаг $S_o \approx 833$ нм. Штрихи взаимно ортогональны и составляют 45° со сторонами рабочей области гибкой реплики.

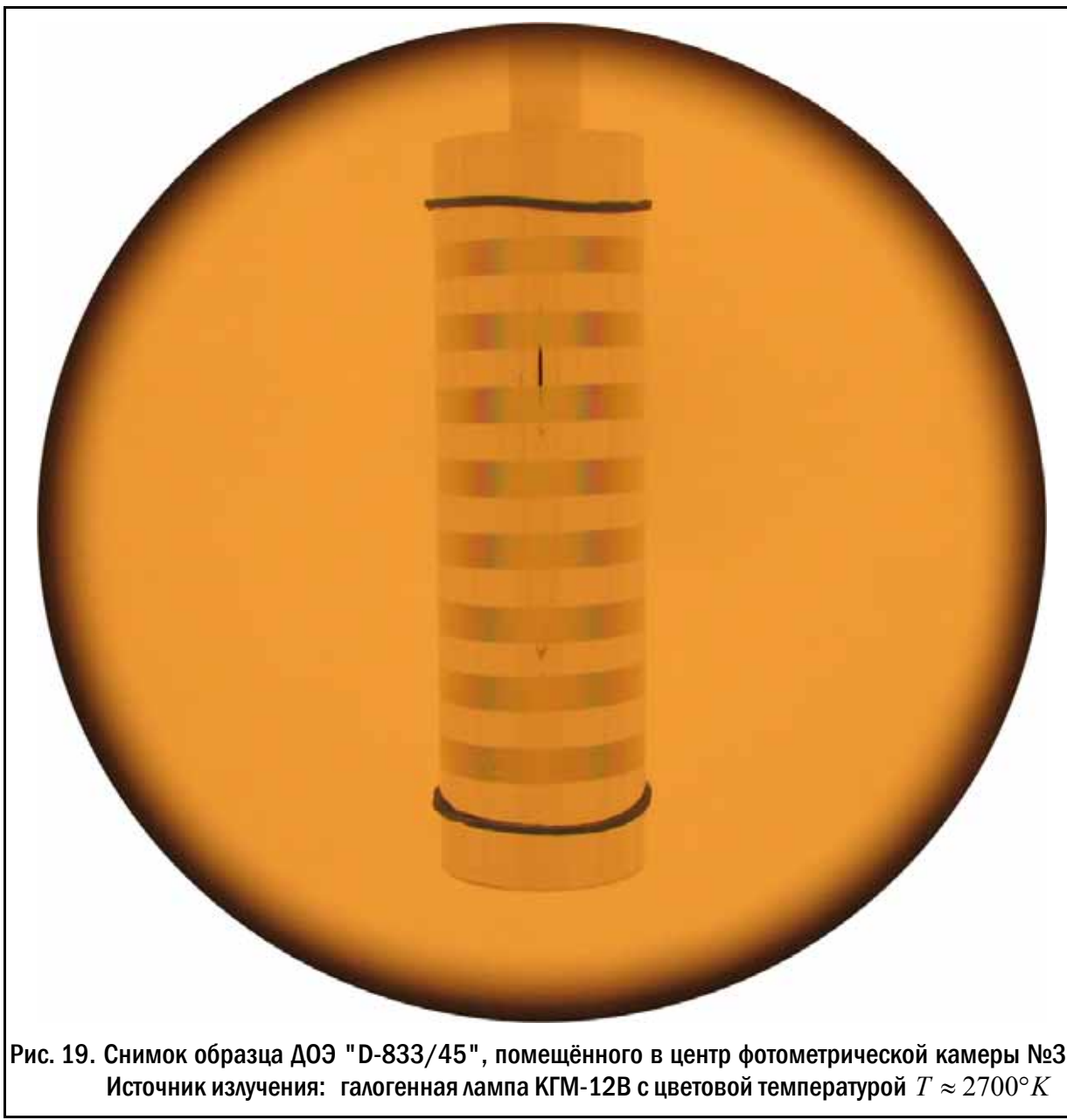

**Рис. 19.** Снимок образца ДОЭ "D-833/45", помещённого в центр фотометрической камеры №3; Источник излучения: галогенная лампа КГМ-12В с цветовой температурой $T \approx 2700°K$

1. **Фотометрическая камера №3:** сферической формы, с внутренним локальным источником первичного светового излучения – см. стр. 5 и далее.

2. **Источник первичного светового излучения:** миниатюрная галогенная лампа накаливания 40 Вт с непрерывным спектром излучения (цветовая температура $T \approx 2700°K$).

3. **Регистрирующая аппаратура:** компактный фотоаппарат Canon™ «**PowerShot S2 IS**».

4. **Размер, форма и положение дифракционного оптического образца (ДОЭ):** гибкая реплика дифракционной решётки закреплена на опорном цилиндре с диаметром 30 мм. Цилиндр помещён в центре камеры, расстояние от его оси до фотообъектива L=210 мм.

5. **Параметры дифракционной решётки ДОЭ:** фазовая, отражающая, двумерная, с квазисинусоидальным профилем штрихов, имеющих шаг $S_o \approx 833$ нм. Штрихи взаимно ортогональны и составляют 45° со сторонами рабочей области гибкой реплики.

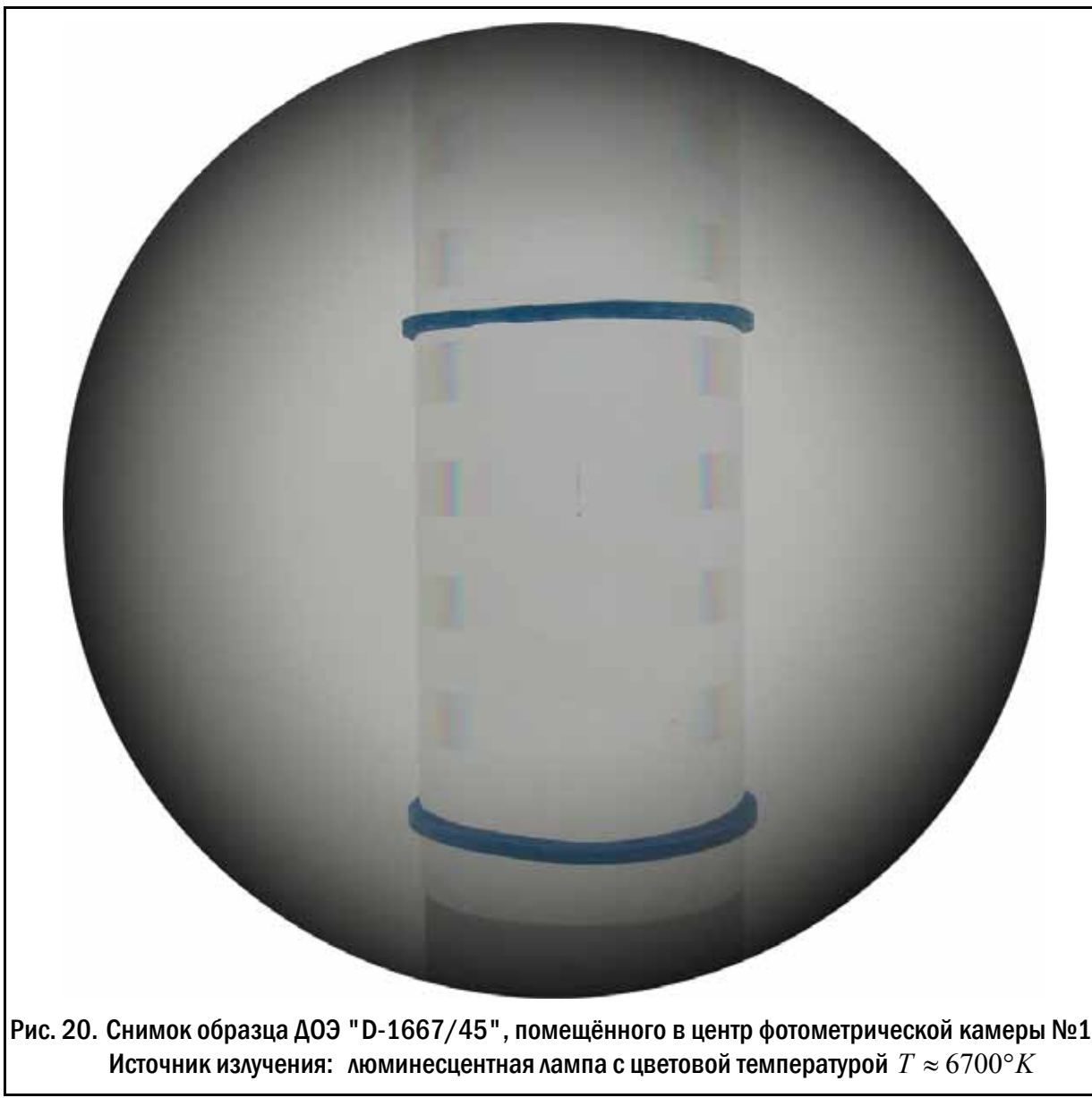

**Рис. 20.** Снимок образца ДОЭ "D-1667/45", помещённого в центр фотометрической камеры №1; Источник излучения: люминесцентная лампа с цветовой температурой $T \approx 6700°K$

## Параметры проведения эксперимента

1. **Фотометрическая камера №1:** цилиндрической формы, с внешним локальным источником первичного светового излучения – см. стр. 5 и далее.

2. **Источник первичного светового излучения:** люминесцентная газоразрядная лампа 24 Вт с "трёхполосным" спектром излучения (цветовая температура $T \approx 6700°K$).

3. **Регистрирующая аппаратура:** ультракомпактный фотоаппарат Canon™ «Digital Ixus 90 IS».

4. **Размер, форма и положение дифракционного оптического образца (ДОЭ):** гибкая реплика дифракционной решётки закреплена на опорном цилиндре с диаметром 30 мм. Цилиндр помещён в центре камеры, расстояние от его оси до фотообъектива L=132 мм.

5. **Параметры дифракционной решётки ДОЭ:** фазовая, отражающая, двумерная, с квазисинусоидальным профилем штрихов, имеющих шаг $S_o \approx 1667$ нм. Штрихи взаимно ортогональны и составляют 45° со сторонами рабочей области гибкой реплики.

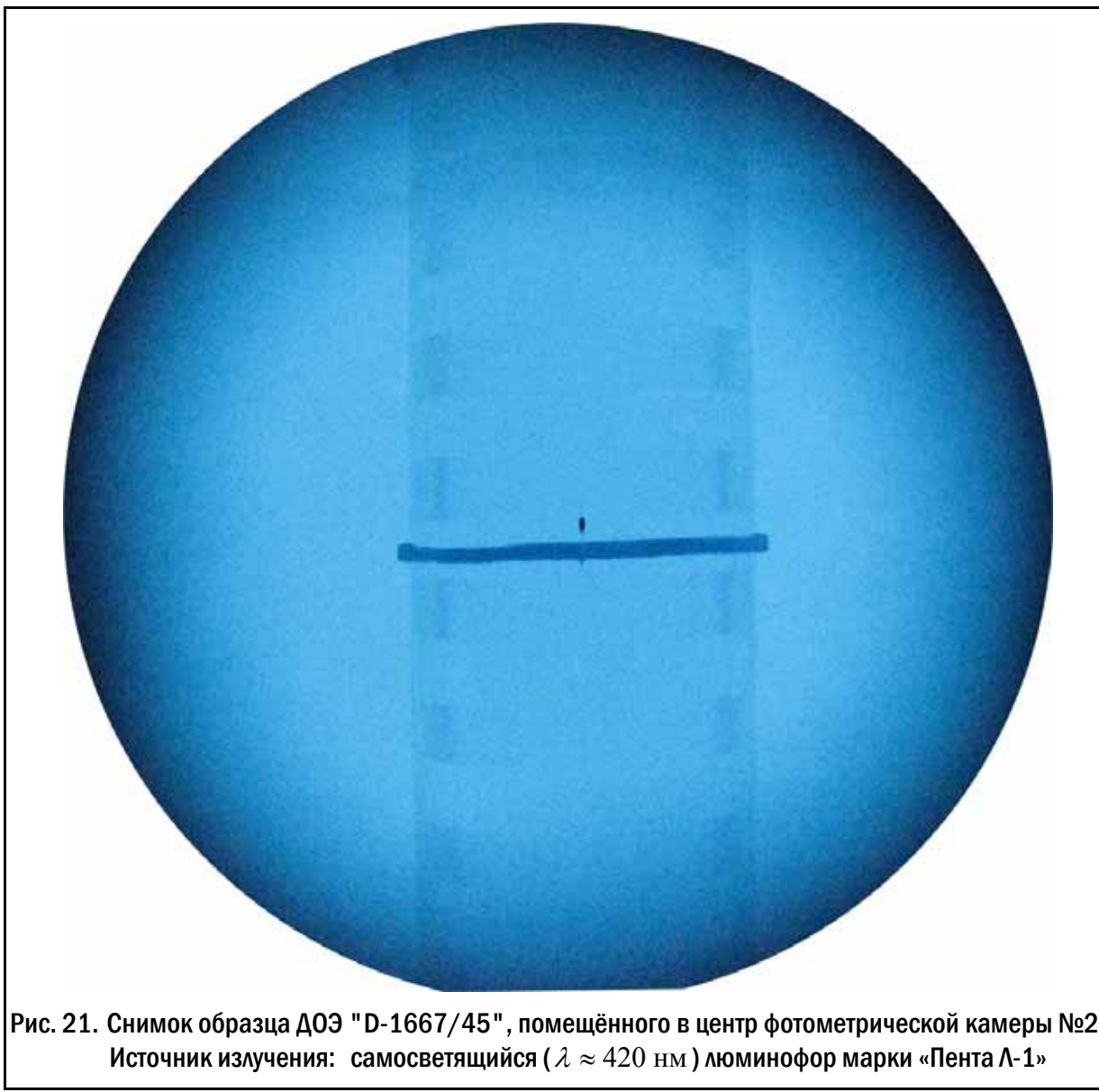

**Рис. 21.** Снимок образца ДОЭ "D-1667/45", помещённого в центр фотометрической камеры №2; Источник излучения: самосветящийся ($\lambda \approx 420$ нм) люминофор марки «Пента Л-1»

### Параметры проведения эксперимента

1. **Фотометрическая камера №2:** цилиндрической формы, с распределённым по внутренней поверхности источником первичного светового излучения – см. стр. 5 и далее.

2. **Источник первичного светового излучения:** светорассеивающая подложка, покрытая слоем твёрдого прозрачного раствора самосветящегося люминофора ($\lambda \approx 420$ нм).

3. **Регистрирующая аппаратура:** ультракомпактный фотоаппарат Canon™ «Digital Ixus 90 IS».

4. **Размер, форма и положение дифракционного оптического образца (ДОЭ):** гибкая реплика дифракционной решётки закреплена на опорном цилиндре с диаметром 30 мм. Цилиндр помещён в центре камеры, расстояние от его оси до фотообъектива L=112 мм.

5. **Параметры дифракционной решётки ДОЭ:** фазовая, отражающая, двумерная, с квазисинусоидальным профилем штрихов, имеющих шаг $S_o \approx 1667$ нм. Штрихи взаимно ортогональны и составляют 45° со сторонами рабочей области гибкой реплики.

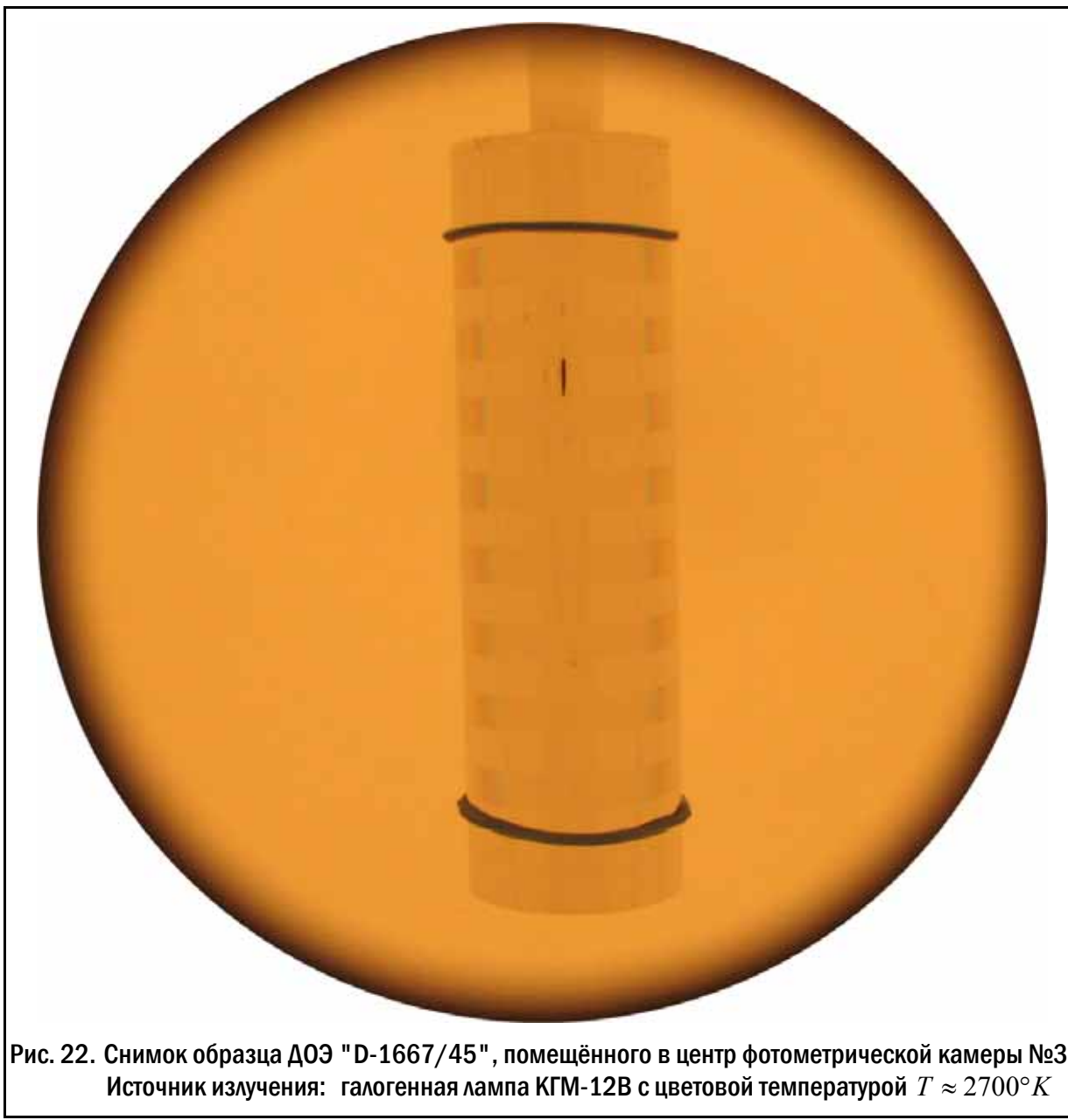

**Рис. 22.** Снимок образца ДОЭ "D-1667/45", помещённого в центр фотометрической камеры №3; Источник излучения: галогенная лампа КГМ-12В с цветовой температурой $T \approx 2700°K$

1. **Фотометрическая камера №3:** сферической формы, с внутренним локальным источником первичного светового излучения — см. стр. 5 и далее.

2. **Источник первичного светового излучения:** миниатюрная галогенная лампа накаливания 40 Вт с непрерывным спектром излучения (цветовая температура $T \approx 2700°K$).

3. **Регистрирующая аппаратура:** компактный фотоаппарат Canon™ **«PowerShot S2 IS»**.

4. **Размер, форма и положение дифракционного оптического образца (ДОЭ):** гибкая реплика дифракционной решётки закреплена на опорном цилиндре с диаметром 30 мм. Цилиндр помещён в центре камеры, расстояние от его оси до фотообъектива L=210 мм.

5. **Параметры дифракционной решётки ДОЭ:** фазовая, отражающая, двумерная, с квазисинусоидальным профилем штрихов, имеющих шаг $S_o \approx 1667$ нм. Штрихи взаимно ортогональны и составляют 45° со сторонами рабочей области гибкой реплики.

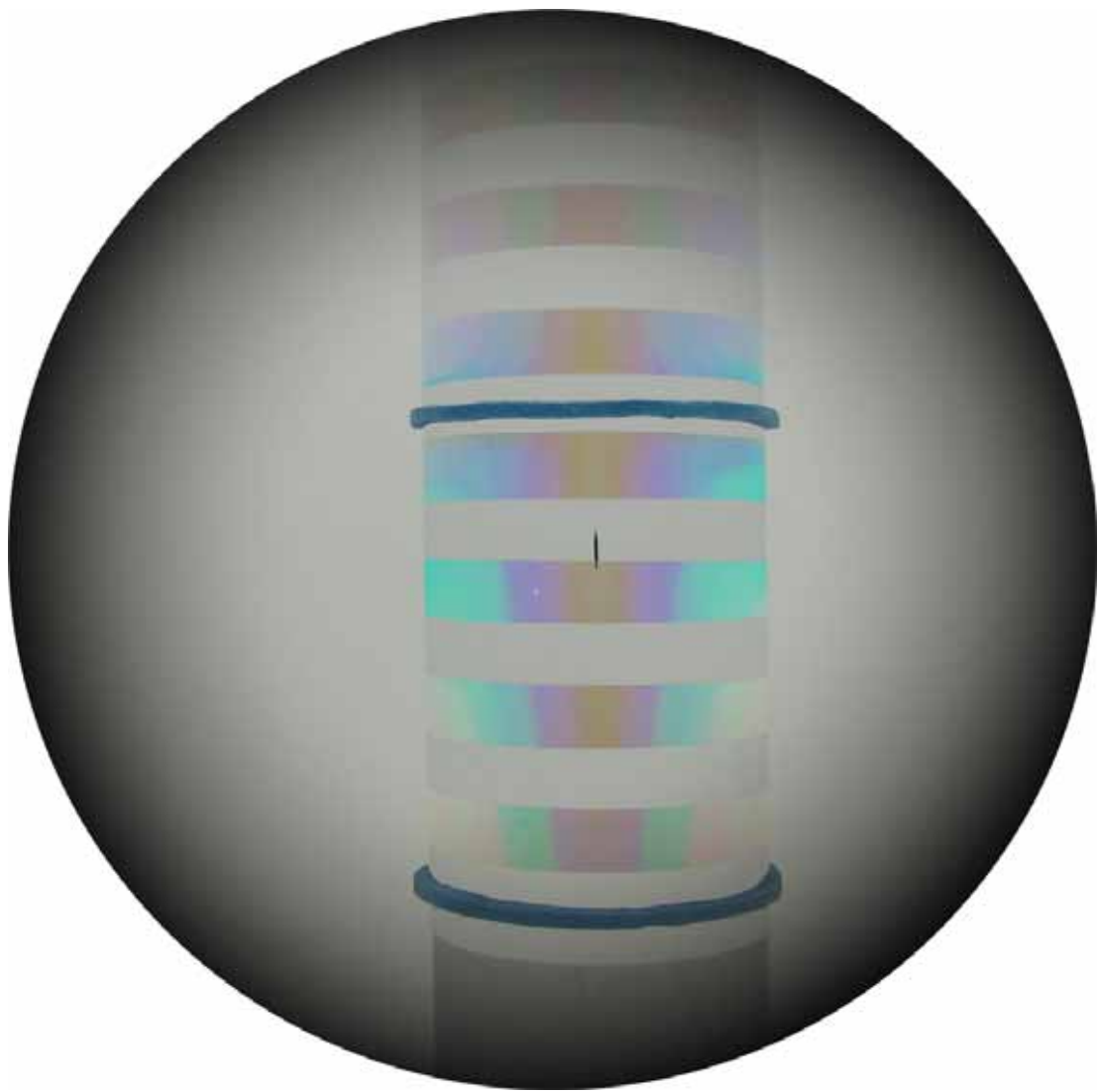

**Рис. 23.** Снимок образца ДОЭ "D-417/45", помещённого в центр фотометрической камеры №1; Источник излучения: люминесцентная лампа с цветовой температурой $T \approx 6700°K$

## Параметры проведения эксперимента

1. **Фотометрическая камера №1:** цилиндрической формы, с внешним локальным источником первичного светового излучения − см. стр. 5 и далее.

2. **Источник первичного светового излучения:** люминесцентная газоразрядная лампа 24 Вт с "трёхполосным" спектром излучения (цветовая температура $T \approx 6700°K$).

3. **Регистрирующая аппаратура:** ультракомпактный фотоаппарат Canon™ «Digital Ixus 90 IS».

4. **Размер, форма и положение дифракционного оптического образца (ДОЭ):** гибкая реплика дифракционной решётки закреплена на опорном цилиндре с диаметром 30 мм. Цилиндр помещён в центре камеры, расстояние от его оси до фотообъектива L=132 мм.

5. **Параметры дифракционной решётки ДОЭ:** фазовая, отражающая, двумерная, с квазисинусоидальным профилем штрихов, имеющих шаг $S_o \approx 417$ нм. Штрихи взаимно ортогональны и составляют 45° со сторонами рабочей области гибкой реплики.

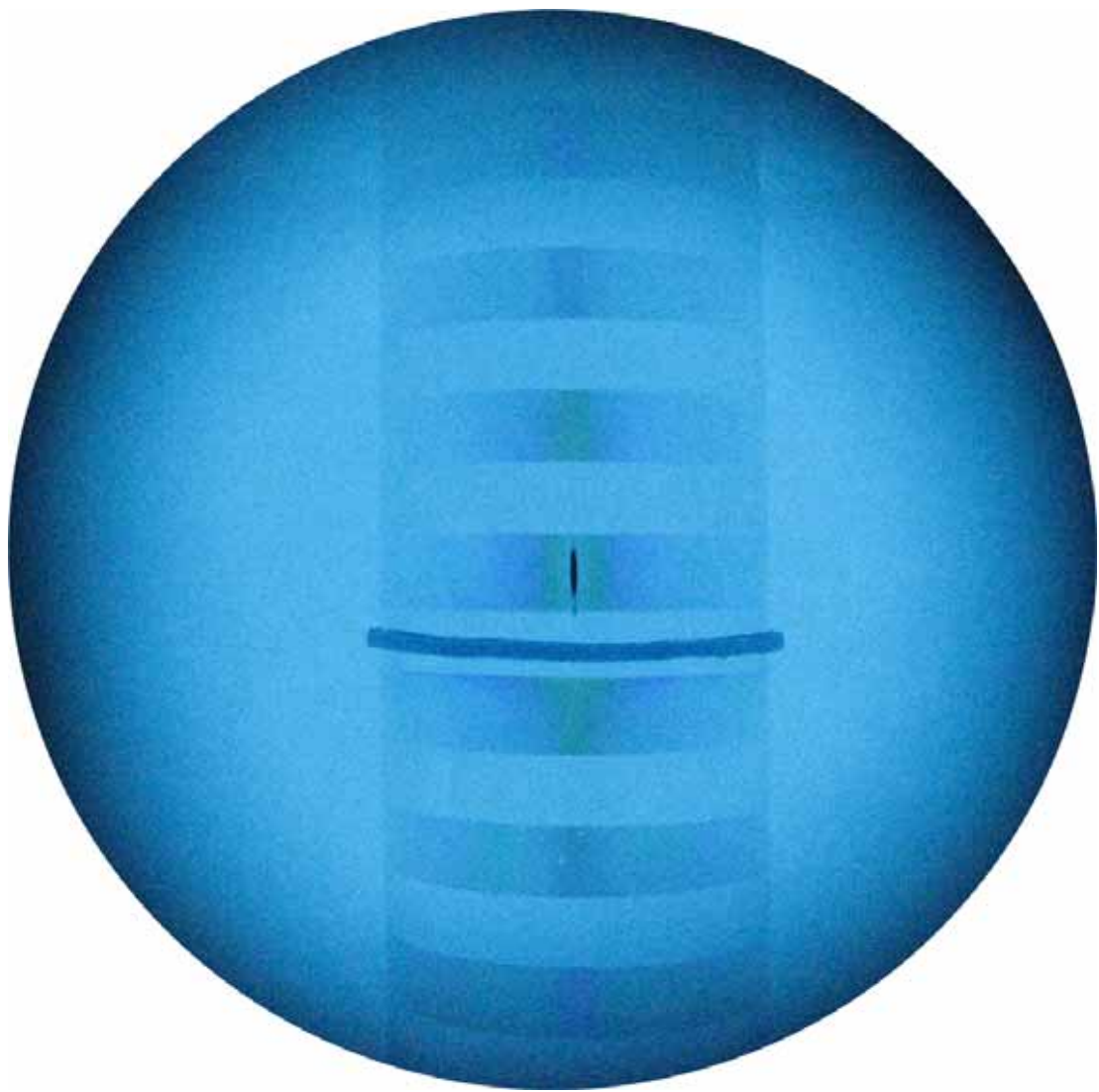

**Рис. 24.** Снимок образца ДОЭ "D-417/45", помещённого в центр фотометрической камеры №2; Источник излучения: самосветящийся ($\lambda \approx 420$ нм) люминофор марки «Пента Л-1»

## Параметры проведения эксперимента

1. **Фотометрическая камера №2:** цилиндрической формы, с распределённым по внутренней поверхности источником первичного светового излучения — см. стр. 5 и далее.

2. **Источник первичного светового излучения:** светорассеивающая подложка, покрытая слоем твёрдого прозрачного раствора самосветящегося люминофора ($\lambda \approx 420$ нм).

3. **Регистрирующая аппаратура:** ультракомпактный фотоаппарат Canon™ «Digital Ixus 90 IS».

4. **Размер, форма и положение дифракционного оптического образца (ДОЭ):** гибкая реплика дифракционной решётки закреплена на опорном цилиндре с диаметром 30 мм. Цилиндр помещён в центре камеры, расстояние от его оси до фотообъектива L=112 мм.

5. **Параметры дифракционной решётки ДОЭ:** фазовая, отражающая, двумерная, с квазисинусоидальным профилем штрихов, имеющих шаг $S_o \approx 417$ нм. Штрихи взаимно ортогональны и составляют 45° со сторонами рабочей области гибкой реплики.

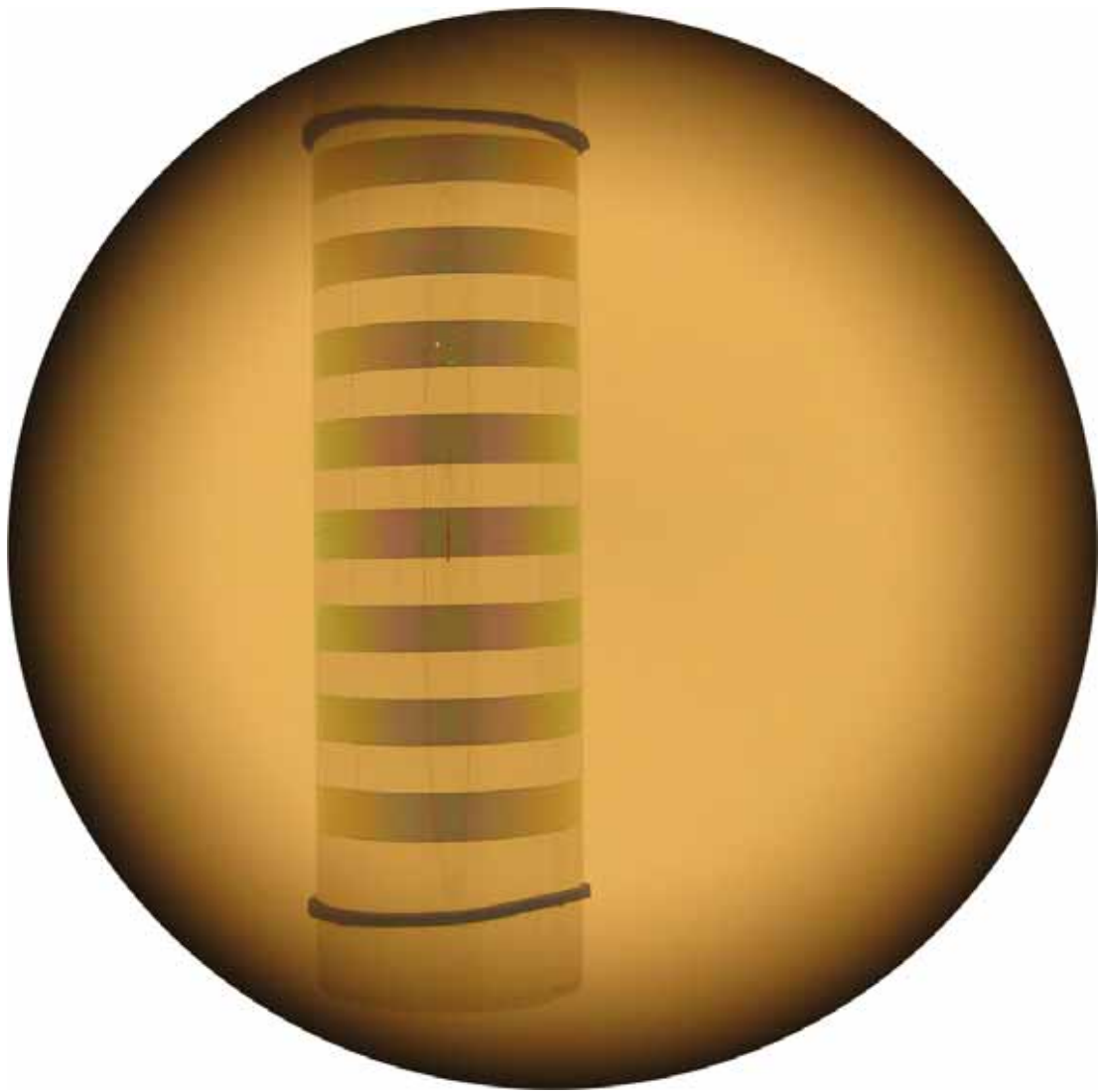

**Рис. 25.** Снимок образца ДОЭ "D-417/45", помещённого в центр фотометрической камеры №3; Источник излучения: галогенная лампа КГМ-12В с цветовой температурой $T \approx 2700°K$

## Параметры проведения эксперимента

1. **Фотометрическая камера №3:** сферической формы, с внутренним локальным источником первичного светового излучения – см. стр. 5 и далее.

2. **Источник первичного светового излучения:** миниатюрная галогенная лампа накаливания 40 Вт с непрерывным спектром излучения (цветовая температура $T \approx 2700°K$).

3. **Регистрирующая аппаратура:** ультракомпактный фотоаппарат Canon™ «Digital Ixus 90 IS».

4. **Размер, форма и положение дифракционного оптического образца (ДОЭ):** гибкая реплика дифракционной решётки закреплена на опорном цилиндре с диаметром 30 мм. Цилиндр помещён в центре камеры, расстояние от его оси до фотообъектива L=210 мм.

5. **Параметры дифракционной решётки ДОЭ:** фазовая, отражающая, двумерная, с квазисинусоидальным профилем штрихов, имеющих шаг $S_o \approx 417$ нм. Штрихи взаимно ортогональны и составляют 45° со сторонами рабочей области гибкой реплики.

**Гипотеза о наличии латентных каналов когерентного рассеяния у диагональных субрешёток**

Вышеприведённый экспериментальный материал свидетельствует о том, что на поверхности двумерных отражательных дифракционных решёток с квазисинусоидальным профилем действительно происходит анизотропное рассеяние частиц изначально однородного (диффузного) фотонного газа. Угловые направления, в которых наблюдается анизотропия яркости поверхности решёток, в основном лежат в диагональных плоскостях, в то время как в ортогональных плоскостях данные аномалии отсутствуют.

Очевидно, что было бы весьма полезным предоставление простой и практически значимой модели обнаруженного эффекта. И хотя, по мнению автора, методологически некорректно пытаться сделать это, опираясь на аппарат классической волновой оптики, наличие такого рода эмпирической модели может до некоторой степени способствовать пониманию механизма возникновения исследуемого явления.

Ещё раз вспомним, как выглядит участок микрорельефа двумерной синусоидальной решётки (см. рис. 4 на стр. 9, - справа). Дифракционная решётка, имеющая такой микрорельеф, фактически является не двумерной, а четырёхмерной, − см. рис. 26.

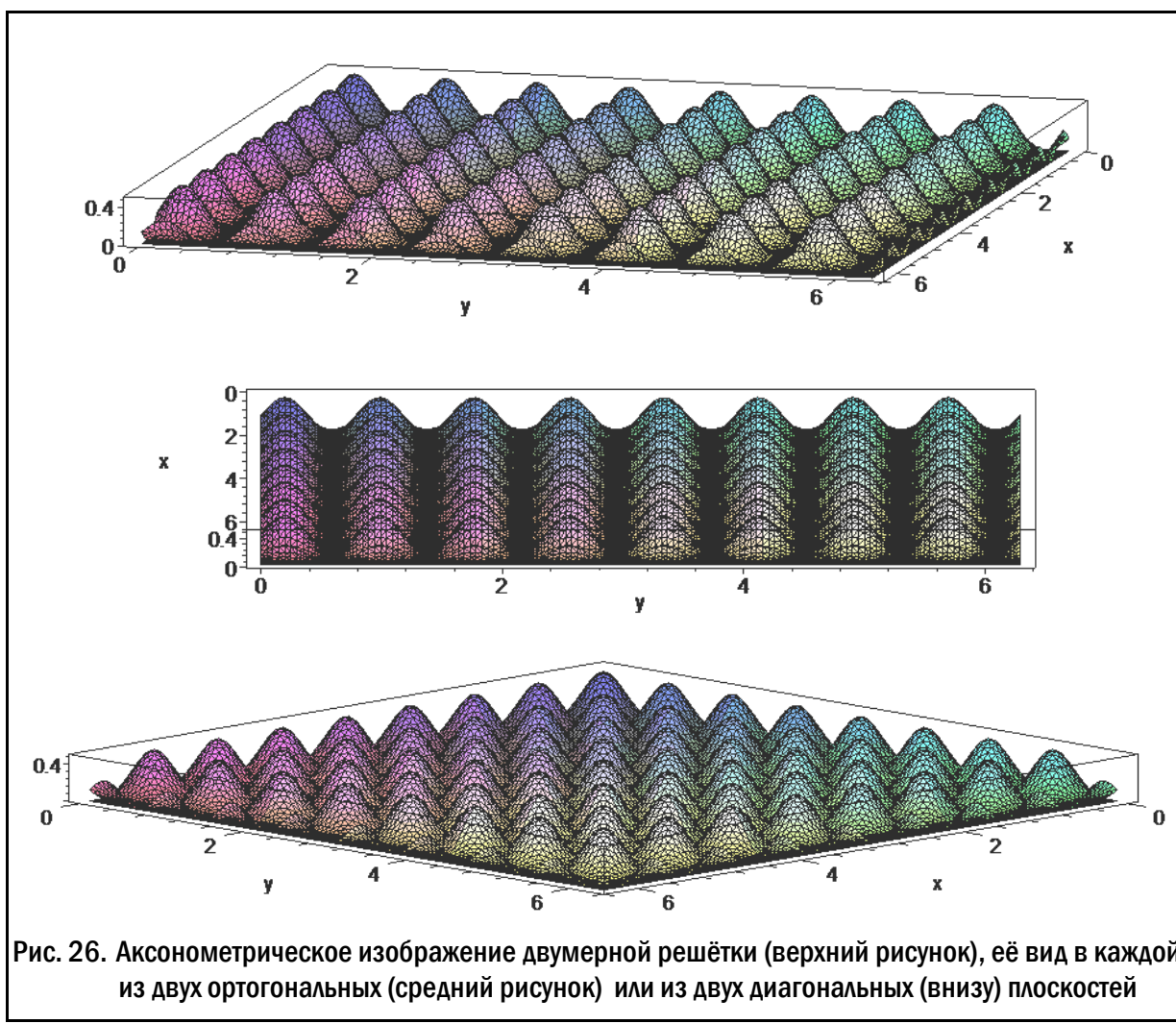

Рис. 26. Аксонометрическое изображение двумерной решётки (верхний рисунок), её вид в каждой из двух ортогональных (средний рисунок) или из двух диагональных (внизу) плоскостей

Действительно, элементы исходной двумерной синусоидальной решётки, чередующиеся подобно чёрным и белым полям шахматной доски (см. рис. 27 на стр. 31), образуют две геометрически правильные двумерные синусоидальные субрешётки, но уже с диагональным положением штрихов (эти субрешётки находятся "одна в другой").

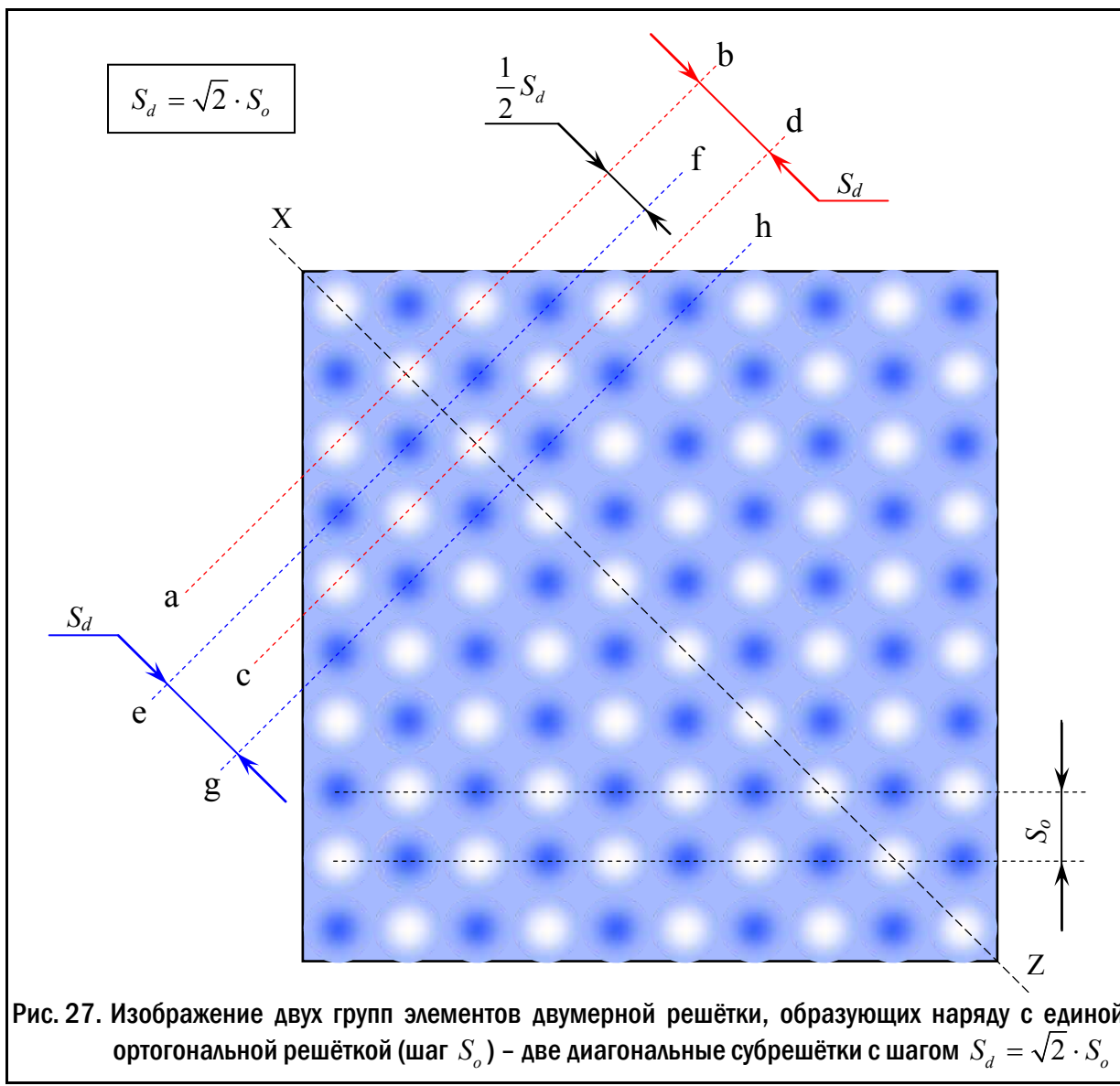

**Рис. 27.** Изображение двух групп элементов двумерной решётки, образующих наряду с единой ортогональной решёткой (шаг $S_o$) – две диагональные субрешётки с шагом $S_d = \sqrt{2} \cdot S_o$

В ортогональных плоскостях элементы обеих субрешёток располагаются строго друг за другом. Благодаря этому они синфазно работают вместе, образуя единую дифракционную двумерную решётку с шагом $S_o$, теория которой хорошо изучена.

На диагональных направлениях, например, таких, как X-Z, элементы каждой отдельной субрешётки по-прежнему следуют друг за другом. Но элементы, принадлежащие *разным* субрешёткам, теперь находятся на разных диагональных линиях, что позволяет каждой отдельной субрешётке самостоятельно и независимо формировать свои собственные каналы когерентного рассеяния в надлежащей диагональной плоскости.

Как видно из рис. 27, диагональный шаг каждой из субрешёток равен $S_d = \sqrt{2} \cdot S_o$, а их взаимное смещение в диагональных направлениях равно половине данного шага. Последнее обстоятельство приводит к тому, что все однотипные каналы когерентного рассеяния обеих субрешёток (кроме нулевого) "запираются" взаимной интерференцией.

Таким образом, эти каналы, соответствующие главным нечётным максимумам диагональных субрешёток, являются латентными (скрытыми). Если, например, отвесно направить лазерный луч на поверхность двумерной фазовой решётки (см. рис. 28), то мы увидим только "классические" дифракционные максимумы, угловые направления которых будут принадлежать ортогональным (синим) плоскостям. В диагональных же плоскостях (красных) никаких порождённых субрешётками максимумов – не будет.

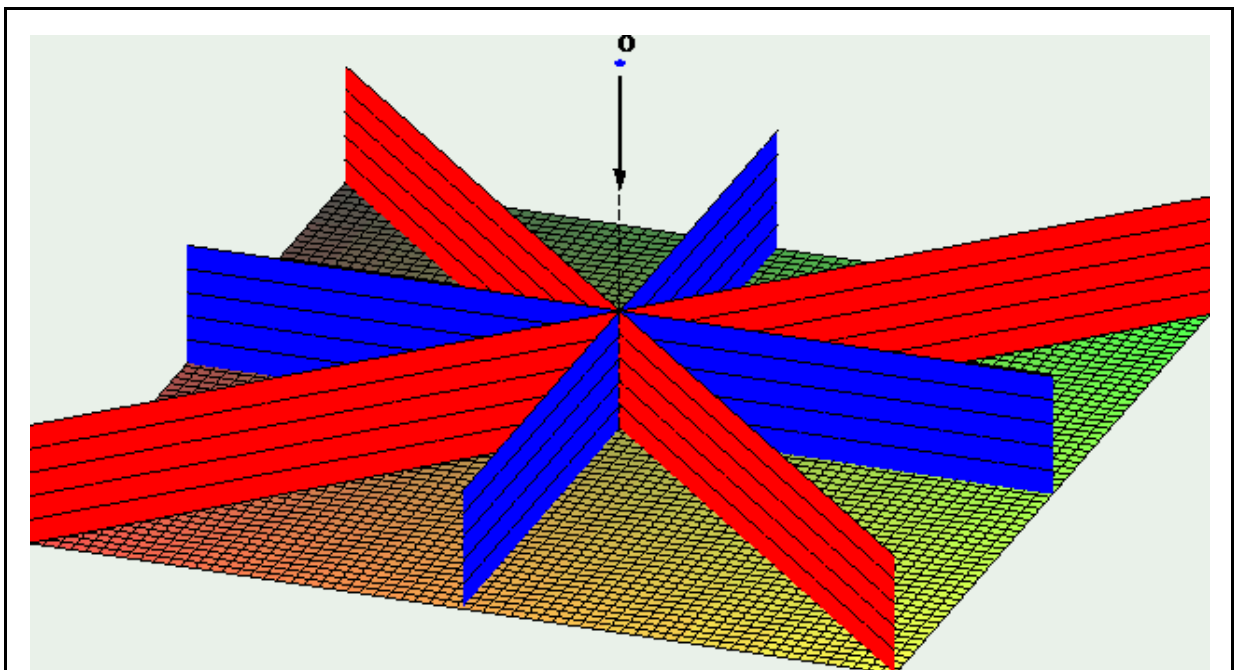

**Рис. 28.** Поверхность двумерной дифракционной решётки, на которую по нормали падает фотон; рассеяние нечётных порядков разрешено в синих плоскостях и запрещено – в красных

Тем не менее, описанные латентные каналы непосредственно влияют на характер вероятностного распределения фотонов по "разрешённым" направлениям дифракционного рассеяния. Дело в том, что каждая из диагональных субрешёток формирует свою квоту фотонов для латентного канала, заранее "не зная", что этот канал будет "заперт" *другой* субрешёткой[1]. Поэтому та доля фотонов, которая предназначалась для запертого интерференцией латентного канала, просто перераспределяется между разрешёнными каналами когерентного рассеяния, – в соответствии с их пропускной способностью[2].

Единственным разрешённым каналом рассеяния, лежащим в диагональной плоскости, является зеркальный максимум нулевого порядка. Для решётки с незеркальной поверхностью есть и другие открытые каналы, которые, однако, лежат вне диагональных плоскостей. Следовательно, перераспределение "латентной квоты" излучения в пользу разрешённых каналов неизбежно связано с потерей диагональной плоскостью соответствующей доли фотонов, поскольку зеркальный максимум, лежащий в такой диагональной плоскости, получит не все фотоны латентных каналов, а только их часть.

---

[1] Это принципиально отличает ситуацию от случая, когда, например, дифракционный минимум (т. е. интерференционный запрет на рассеяние в выделенном направлении) и распределение по максимумам – формируются элементами одной и той же решётки.

[2] Как в соответствующих задачах по гидравлике или при расчёте электрических цепей.

Рассмотренный механизм рассеяния частично аналогичен тому, который реализуется при формировании аномалий Рэлея-Вуда [7, 8]. Роберт Вуд ещё в 1902 году обнаружил минимумы и максимумы спектра отражения металлической решётки. Объяснение таким явлениям дал Рэлей, который выявил сингулярности электромагнитного поля для случая, когда один из каналов дифракционного рассеяния становится параллельным макроповерхности проводящей решётки. При окончательном исчезновении этого порядка дифракции возникает скачкообразное перераспределение амплитуд вероятности рассеяния между существующими открытыми каналами.

В настоящей работе при формировании дифракционных аномалий в качестве "исчезающего порядка дифракции" используется не открытый канал когерентного рассеяния, как это происходит при возникновении аномалий Рэлея-Вуда, а латентный канал, запертый взаимной интерференцией диагональных субрешёток. Это исключает необходимость поглощения фотонов решёткой при аномальном режиме их дифракционного рассеяния. Дифракция здесь остаётся изоэнергетическим процессом углового перераспределения фотонов (см. рис. 29), что важно с точки зрения цели данных исследований.

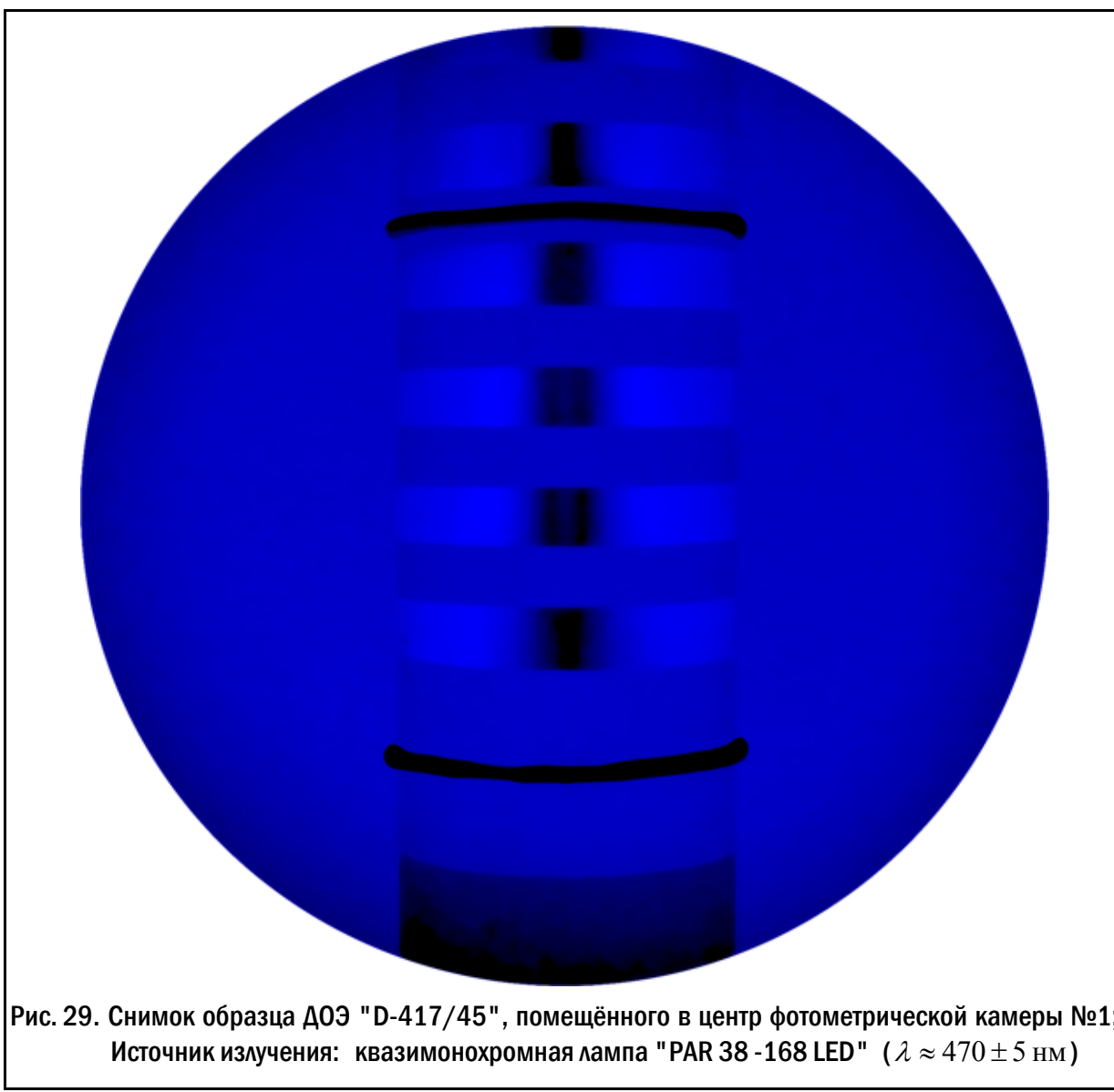

Рис. 29. Снимок образца ДОЭ "D-417/45", помещённого в центр фотометрической камеры №1; Источник излучения: квазимонохромная лампа "PAR 38 -168 LED" ($\lambda \approx 470 \pm 5$ нм)

### Проверка предположения о существовании латентных каналов когерентного рассеяния

Гипотеза о существовании латентных каналов носит умозрительный характер. Без объективного доказательства своей справедливости она так и останется изощрённой попыткой объяснить реальные физические явления на уровне натуральной философии.

Попробуем теперь количественно оценить масштаб потери фотонов диагональной плоскостью двумерной решётки, и сравним полученное с результатами экспериментов. Разумное соответствие между фотометрическими данными и математической моделью обнаруженного явления позволит более серьёзно отнестись к выдвинутой гипотезе.

Имеется двумерная фазовая решётка с синусоидальным профилем микрорельефа, ортогональный шаг которой $S_o$, а диагональный шаг $S_d = \sqrt{2} \cdot S_o$. Пусть на поверхность этой решётки в её диагональной плоскости падает фотон, причём угол его падения равен $\alpha$, а длина волны — не превышает величины диагонального шага: $\lambda \leq S_d$.

Рассмотрим структуру области допустимых значений угла рассеяния $\beta$, лежащего в той же диагональной плоскости, что и угол падения $\alpha$. Вся эта область занимает открытый интервал от $-\pi/2$ до $+\pi/2$. В этом интервале выделим зону $\beta_{\min} \leq \beta \leq \beta_{\max}$ индикатрисы, "заполненную" её главным зеркальным максимумом нулевого порядка:

**левая граница (минимум порядка -1):** $\quad \beta_{\min} = -\arcsin\left(\sin(\alpha) + \frac{1}{2} \cdot \frac{\lambda}{S_d}\right)$ или $\beta_{\min} = -\frac{\pi}{2}$; (6)

**правая граница (минимум порядка +1):** $\quad \beta_{\max} = -\arcsin\left(\sin(\alpha) - \frac{1}{2} \cdot \frac{\lambda}{S_d}\right)$ или $\beta_{\max} = +\frac{\pi}{2}$. (7)

предельные значения $\beta_{\min} = -\pi/2$ и (или) $\beta_{\max} = +\pi/2$ выбираются в том случае, когда модуль величины аргумента арксинуса превышает единицу, т. е. когда не существует вещественного значения угла $\beta_{\min}$ или $\beta_{\max}$ для нужного минимума первого порядка.

Введём теперь безразмерную функцию, которая для отдельной диагональной субрешётки равняется отношению суммарной вероятности рассеяния в максимумы нечётных порядков к аналогичной вероятности рассеяния в область зеркального максимума:

$$\xi(\alpha, \lambda, S_d, N) = \frac{\int\limits_{-\pi/2}^{\beta_{\min}} I_m(\alpha, \beta, \lambda, S_d, N) \cdot d\beta + \int\limits_{\beta_{\max}}^{+\pi/2} I_m(\alpha, \beta, \lambda, S_d, N) \cdot d\beta}{\int\limits_{\beta_{\min}}^{\beta_{\max}} I_m(\alpha, \beta, \lambda, S_d, N) \cdot d\beta} \qquad (8)$$

где функция $I_m(\alpha, \beta, \lambda, S_d, N)$ — плотность вероятности, аналогичная представленной формулой (5) на стр. 16. Разница состоит лишь в том, что здесь вместо ортогонального шага полной решётки $S_o$ указан диагональный шаг $S_d$ для её субрешёток.

Из ранее сказанного очевидно, что чем выше значение интегрального функционала $\xi(\alpha, \lambda, S_d, N)$, тем ме́ньшая доля фотонов, падающих в диагональной плоскости под углом $\alpha$ на поверхность решётки, попадёт в область нулевого зеркального максимума. Но, в отличие от рассеяния в ортогональной плоскости (см. стр. 16 и далее), максимумы нечётных порядков не смогут возместить эту потерю при иных значениях угла падения по причине "выдавливания" части своих фотонов за пределы диагональной плоскости.

Для проверки наличия корреляции между функцией $\xi(\alpha, \lambda, S_d, N)$ и уровнем яркости поверхности двумерной решётки, наблюдаемой в диагональной плоскости под разными углами, надо выбрать результаты эксперимента со следующими параметрами:

– спектральный состав используемого излучения должен включать в себя все участки RGB-диапазона ("красный", "зелёный" и "синий"), причём интенсивность этих трёх компонент излучения должна быть примерно одинакова и достаточно велика;

– наиболее существенные изменения яркости поверхности ДОЭ должны быть сосредоточены в диапазоне углов отражения небольшой протяжённости, что позволит легко локализовать угловую зону с самым минимальным значением уровня яркости.

В качестве подходящего объекта анализа были выбраны результаты эксперимента на образце "D-833/45" с галогенным источником излучения ($T \approx 3000°K$). На рис. 30 даётся снимок ДОЭ "D-833/45", помещённого в фотометрическую камеру №1, а также спектральные снимки анализируемого сегмента его поверхности. Снимок анализируемого сегмента содержит 814×138=112332 пиксела (по 3 sRGB-субпиксела). Этот сегмент делится на 200 групп (участков наблюдения) по $\approx 562$ пиксела в каждом участке.

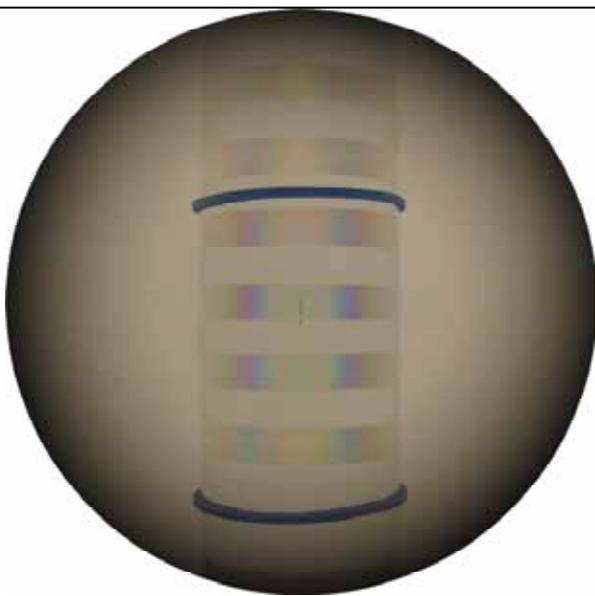

Рис. 30. Фотографические изображения дифракционного оптического элемента "D-833/45" и анализируемого сегмента его поверхности, – в диффузном световом поле камеры №1. Источник излучения: галогенная лампа 35 Ватт с цветовой температурой $T \approx 3000°K$

Далее, на каждом из рис. 31, рис. 32 и рис. 33 (стр. 36-38), – приводятся по три графика, содержащих результаты обработки спектральных изображений сегмента:

– **Верхний график:** угловая зависимость относительной яркости сегмента "D-833/45" (математическое ожидание и дисперсия) для выбранного спектрального диапазона;

– **Средний график:** отношение вероятностей $\xi(\alpha, \lambda, S_d, N)$, определяемое формулой (8);

– **Нижний график:** индикатриса рассеяния для $\alpha \to \min(\xi(\alpha, \lambda, S_d, N))$; для сравнения изображена индикатриса, соответствующая случаю $\alpha = 0$; угол $\beta$ - дан в радианах.

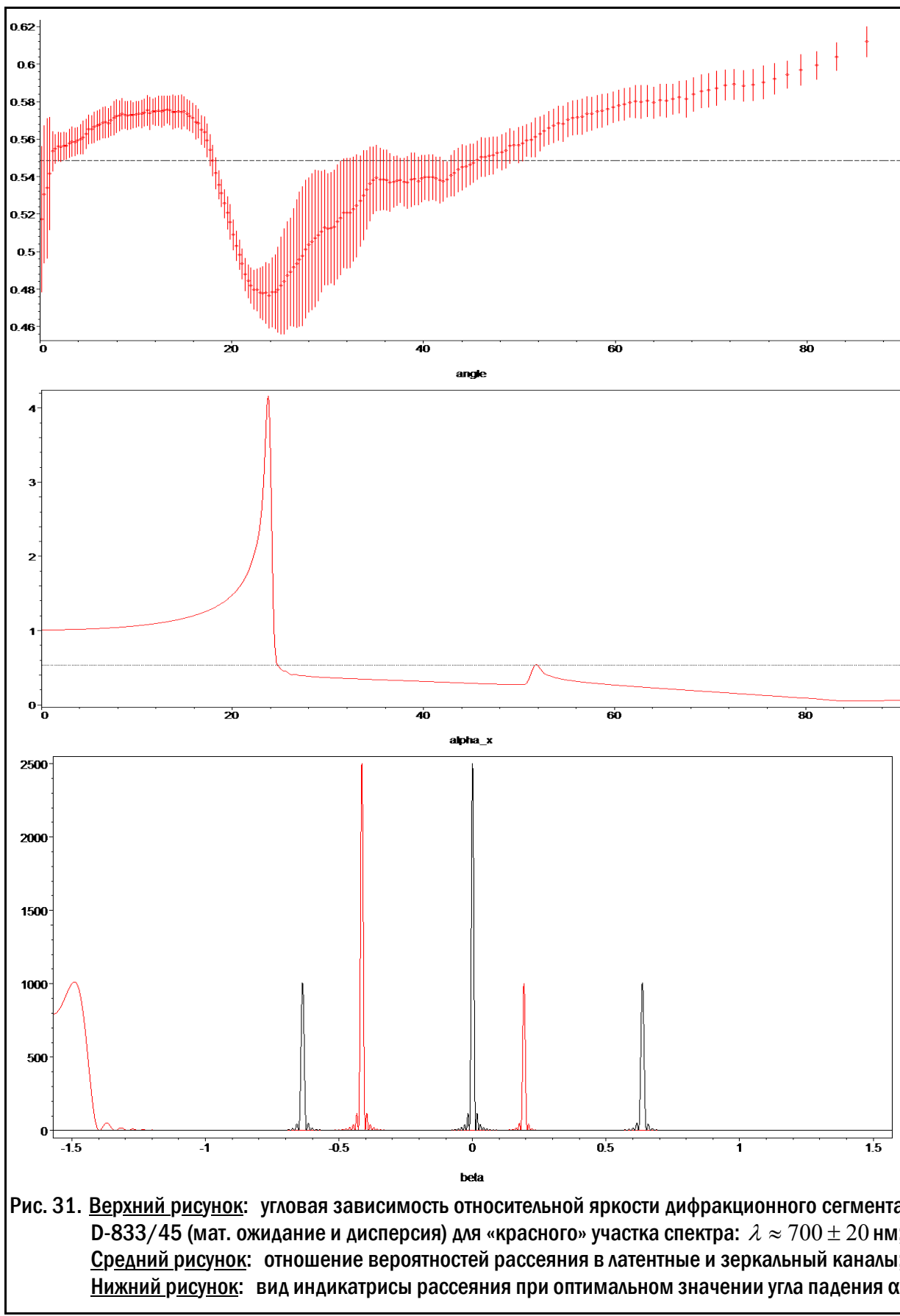

Рис. 31. <u>Верхний рисунок</u>: угловая зависимость относительной яркости дифракционного сегмента D-833/45 (мат. ожидание и дисперсия) для «красного» участка спектра: $\lambda \approx 700 \pm 20$ нм;
<u>Средний рисунок</u>: отношение вероятностей рассеяния в латентные и зеркальный каналы;
<u>Нижний рисунок</u>: вид индикатрисы рассеяния при оптимальном значении угла падения α

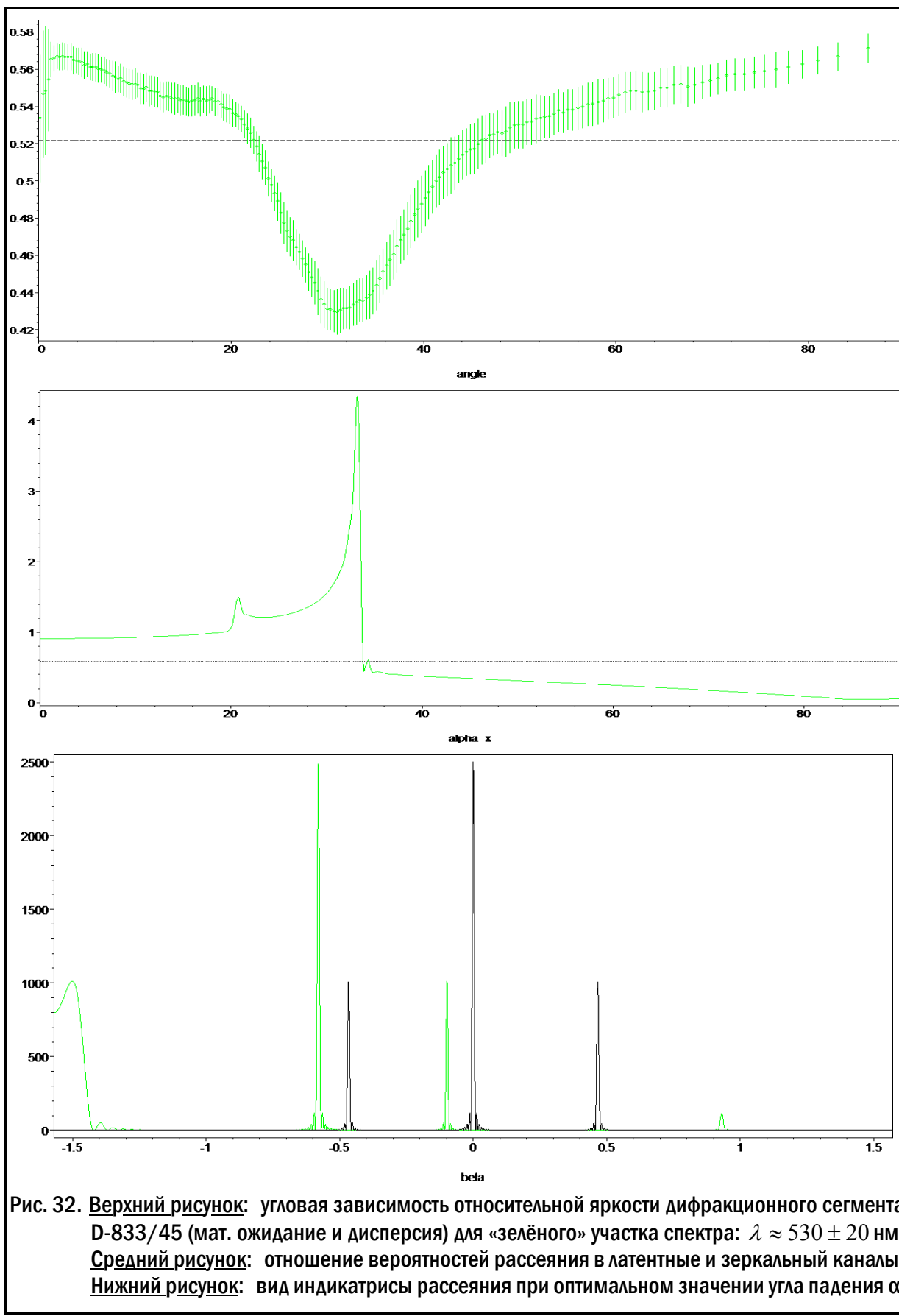

Рис. 32. <u>Верхний рисунок</u>: угловая зависимость относительной яркости дифракционного сегмента D-833/45 (мат. ожидание и дисперсия) для «зелёного» участка спектра: $\lambda \approx 530 \pm 20$ нм;
<u>Средний рисунок</u>: отношение вероятностей рассеяния в латентные и зеркальный каналы;
<u>Нижний рисунок</u>: вид индикатрисы рассеяния при оптимальном значении угла падения α

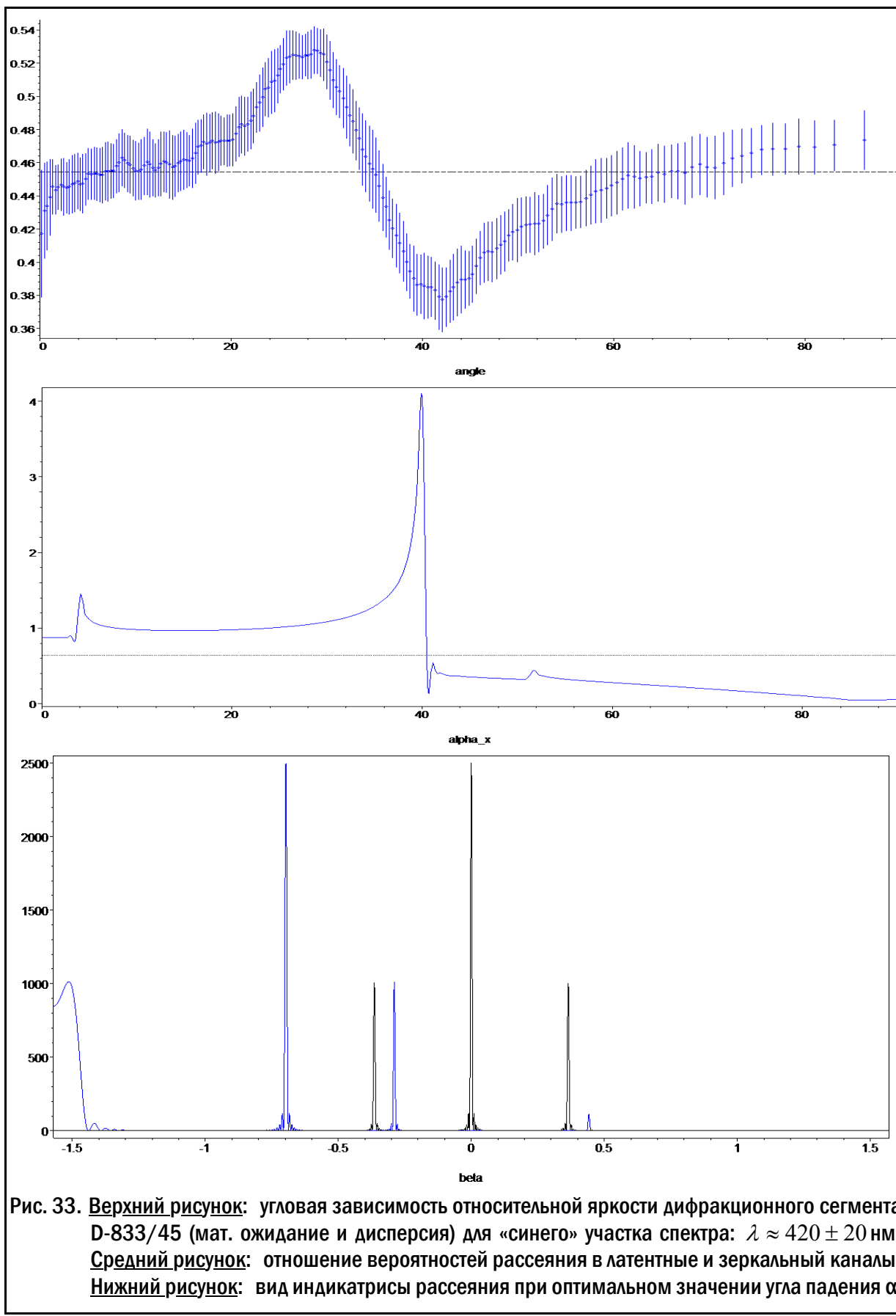

Рис. 33. <u>Верхний рисунок</u>: угловая зависимость относительной яркости дифракционного сегмента D-833/45 (мат. ожидание и дисперсия) для «синего» участка спектра: $\lambda \approx 420 \pm 20$ нм;
<u>Средний рисунок</u>: отношение вероятностей рассеяния в латентные и зеркальный каналы;
<u>Нижний рисунок</u>: вид индикатрисы рассеяния при оптимальном значении угла падения α

В таблице 2 приводится информация анализа, характеризующая степень соответствия экспериментальных данных расчётным[1]. Полученное расхождение приемлемо, поскольку для фотометрии погрешность измерений в 5% считается вполне допустимой.

**Таблица 2.** Сравнение экспериментальных и расчётных значений угла отражения $\beta$, для которого:
$$\beta = -\alpha \to \min\left(\xi(\alpha, \lambda, S_d, N)\right)$$
(ДОЭ "D-833/45", установка №1, галогенная лампа с цветовой температурой $T \approx 3000°K$)

| Длина волны (нм) | Эксперимент | Теор. расчёт | Абс. погрешность | % погрешности |
|---|---|---|---|---|
| ● Red:    $\lambda \approx 700 \pm 20$ | 23.8764° | 23.7443° | −00.1321° | 00.1468 % |
| ● Green: $\lambda \approx 530 \pm 20$ | 30.9998° | 33.2130° | +02.2132° | 02.4591 % |
| ● Blue:   $\lambda \approx 420 \pm 20$ | 42.0654° | 39.9305° | −02.1349° | 02.3721 % |

Для дополнительной проверки наличия предполагаемой корреляции всё-таки проанализируем результаты уже другого эксперимента – с образцом "D-1667/45", имеющим более крупный микрорельеф. На рис. 34 даётся снимок ДОЭ "D-1667/45", помещённого в фотометрическую камеру №1, а также спектральные снимки анализируемого сегмента его поверхности. В данном случае исследуемый сегмент изображения образца содержит 804×113=90852 пиксела (по 3 sRGB-субпиксела). Этот сегмент делится на 200 групп (участков наблюдения) по $\approx 454$ пиксела в каждом участке.

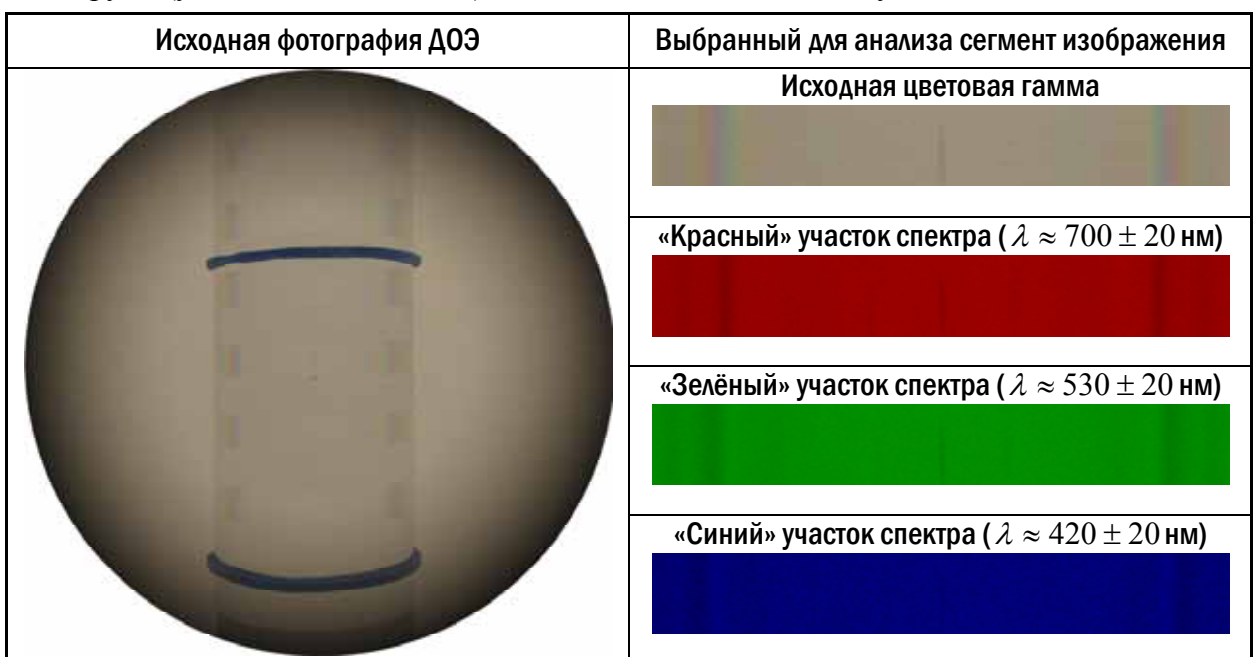

**Рис. 34.** Фотографические изображения дифракционного оптического элемента "D-1667/45" и анализируемого сегмента его поверхности, – в диффузном световом поле камеры №1. Источник излучения: галогенная лампа 35 Ватт с цветовой температурой $T \approx 3000°K$

---

[1] Расчёт значений индикатрисы (5) и функции (8) производился для значения числа штрихов дифракционной решётки $N = 50$. Эта величина уже достаточно велика, чтобы значения содержащих её функций стали близки к пределу $N \to +\infty$, но при этом ещё можно визуально оценить, например, ширину максимумов графика индикатрисы.

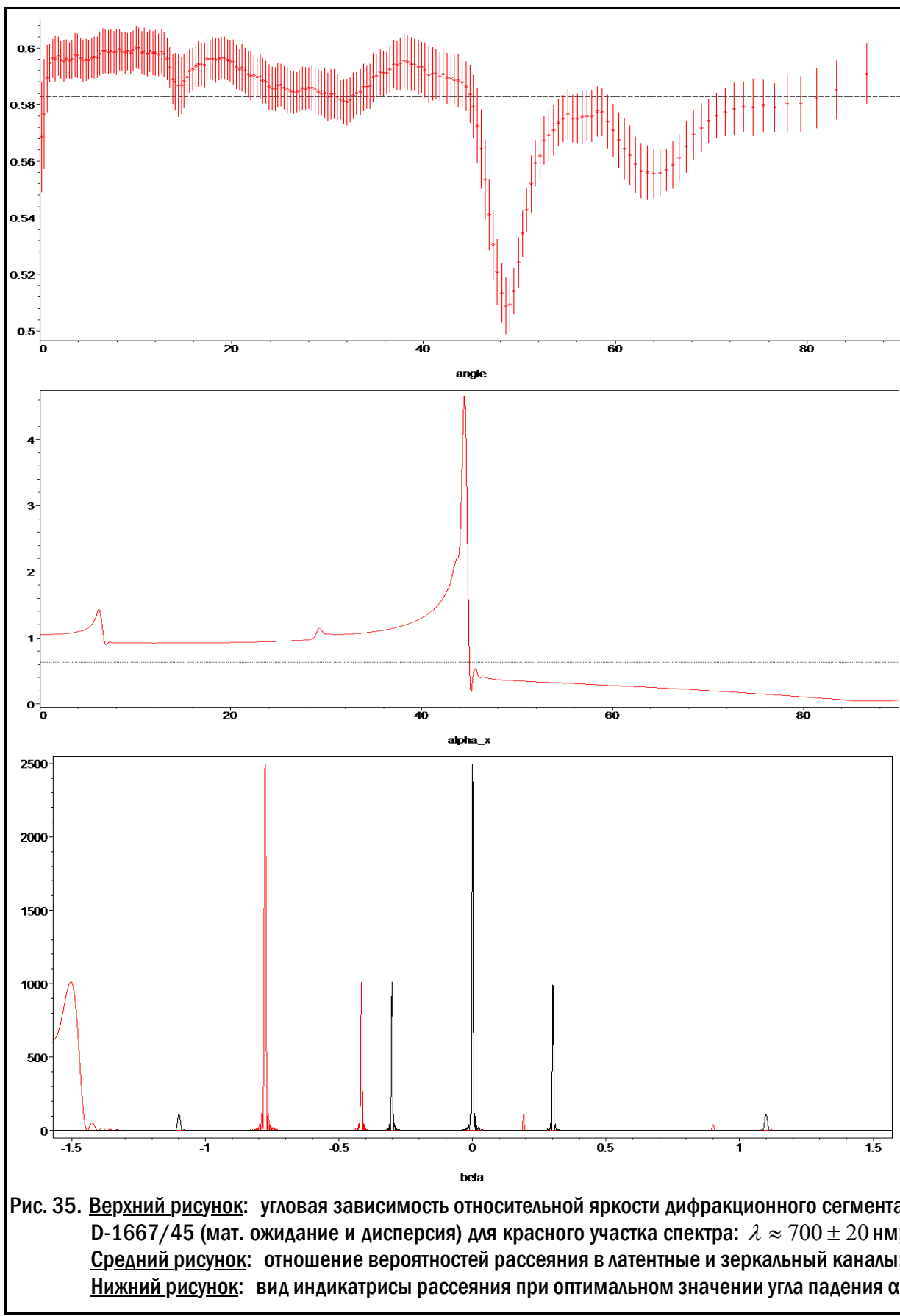

**Рис. 35.** <u>Верхний рисунок</u>: угловая зависимость относительной яркости дифракционного сегмента D-1667/45 (мат. ожидание и дисперсия) для красного участка спектра: $\lambda \approx 700 \pm 20$ нм;
<u>Средний рисунок</u>: отношение вероятностей рассеяния в латентные и зеркальный каналы;
<u>Нижний рисунок</u>: вид индикатрисы рассеяния при оптимальном значении угла падения α

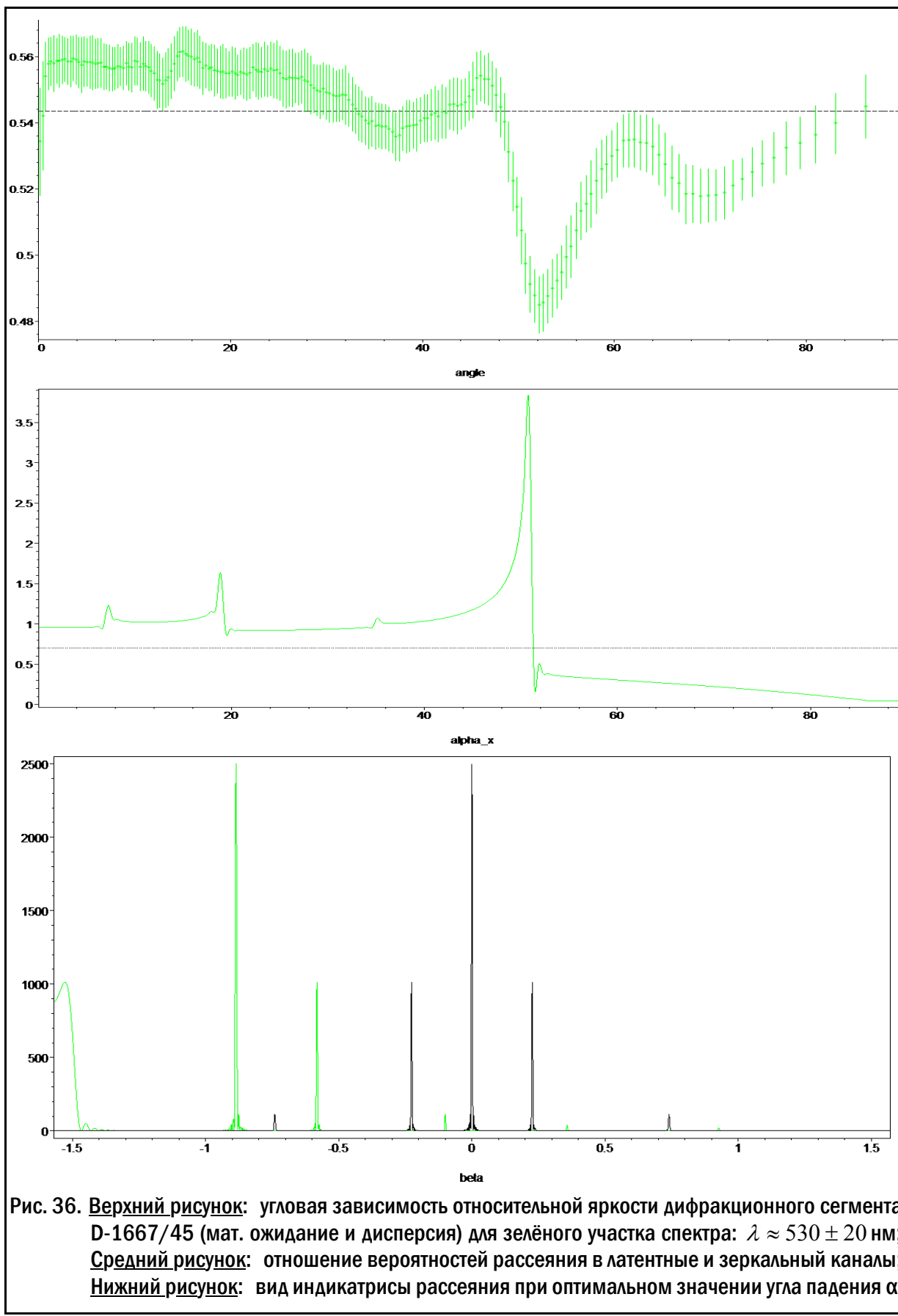

Рис. 36. <u>Верхний рисунок</u>: угловая зависимость относительной яркости дифракционного сегмента D-1667/45 (мат. ожидание и дисперсия) для зелёного участка спектра: $\lambda \approx 530 \pm 20$ нм;
<u>Средний рисунок</u>: отношение вероятностей рассеяния в латентные и зеркальный каналы;
<u>Нижний рисунок</u>: вид индикатрисы рассеяния при оптимальном значении угла падения α

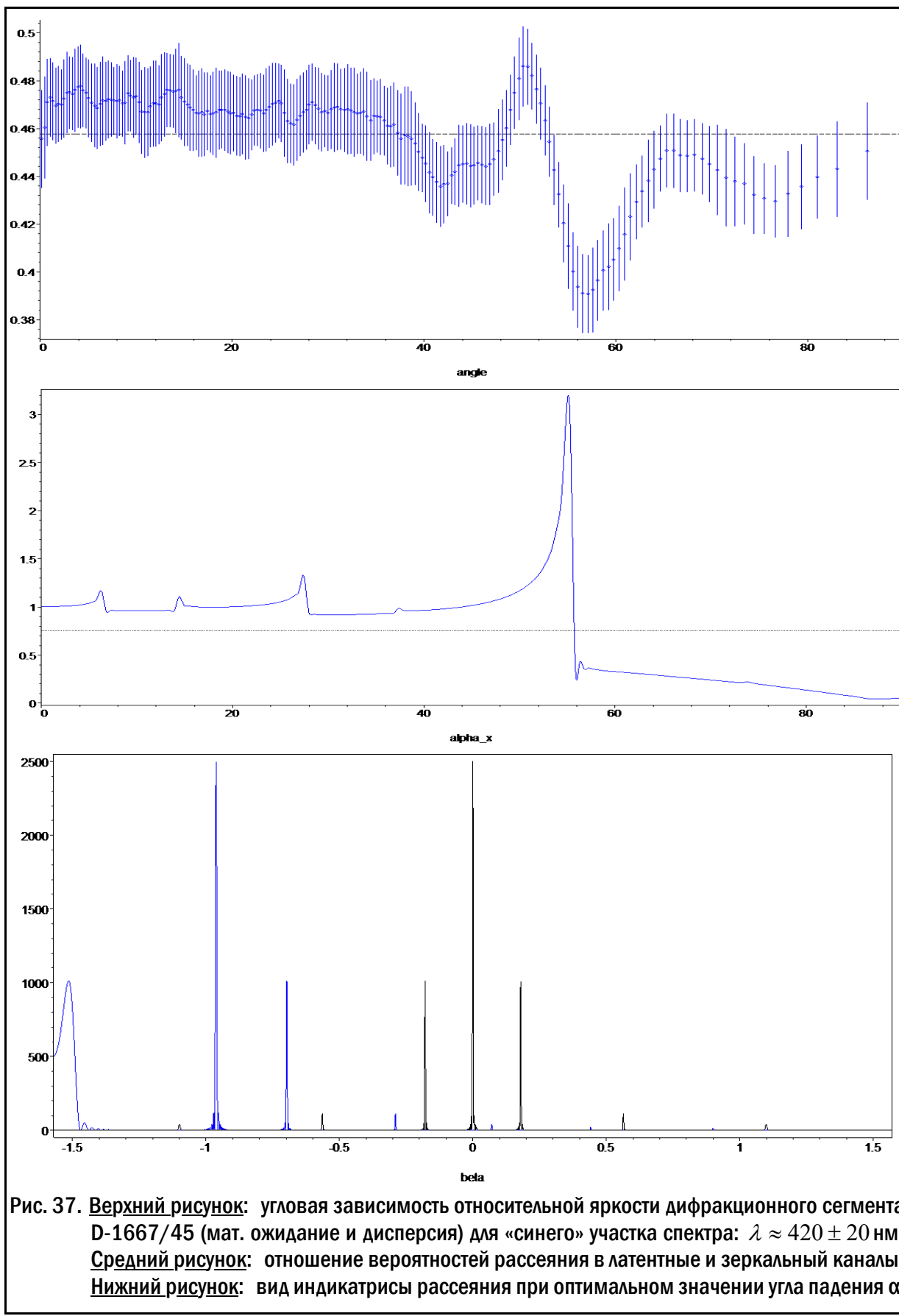

Рис. 37. <u>Верхний рисунок</u>: угловая зависимость относительной яркости дифракционного сегмента D-1667/45 (мат. ожидание и дисперсия) для «синего» участка спектра: $\lambda \approx 420 \pm 20$ нм;
<u>Средний рисунок</u>: отношение вероятностей рассеяния в латентные и зеркальный каналы;
<u>Нижний рисунок</u>: вид индикатрисы рассеяния при оптимальном значении угла падения α

Рисунки 35-37 (см. стр. 40-42), – содержат результаты обработки спектральных изображений сегмента "D-1667/45", аналогичные ранее приведённым для "D-833/45".

В таблице 3 приводится информация анализа, характеризующая степень соответствия экспериментальных данных расчётным. И здесь полученное расхождение находится на приемлемом уровне и не превышает допустимых 5%.

| Таблица 3. Сравнение экспериментальных и расчётных значений угла отражения $\beta$, для которого: $\beta = -\alpha \rightarrow \min\left(\xi\left(\alpha, \lambda, S_d, N\right)\right)$ (ДОЭ "D-1667/45", установка №1, галогенная лампа с цветовой температурой $T \approx 3000°K$) | | | | |
|---|---|---|---|---|
| Длина волны (нм) | Эксперимент | Теор. расчёт | Абс. погрешность | % погрешности |
| ● Red:    $\lambda \approx 700 \pm 20$ | 48.5650° | 44.4959° | -04.0691° | 04.5212 % |
| ● Green: $\lambda \approx 530 \pm 20$ | 52.1389° | 50.7401° | -01.3988° | 01.5542 % |
| ● Blue:   $\lambda \approx 420 \pm 20$ | 57.0528° | 55.1125° | -01.9403° | 02.1559 % |

Проверка воспроизводимости корреляции между функцией $\xi\left(\alpha, \lambda, S_d, N\right)$ и уровнем яркости поверхности двумерной решётки, наблюдаемой в диагональной плоскости под разными углами, была реализована для всех конструктивно выполнимых сочетаний каждого отдельного образца ДОЭ "D-417/45", "D-833/45" или "D-1667/45" – со всеми имеющимися фотометрическими камерами и источниками излучения.

Для образцов "D-833/45" и "D-1667/45" относительное расхождение между результатами экспериментов и теоретических расчётов оказалось умеренным (1-3%) и не превышало граничного уровня в 5%. Надо заметить, что все работы с данными образцами проводились при длине волн излучения, существенно меньших, чем размеры диагонального шага дифракционных решёток ДОЭ: $\lambda \ll S_d$.

Для образца "D-417/45" по мере приближения длины волны излучения к размеру диагонального шага ($\lambda \approx S_d$) величина вышеуказанной погрешности быстро возрастала до уровня в 10-15%; при этом оптимальный угол $\beta \rightarrow +0°$. Когда же длина волны превышала диагональный шаг ($\lambda > S_d$), то угловая локализация наблюдаемого эффекта становилась неопределённой, и теоретический расчёт его местоположения терял смысл.

В настоящей главе были приведены результаты проверки теоретического предположения о существовании латентных каналов когерентного рассеяния. Изложенный материал даёт объективное основание для положительного ответа на данный вопрос:

*Латентные каналы когерентного рассеяния действительно существуют, и их роль*[1] *в анизотропном перераспределении фотонов диффузного газа внутри геометрического пространства соответствует ранее выдвинутым теоретическим предположениям.*

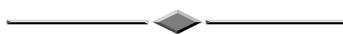

---

[1] На понятийном уровне можно сказать, что дифракционная решётка на поверхности изучаемых ДОЭ играет роль своеобразного "фотонного клапана". Решётка принимает фотоны из окружающего её диффузного газа с любого направления, но минимизирует обратное движение частиц внутри угловых каналов с $\beta = -\alpha \rightarrow \min\left(\xi\left(\alpha, \lambda, S_d, N\right)\right)$.

## Рекомендации по выбору оптимальных параметров физических установок

Первая рекомендация относится к методу расчёта угловой локализации наблюдаемого оптического эффекта. В предыдущей главе для этого непосредственно использовалась зависимость (8) интегрального функционала $\xi(\alpha, \lambda, S_d, N)$ от угла падения[1] $\alpha$, вычислялись значения этого функционала для всего диапазона варьирования параметра $\alpha \in [0, \pi/2)$, и по положению глобального максимума на полученном графике находилось оптимальное значение угла рассеяния $\beta$: $\beta = |\alpha| \to \min(\xi(\alpha, \lambda, S_d, N))$.

Очевидно, что описанный способ связан с большими трудозатратами[2]. Для получения простой эмпирической формулы расчёта оптимального значения угла $\beta$ можно использовать ранее выявленную особенность индикатрисы рассеяния, а именно: искомый эффект максимально проявляется в тот момент, когда угловая ориентация какого-либо из латентных каналов начинает становиться "скользящей" ($\beta \to \pm 90°$) и "удельный вес" данного канала резко возрастает. Это обстоятельство хорошо заметно на нижних графиках рис. 31-33 (стр. 36-38) и рис. 35-37 (стр. 40-42), где изображены сравнительные индикатрисы для случаев нормального и оптимального углов падения.

Для рассматриваемых решёток с квазисинусоидальным профилем максимумы чётных порядков отсутствуют (кроме нулевого, зеркального). Удельный вес максимумов третьего и более высоких порядков – существенно меньше, чем у максимумов первого порядка. Поэтому будем считать, что упомянутый латентный канал, "скользящее" положение которого обуславливает минимальную яркость зеркальной компоненты рассеяния, – это латентный канал когерентного рассеяния, имеющий порядок ±1.

Из вышеуказанных графиков видно, что модуль оптимального значения угла скольжения для латентного канала чуть меньше "полностью горизонтальной" величины $\pi/2$ и всегда близок к значению примерно 3/2 радиана ($\approx 85.94°$). Исходя из известной формулы дифракционной решётки:

$$S_d \cdot (\sin(\alpha) + \sin(\beta)) = m \cdot \lambda \qquad (9)$$

где $\alpha$ – угол падения фотона на решётку: $\forall \alpha \in (-\pi/2, +\pi/2)$;
$\beta$ – угол отражения фотона от решётки: $\forall \beta \in (-\pi/2, +\pi/2)$;
$\lambda$ – длина волны рассеиваемого излучения: $\lambda > 0$;
$m$ – порядок максимума, наблюдаемого под углом $\beta$: $m = \pm 1$ (в данном случае);
$S_d$ – шаг (период) решётки вдоль её диагонального направления: $S_d \geq \lambda$.

получаем эмпирическую зависимость для вычисления оптимального угла $\beta_{opt}$:

$$\beta_{opt} \approx \pm \arcsin\left(\sin\left(\frac{3}{2}\right) - \frac{|m| \cdot \lambda}{S_d}\right) \quad \text{для} \quad \lambda < \left(\sin\left(\frac{3}{2}\right) + 1\right) \cdot \frac{S_d}{|m|}, \quad \text{где:} \quad m = \pm 1 \qquad (10)$$

---

[1] Параметры $\lambda, S_d$ и $N$ – являлись неизменными в каждом отдельном эксперименте.

[2] Интегралы в формуле (8) не имеют аналитического представления в замкнутой форме. Поэтому расчёт одного-единственного значения $\xi(\alpha, \lambda, S_d, N)$ требовал использования весьма особых численных методов (осциллирующая подынтегральная функция) с высоким порядком точности ("острый" максимум результирующего функционала).

Поскольку при $\lambda > S_d$ используемый метод расчёта $\beta_{opt}$ теряет свою значимость, то можно записать проще:

$$\beta_{opt} \approx \arcsin\left(Const - \frac{\lambda}{S_d}\right) \text{ для } \lambda \leq Const \cdot S_d, \text{ где: } Const = \sin\left(\frac{3}{2}\right) \cong 0.997495 \approx 1; \quad (11)$$

Точность предложенной аппроксимации можно оценить по графикам на рис. 38. На этом же рисунке для справки приведены результаты ранее выполненного экспериментального исследования оптимального угла $\beta_{opt}$ для ДОЭ "D-833/45" и "D-1667/45".

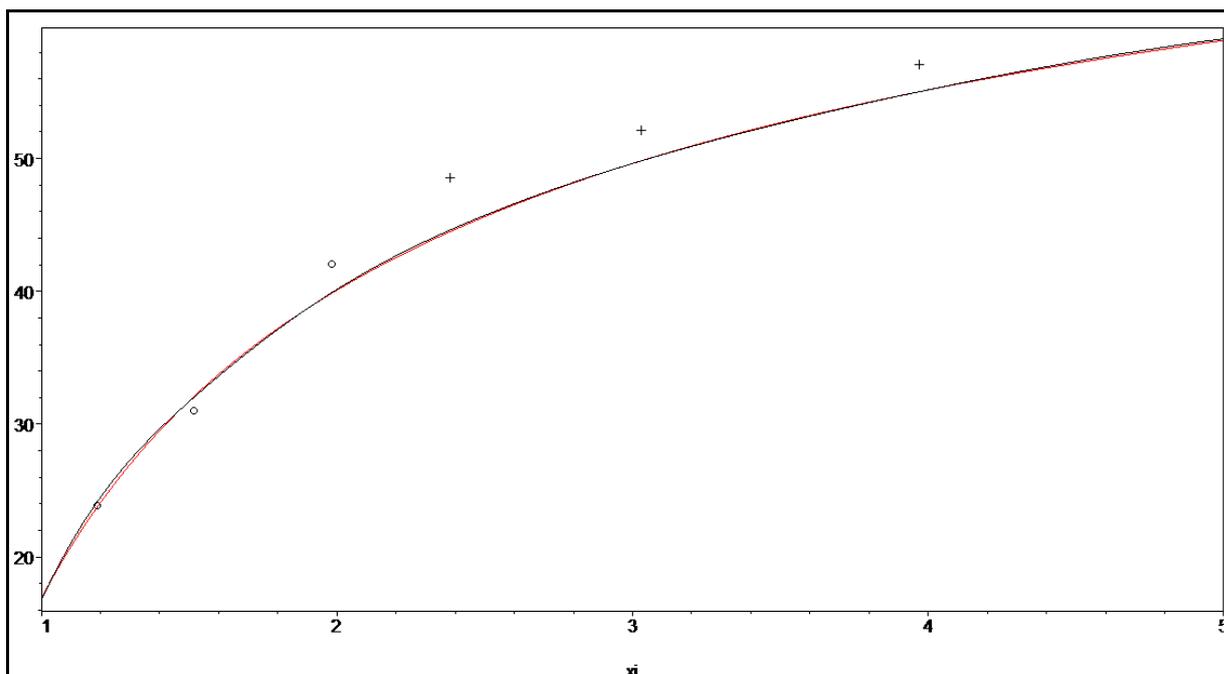

**Рис. 38.** Сравнение графиков оптимального угла $\beta$, вычисленных прямым способом на основе анализа функции (8) — чёрный цвет, и эмпирическим способом (11) — красный цвет;
○ — экспериментальные данные по исследованию образца "D-833/45" (см. табл. 2);
+ — экспериментальные данные по исследованию образца "D-1667/45" (см. табл. 3);
по оси абсцисс представлено соотношение $S_d/\lambda$, по оси ординат — угол $\beta$ (в градусах)

Вторая рекомендация относится к оптимальному выбору конкретного соотношения между шагом дифракционной решётки и доминирующей длиной волны излучения. Здесь настоятельно рекомендуется обратить внимание на следующее простое условие:

$$\text{Оптимально: } S_d = \sqrt{2} \cdot S_o = \lambda, \text{ при этом получается: } \beta_{opt} \approx 0° \quad (12)$$

При выполнении условия (12) проявление изучаемого эффекта становится чрезвычайно интенсивным. Для отвесного падении фотонов на поверхность такой решётки значение функционала (8) достигает глобального максимума, который значительно превышает его же локальные максимумы, реализуемые при других целочисленных кратностях[1] размера диагонального шага решётки и длины волны: $S_d = 2\lambda, 3\lambda, ...$

---

[1] Соотношение между $S_d$ и $\lambda$ не обязано быть строго оптимальным (целочисленным).

На рис. 39 дан график, поясняющий смысл изложенной рекомендации. На основании полученного опыта можно сказать, что когда анализируемый функционал достигает значения 3, изучаемый эффект становится уже заметен, при значении, равном 4, он заметен хорошо, а когда $\xi(\alpha, \lambda, S_d, N) \geq 5$ – эффект становится силён необычайно.

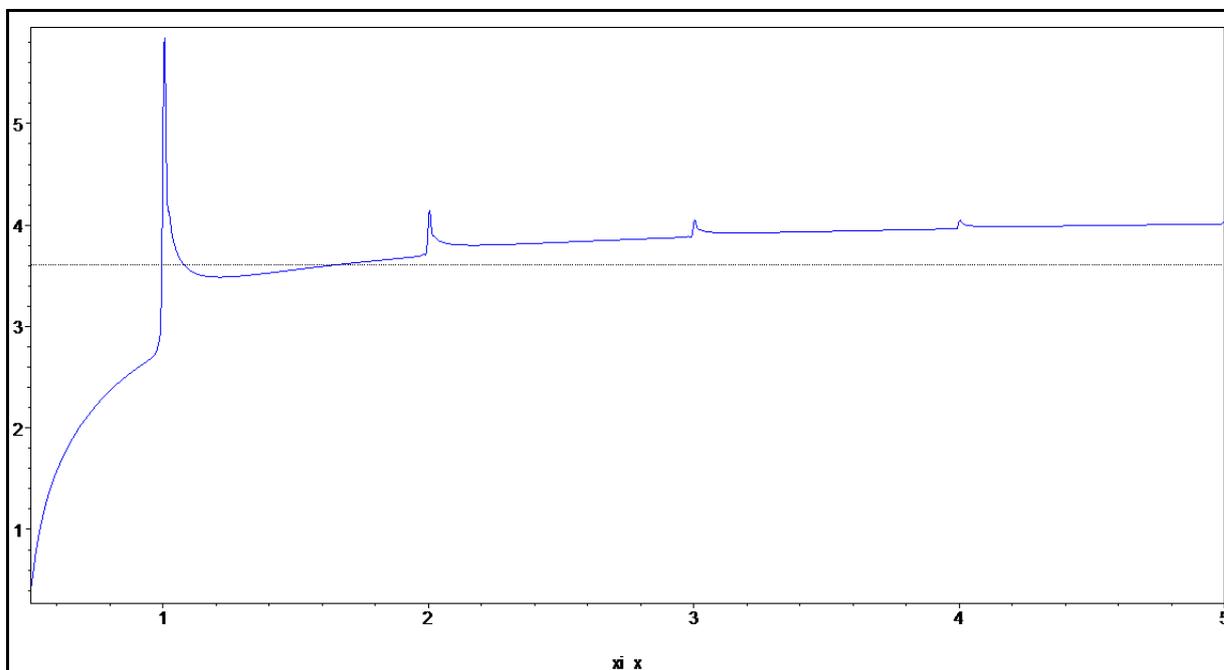

**Рис. 39.** По оси ординат – самые оптимальные значения функционала $\xi(\alpha, \lambda, S_d, N)$ – см. (8)
По оси абсцисс – представлено соотношение диагонального шага и длины волны $S_d/\lambda$

Причина наличия такого глобального максимума у функционала $\xi(\alpha, \lambda, S_d, N)$ ясна из вида индикатрисы рассеяния фотона на оптимальной решётке (см. рис. 40).

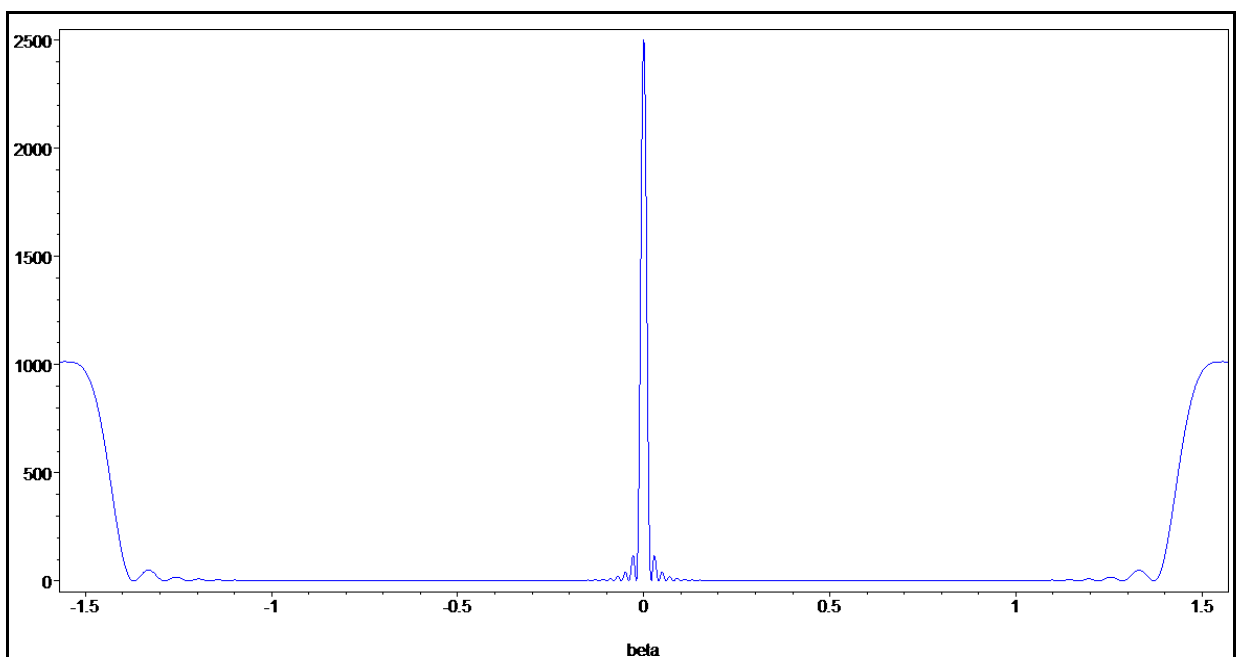

**Рис. 40.** Индикатриса рассеяния при нормальном ($\alpha = 0$) падении фотона на решётку с $S_d = \lambda$

Надо заметить, что для оптимальной ($S_d = \lambda$) решётки при нормальном ($\alpha = 0$) падении излучения на её поверхность, — число одновременно запертых латентных каналов отнюдь не удваивается, как это может показаться на основании только рис. 40. Суммарное количество этих каналов становится больше в четыре раза, поскольку при отвесном падении фотонов на дифракционную решётку обе её диагональные плоскости (см. рис. 28 на стр. 32) находятся в равном положении и "работают" одновременно.

Наглядный пример того, как выглядит визуальное проявление вышеописанного оптимального режима – можно увидеть на рис. 25 (стр. 29). Для образца "D-417/45" оптимальное условие $S_d = \lambda \approx 590$ нм (жёлтый цвет излучения) как раз реализовался[1] в проведённом эксперименте. Для сравнения можно посмотреть на результат другого, "неоптимального" ($S_d \approx 4\lambda$) эксперимента, — см. рис. 22 на стр. 26.

Что же касается рекомендаций по практическому использованию выявленного эффекта, то предварительный материал на этот счёт[2] можно найти в [6].

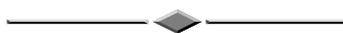

---

[1] Вид зависимости на рис. 39 говорит о том, что условие $S_d = \lambda$ надо соблюдать тонко.

[2] Для области физики полупроводников.

## Исследования второго плана

Кроме описанных работ с двумерными решётками, предназначенными для видимой части спектра, в рамках проекта **«Pith Fleck»** осуществлялись особые тесты на одномерной решётке "M-833/45", рассчитанной для видимого излучения, а также на двумерной решётке "D-10000/45" для дальней инфракрасной области ($\lambda \approx 8 \div 14$ мкм).

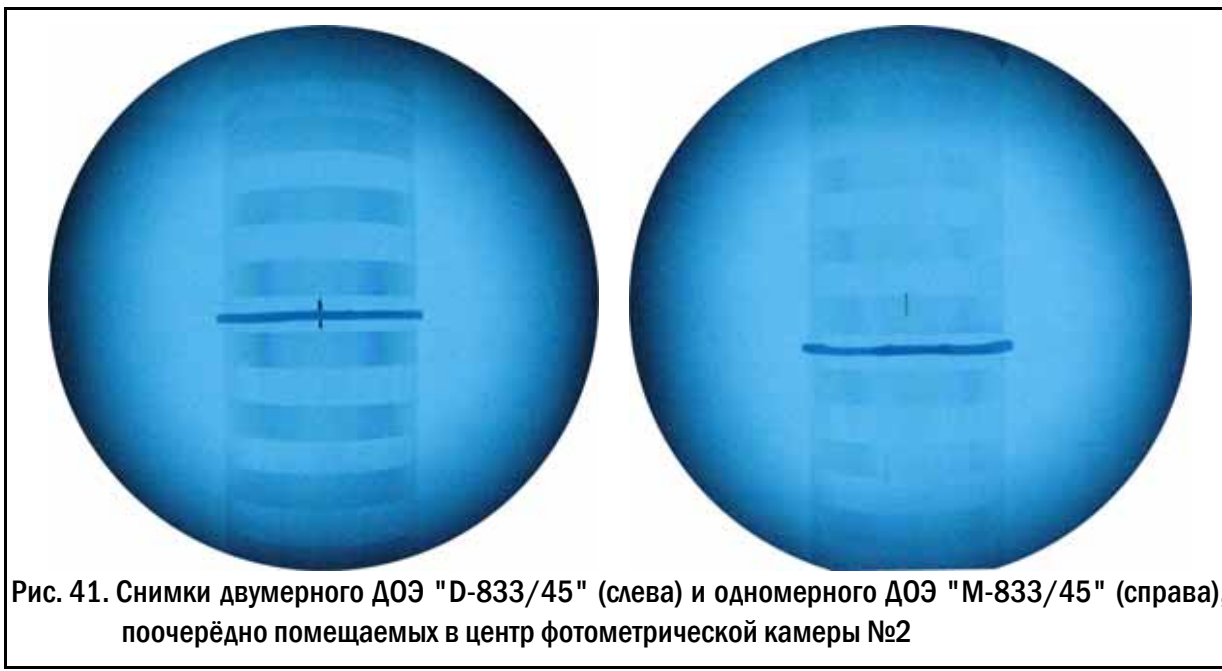

**Рис. 41.** Снимки двумерного ДОЭ "D-833/45" (слева) и одномерного ДОЭ "M-833/45" (справа), поочерёдно помещаемых в центр фотометрической камеры №2

Тесты на одномерной решётке "M-833/45" выявили воспроизводимое, но слабое подобие того эффекта, который уверенно проявляется на двумерных ДОЭ (см. рис. 41). Типичный же тепловизионный снимок образца "D-10000/45" – приводится на рис. 42.

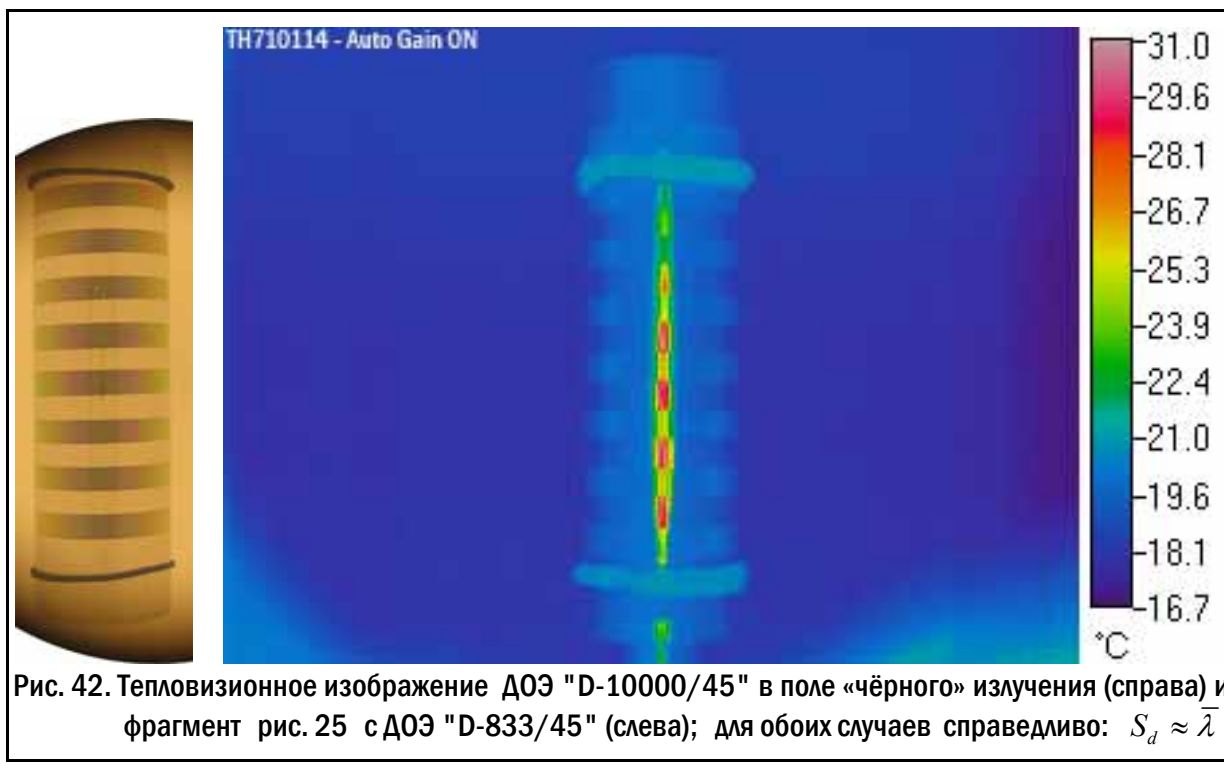

**Рис. 42.** Тепловизионное изображение ДОЭ "D-10000/45" в поле «чёрного» излучения (справа) и фрагмент рис. 25 с ДОЭ "D-833/45" (слева); для обоих случаев справедливо: $S_d \approx \overline{\lambda}$

Тёмная широкая симметричная область вдоль боковой поверхности "D-10000/45" аналогична видимой и на фотоснимке "D-833/45" при похожем[1] соотношении $S_d \approx \overline{\lambda}$. Яркая вертикальная полоса по оси опорного цилиндра является отражением излучения объектива тепловизора **NEC TH7102MX** (см. рис. 43), которое так и не удалось исключить.

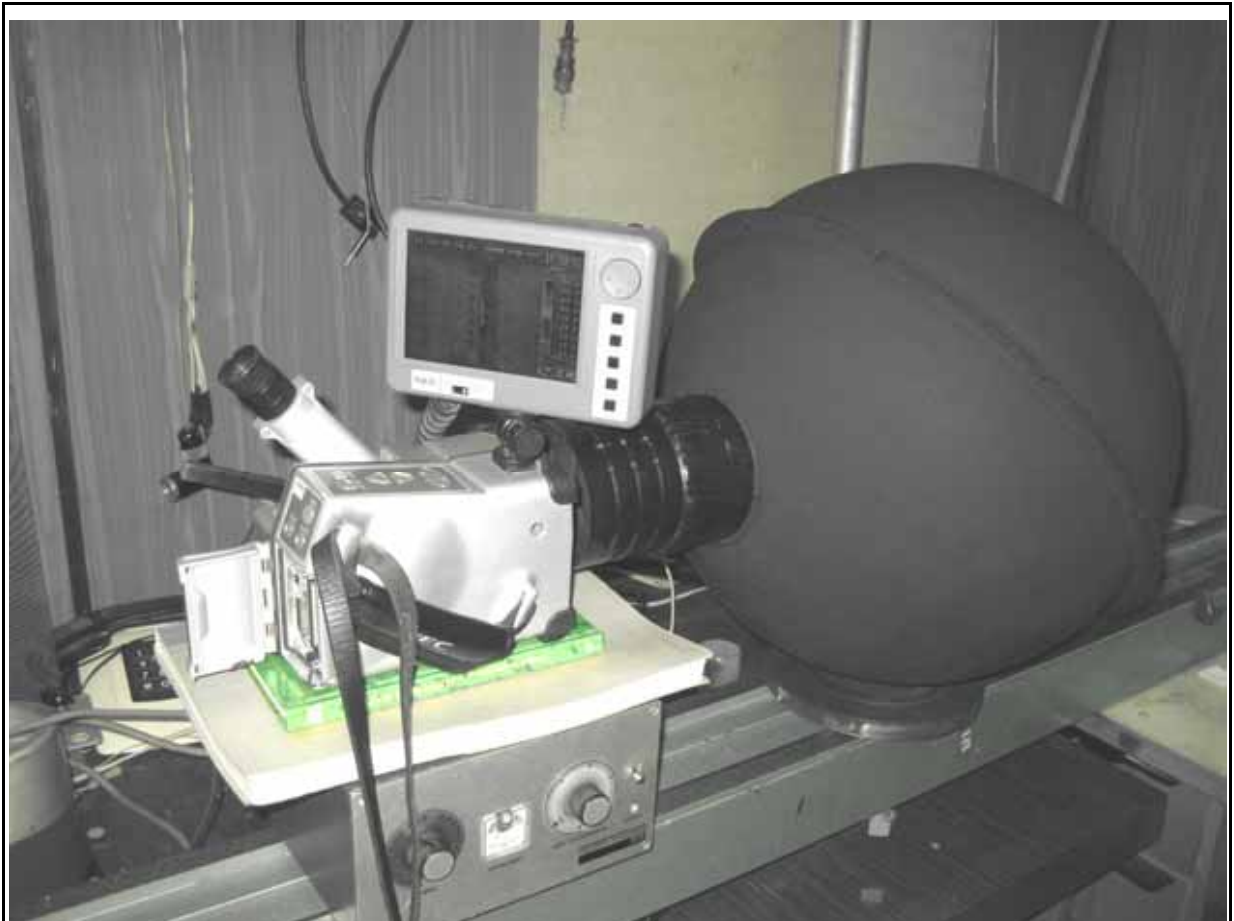

**Рис. 43. Внешний вид установки для изучения рассеяния планковского («чёрного») радиационного излучения поверхностью ДОЭ "D-10000/45" — на основе специальной камеры №4**

Однако все эффекты, выявленные при работе с ДОЭ "М-833/45" и "D-10000/45", крайне слабо выражены. Поэтому природа данных эффектов остаётся под вопросом[2].

На основании вышесказанного было решено в будущем сконцентрировать усилия на исследовании взаимодействия диффузных световых полей с квазисинусоидальными двумерными дифракционными решётками фазового типа (особенно важно для $S_d \approx \lambda$).

---

[1] Для уровня температуры, при котором проводился радиометрический эксперимент ($T \cong 290°K$), максимум плотности вероятности соответствует фотонам с $\lambda \approx 10$ мкм.

[2] Вероятно, что даже неоднородное излучение с планковским спектром, соответствующим определённой температуре, не способно создавать градиентные изображения на неохлаждаемой (т. е. с той же температурой) болометрической матрице тепловизора. По этой причине нельзя, например, концентрируя поток солнечного света с помощью параболического зеркала любого размера, нагреть помещённый в его фокус объект до температуры, превышающей спектральную температуру поверхности солнца: $6500°K$.

## Выводы

Эксперименты выявили значимое отклонение от закона Ламберта при дифракции диффузного фотонного газа на поверхности решётки, которое вызывало угловую анизотропию потоков излучения в изначально однородном световом поле. Эти результаты дают объективное основание для ревизии существующего понятия о наиболее вероятном макроскопическом состоянии замкнутой системы, поскольку очевидно, что такое состояние может не соответствовать определению термодинамического равновесия.

Действительно, в экспериментально исследованных квазизамкнутых фотометрических системах инициируемое наиболее вероятное макроскопическое состояние не может быть осуществлено на уровне детального равновесия: фотоны диффузного газа "прибывают" на поверхность дифракционной решётки с угловой плотностью вероятности, определяемой законом Ламберта, а покидают эту поверхность уже в соответствии с иной индикатрисой *изоэнергетического* (что важно) рассеяния. В итоге реализуется такое наиболее вероятное стационарное макросостояние фотонного газа, при котором его изначальная изотропность сменяется устойчивым дисбалансом встречных световых потоков вдоль некоторых направлений в геометрическом пространстве.

Это обусловлено тем, что фазовая траектория каждой квантовой частицы[1] не является непрерывной на уровне подпространства импульсов. Более того, такая траектория не может быть описана даже разрывной функциональной зависимостью, подчиняющейся условиям Дирихле: если интерпретировать каждый акт индетерминированного (дифракционного) рассеяния квантовой частицы, как разрыв первого рода в подпространстве импульсов её фазового $\mu$-пространства, то, например, однозначно определённому[2] чистому состоянию частицы до рассеяния будет соответствовать некое вероятностное множество состояний после рассеяния. Причём, согласно воззрениям копенгагенской школы, принципиально не существует каких-либо "скрытых параметров", позволяющих исключить указанную неоднозначность траекторной функции.

Описанная способность квантовых частиц «исчезать» и «появляться» в различных частях доступного им фазового пространства открывает возможность существования в этом пространстве *источников* и *стоков* фазовых траекторий, имеющих не одинаковую плотность в одних и тех же локальных участках фазового объёма. В частности, особый механизм изоэнергетического рассеяния фотонного газа на поверхности многомерной дифракционной решётки обеспечил устойчивое наличие ненулевой дивергенции потока фазовых траекторий в конкретных частях фазового пространства системы. Это делает такую систему неэргодичной[3] и не сообразной с аксиоматикой статистической физики.

---

[1] Применительно к квантовым частицам понятие фазовой траектории может быть сохранено путём его переопределения на основании теоремы Эренфеста, описывающей динамику центра тяжести нормированного к единице объёма «жидкости вероятности» с плотностью $\rho = |\Psi|^2$ [9]. При этом обычно говорят не столько о *линии* фазовой траектории, сколько о *траекторной трубке*, ширина которой зависит от вероятностной «размытости» соответствующих сопряжённых параметров частицы.

[2] Разумеется, лишь с точностью до неопределённостей Гейзенберга.

[3] Свойство эргодичности предполагает достоверность микроканонической гипотезы статистической физики о тождественности результатов усреднения по времени и фазового усреднения при вычислении значений макроскопических параметров системы.

## Список использованных источников

1. ***Gaede W.*** The external friction of gases // Annalen der Physik, 1913, B. **41**, 289-336.

2. ***Epstein P. S.*** On the Resistance Experienced by Spheres in their Motion through Gases // Physical Review, 1924, v. **23**, 710-733.

3. ***Clausing P.*** Cosine law of reflection as a result of the second main theorem of thermodynamics // Annalen der Physik, 1930, B. **4**, 533-566.

4. ***Goodman F. O., Wachman H. Y.*** Dynamics of Gas-Surface Scattering // Academic Press, New York, 1976, 423 p.

    Имеется перевод: ***Гудман Ф., Вахман Г.*** Динамика рассеяния газа поверхностью. – М.: Мир, 1980. – 423 с.

5. ***Ramsey N. F.*** Molecular Beams // Clarendon Press, Oxford, 1956, 466 p.

    Имеется перевод: ***Рамзей Н.*** Молекулярные пучки. – М.: Издательство иностранной литературы, 1960. – 411 с.

6. ***Савуков В. В.*** Уточнение аксиоматических принципов статистической физики. // Деп. ВИНИТИ № **1249-В2004,** – 177 с.[1]

7. ***Wood R.W.*** On a remarkable case of uneven distribution of light in a diffraction grating spectrum // Philosophical Magazine, 1902, v. **4**, 396-402.

8. ***Lord Rayleigh.*** On the Dynamical Theory of Gratings // Proceedings of the Royal Society of London, 1907, Series A 79, 399-416.

9. ***Louis de Broglie.*** Heisenberg's Uncertainties and the Probabilistic Interpretation of Wave Mechanics with Critical Notes of the Author // Kluwer Academic Publishers, Dordrecht and London, 1990, 302p.

    Имеется перевод: ***Луи де Бройль.*** Соотношения неопределённостей Гейзенберга и вероятностная интерпретация волновой механики. – М.: Мир, 1986. – 340 с.

10. ***Savukov V.V.*** Breakdown of the isotropy of diffuse radiation as a consequence of its diffraction at multidimensional regular structures // Journal of Optical Technology, 2010, v. **77**, 74-77.[2]

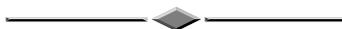

---

[1] URL: http://www.savukov.ru/viniti_1249_b2004_full_rus.pdf

[2] URL: http://arxiv.org/abs/1003.0706